# Electro-Chemo-Mechanical Modeling of Solid-State Batteries

Ting Hei Wan[a], and Francesco Ciucci[a,b,†]

[a] Department of Mechanical and Aerospace Engineering, The Hong Kong University of Science and Technology, Hong Kong, China

[b] Department of Chemical and Biological Engineering, The Hong Kong University of Science and Technology, Hong Kong, China

[†]Corresponding author: francesco.ciucci@ust.hk

# Abstract

Solid-state batteries (SSBs) have recently been proposed as promising alternatives to conventional Li-ion batteries because of their high level of safety and power density. In this article, continuum-scale simulations are chosen as the modeling framework to study such properties. A comprehensive continuum model is constructed for the simulation of the electro-chemo-mechanical (ECM) response of an SSB that resolves the bulk transportation of charged species and their interfacial transfer kinetics. It also studies the formation of space charge layers (SCLs) at interfaces and the development of interfacial stresses. The results suggest that the SCLs and the charge transfer kinetics are intertwined. The emergence of the SCLs and the depletion of reactants increases the charge transfer overpotential. We have also studied the coupling between electrochemistry and mechanics at interfaces, the results of which indicate that the strong electric fields originating at interfaces yield significant stresses. We, thereby, highlight the necessity of considering the ECM coupling in the SCLs when modeling an SSB.

# 1. Introduction

Driven by the demand for storing energy from renewable energy sources, the research and industrial interest on the next generation of batteries has been growing substantially in recent years [1]. In particular, solid-state batteries (SSBs) have gained significant attention [2, 3]. An SSB has a solid ionic conductor as its electrolyte. The utilization of a solid electrolyte (SE) intrinsically mitigate the safety issues associated with the flammable organic liquids typically used as electrolytes in conventional Li-ion batteries [4-6]. Various authors have also suggested that SEs may suppress the formation of Li dendrites at the anode|electrolyte interface, and prevent the short-circuiting of the battery [7-9]. Although such a claim has been recently challenged [10-12], the potential of avoiding Li dendrite growth with SEs remains attractive for the use of high capacity Li metal anodes [13-15]. Another attractive feature of SSBs is that they can be manufactured by bipolar stacking, a method that can significantly increase the energy density [14-16].

Continuum-scale simulations are the method chosen for rationally designing SSBs [17-22]. One notable article written by the Notten group [17] identified the most significant overpotential losses within thin-film SSBs including 1) the charge transfer at the interface; 2) the mass-transport in the electrolyte; and 3) the diffusion in the cathode. In a later article, the same group extended this framework to include thermal effects [19]. Based on these articles, 2D [23] and 3D [24] models have been developed to investigate the impact of the geometrical configuration on the discharge potential, Li concentration, and heat generation. However, the early work of the Notten group [17] assumed the SE to be neutral, where the ionization of lattice Li into mobile Li ions and fixed negative charges activates the Li transportation within the SE. While this assumption may hold for $Li_3PO_4$, the electrolyte that the Notten group employed, it is not necessarily correct for other SEs, especially if space charge layers (SCLs) form at the interfaces.

Indeed, interfaces have been regarded as one of the major bottlenecks of the SSB technology [25, 26]. In particular, the poor electrochemical and mechanical compatibility between SE and electrodes is a significant challenge [25, 26]. For example, during operation, SCLs form due to the sharp electric potential variation established at each electrode|electrolyte interface. The SCLs result in the accumulation/depletion of charged species [27], altering the material's local composition [25, 28]. Also, the accumulation/depletion of charged species limits the concentration of sites that can take part in charge transfer reactions across the electrode|electrolyte interfaces. This accumulation/depletion can lead to lower exchange current densities and increased charge transfer resistances [27]. The two sides of the electrode|electrolyte interfaces may also be mechanically incompatible [26, 29]. In particular, microscopic gaps can be present at the interfaces [30, 31]. Additionally, variation in the concentration of defects during operation may induce volumetric changes near interfaces. These concentration changes may, in turn, yield substantial stresses and lead to the delamination of the electrode from the SE [32].

The issues mentioned above highlight the importance of studying SCLs from an electro-chemo-mechanical (ECM) point of view. To model the SCLs, Braun *et al*. [20] developed a thermodynamically consistent formalism. These authors derived a semi-analytical solution to the SE with blocking electrodes based on rational thermodynamics [33]. To understand the characteristic of the SCLs, Braun, and co-authors parametrically studied the influence of the applied potentials and dielectric susceptibilities by including hydrostatic pressures into their formulation. However, their model considered a blocking electrode configuration only and did not take into account the impact of discharge currents on the SCLs. Landstorfer *et al*. [21, 34] investigated the evolution of the SCLs during discharge by assuming a constant Li current and by including a Stern-layer type of boundary condition at the interface. Nonetheless, the model of Landstorfer *et al*. did not link the evolution of the SCLs with the variation of electric potential due to charge transfer. Such a connection has been suggested in the literature of conventional Li batteries [35, 36].

Moreover, the works of Braun *et al*. and Landstorfer *et al*. focused primarily on the SCL in the electrolyte side but did not match that SCL to the one developed in the electrode side of the interface. In this regard, de Klerk and Wagemaker computed the SCL profile in both electrode and electrolyte and quantified the interfacial resistance [37]. However, they did not link the SCL to the interfacial kinetics nor discuss the impact of stresses.

While ECM coupling has long been studied in solid mechanics [38, 39], only a handful of works have studied SSBs. Two important articles were recently published by the Carter and Chiang groups [40, 41]. These authors considered the interaction between chemistry and mechanics in a 2D composite electrode composed of randomly distributed Si particles embedded in a rigid SE matrix. The model quantified the chemically induced stresses occurring during cycling. These stresses are crucial because they can cause loss of contact, capacity fading, and the formation of microcracks in the host matrix [40, 41]. However, the authors did not study the impact of SCLs and stresses on the interfacial charge transfer and other electrochemical properties.

## 1.1 Objectives and Outline of the Work

The formation of SCLs in electrodes, the impact of charge transfer kinetics on the SCLs characteristics, and the stresses exerted on the interface due to the SCL are yet to be discussed in the literature. The goal of this article is to bridge this gap by developing a comprehensive continuum-level model that satisfies fundamental laws including 1) charge and mass conservation; 2) Gauss's law (Maxwell's $1^{st}$ law); and 3) conservation of linear momentum (Newton's $2^{nd}$ law).

In section 2, we formulate the general theoretical framework used in this article. We start with non-equilibrium thermodynamics to handle Li transport and comply with Gauss's law. After that, the necessary governing equations and constitutive relations are introduced to resolve the mechanical stresses within SSBs under linear mechanics. Moreover, we derive equations describing the rates at which the interfacial reaction occur at the electrode|electrolyte interfaces. This general theoretical framework allows us to develop sub-models that investigate specific aspects of the SSB interfacial electrochemistry.

In section 3, we specialize the general framework to be electroneutral, where the local charge is zero everywhere. Consequently, we omit, as Danilov et al. [17] and Fabre et al. [18] did, the SCLs formation at the interfaces. This electroneutral model allows us to compute the Li distribution in the cathode and the discharge potentials.

In section 4, we extend the modeling framework by relaxing the assumption that electroneutrality holds at interfaces. Specifically, the governing equations are applied to the two adjacent SCL domains that characterize interfaces. We also introduce the necessary matching conditions for each interface and discuss how the charge transfer equations need to be reformulated for SCLs. Such a model allows the investigation of the evolution of concentrations and electric potentials in SCLs during discharge and the evaluation of the impact of SCLs on the overpotentials.

In section 5, we include mechanics into the model by adding a continuum-level mechanical equilibrium equation to the one developed in section 4. This approach allows us to quantify the stresses exerted at the interfaces due to the accumulation/depletion of charge and the presence of an electric field.

After discussing the model formulation, we solve, in section 6, the three specific models (*i.e.* the electroneutral model, the non-electroneutral model, and the ECM model) developed. Principally, we apply the electroneutral model to fit the experimental discharge curves and identify the key terms contributing to the overpotentials. Starting from the results of the electroneutral model, we utilize the non-electroneutral model to study the SCLs profile at the electrode|electrolyte interfaces. SCLs formed at the interfaces depend on and affect the charge transfer equations as concentration profiles are modified. Specifically, the charge transfer overpotentials increase if SCLs are included in the models. Furthermore, we model the stresses at the electrolyte|cathode interface and study the factors that control them. Our model suggests that stresses perpendicular to the interface are primarily due to the electric fields in the SCLs. Conversely, in-plane stresses result from the combination of electric fields and compositional contraction/expansion due to charge accumulation/depletion.

This article develops a comprehensive modeling framework for the ECM coupling at interfaces and the connection with SCLs. We emphasize that such a framework is broad and flexible. It can be expanded to study other interfacial phenomena taking place in SSBs, including electrochemical and mechanical instabilities [42, 43], and the interplay among stresses, electrochemical reactions, and dendrites [44].

## 2. General Model
### 2.1 Transport of Charged Species

The SSB modeled consists of a Li metal anode, an SE, and a metal oxide cathode. For simplicity, it is assumed that there are no passivation layers at the electrode|electrolyte interfaces, though the formation of such layers has been suggested in the literature [43]. In other words, the only interfacial reaction we will model is the insertion and de-insertion of Li. The bulk transport of charged species is modeled using Poisson-Nernst-Planck (PNP) equations [45, 46]:

$$z_i e \frac{\partial c_i}{\partial t} + \nabla \cdot \boldsymbol{j}_i = 0 \qquad (1a)$$

$$\boldsymbol{j}_i = -\sigma_i \nabla \tilde{\mu}_i^* = -\sigma_i \nabla (\mu_i^* + \phi) \tag{1b}$$

$$-\varepsilon_r \varepsilon_0 \nabla^2 \phi = \sum_i z_i e c_i \tag{1c}$$

where $\phi$ is the electric potential, $e$ is the electron charge, and $\mu_i^*$ and $\tilde{\mu}_i^*$ are, as defined below, the reduced chemical and electrochemical potential, respectively, of the mobile charged species $i$. The species $i$ is characterized by a concentration $c_i$, a charge $z_i e$, and a conductivity $\sigma_i$. The $\sigma_i$ is obtained from the Einstein relation, *i.e.*, $\sigma_i = \frac{D_i (z_i e)^2 c_i}{k_B T}$, where $D_i$ is the diffusion coefficient of $i$. Further, the current density of the species $i$ is $\boldsymbol{j}_i = z_i e \boldsymbol{J}_i$, where $\boldsymbol{J}_i$ is its flux. Additionally, $\varepsilon_r$ and $\varepsilon_0$ are the relative and vacuum permittivities, respectively.

The chemical potential of species $i$ takes the form [47-49]:

$$\mu_i = \mu_i^0 + k_B T \ln(\gamma_i(\tilde{c}_i) \tilde{c}_i) \tag{2}$$

where $\mu_i^0$ is the standard potential, $\gamma_i(\tilde{c}_i)$ is the activity coefficient, and $\tilde{c}_i = c_i / c_i^0$ is the normalized concentration of species $i$ with $c_i^0$ being the standard concentration. To include the electrical energy contribution to the chemical potential one needs to add a $z_i e \phi$ term to (2) [50] to obtain the electrochemical potential, *i.e.*,

$$\tilde{\mu}_i = \mu_i + z_i e \phi \tag{3}$$

A mechanical energy contribution can also be included in the electrochemical potential. This is done by adding a $-\Omega \sigma_h$ term to (3), where $\Omega$ is the partial molar volume and $\sigma_h$ is the hydrostatic stress [51]. This gives the ECM potential, *i.e.*,

$$\tilde{\mu}_{i,\text{ECM}} = \mu_i + z_i e \phi - \Omega \sigma_h \tag{4}$$

The reduced chemical, electrochemical, and ECM potentials of species $i$ are then given by $\mu_i^* = \mu_i / z_i e$, $\tilde{\mu}_i^* = (\mu_i + z_i e \phi)/z_i e$, and $\tilde{\mu}_{i,\text{ECM}}^* = (\mu_i + z_i e \phi - \Omega \sigma_h)/z_i e$, respectively.

The concentration of species near interfaces may vary significantly due to the formation of SCLs. Under these circumstances, interactions among defects, as well as the limited availability of lattice sites (*i.e.* site exclusion), need to be considered [52]. In this regard, we can write $\mu_i$ to be the difference between the chemical potential of a filled phase $\mu_{i,F}$ and a vacant phase $\mu_{i,V}$, *i.e.*,

$$\mu_i = \mu_{i,F} - \mu_{i,V} = \left(\mu_{i,F}^0 + k_B T \ln \tilde{c}_i\right) - \left(\mu_{i,V}^0 + k_B T \ln(\beta_i - \tilde{c}_i)\right)$$
$$= \mu_{i,F}^0 - \mu_{i,V}^0 + k_B T \ln\left(\frac{\tilde{c}_i}{\beta_i - \tilde{c}_i}\right) \tag{5}$$

where $\beta_i = c_i^{max}/c_i^0$ ($c_i^{max}$ is the maximum concentration of species $i$) [52].

While the equations (1) to (5) outlined above are general, we will use specific defect types in our SSB model. For the SE, we will assume that the transportation of Li ions occurs by a vacancy-assisted hopping mechanism [3, 53]. We shall, therefore, assume that Li vacancies, with a concentration $c_v$, are the only mobile defect in the SE. As for the cathode, we will track Li ions and electrons (or holes). The concentration of Li ions and Li atoms are denoted by $c_{Li^+}$ and $c_{Li}$, respectively. Also, the concentration of electrons and holes are denoted by $c_-$ and $c_h$, respectively. In addition, we will consider two global Li transfer interfacial reactions. One such reaction takes place at the electrolyte|cathode interface, *i.e.*,[1]

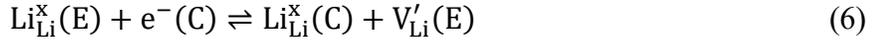
$$\text{Li}_{Li}^x(E) + e^-(C) \rightleftharpoons \text{Li}_{Li}^x(C) + V'_{Li}(E) \tag{6}$$

The other occurs at the anode|electrolyte interface, *i.e.*,

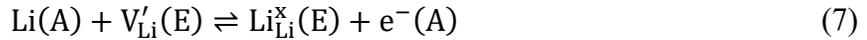
$$\text{Li}(A) + V'_{Li}(E) \rightleftharpoons \text{Li}_{Li}^x(E) + e^-(A) \tag{7}$$

In the last two equations, the Kröger-Vink notation [54] is used, and, E, C, and A denote electrolyte, cathode, and anode phases, respectively. The kinetics corresponding to these two interfacial reactions will be discussed in subsection 2.3.

## 2.2 Mechanics

We note that the accumulation and migration of charged species significantly modify the local volume of materials. Since the casing constrains the volume of the battery, significant mechanical stresses are generated. Moreover, the strong electric fields in the SCLs may lead to body forces that need to be compensated by stresses. Therefore, in addition to the transportation of charged species, our model needs to include a force balance equation. Under the hypothesis of static mechanical equilibrium, we can write that [55]

$$\text{div } \boldsymbol{\sigma} = \left(\sum_i z_i e c_i\right) \nabla \phi \tag{8}$$

where $\boldsymbol{\sigma}$ is the stress tensor. If we insert (1c) into (8), we obtain

$$\text{div } \boldsymbol{\sigma} = -(\varepsilon_r \varepsilon_0 \nabla^2 \phi) \nabla \phi \tag{9}$$

---

[1] Common cathode materials may either be p-type semi-conductors (such as LiCoO$_2$ at a high state of charge or LiFePO$_4$) or metallic conductors (such as LiCoO$_2$ at a low state of charge). For cathode materials that are p-type semi-conductors, the electrolyte|cathode interfacial reaction is $\text{Li}_{Li}^x(E) \rightleftharpoons \text{Li}_{Li}^x(C) + V'_{Li}(E) + h^\bullet(C)$. Without a loss of generality, we use (6) in the article.

For simplicity, it is assumed that the strain is small, and no plastic deformation occurs in the battery. Therefore, we can write, under the hypotheses typical of linear elasticity, that the stress tensor takes the following form [51]:

$$\sigma_{mn} = 2G\epsilon_{mn} + \left(\kappa \epsilon_{kk} - \zeta(c_i - c_i')\right)\delta_{mn} \tag{10}$$

where $G = \frac{Y}{2(1+v)}$, $\kappa = \frac{2vG}{(1-2v)}$, and $\zeta = \frac{\Omega(3\kappa+2G)}{3}$. In these expressions, Y is Young's modulus, $v$ is the Poisson's ratio, $c_i$ is the concentration of species $i$ and $c_i'$ is its stress-free value. Also, $\delta_{mn}$ is the Kronecker delta ($\delta_{mn} = 1$ if $m = n$ and 0 otherwise). Moreover, $\epsilon_{mn}$ is the "small displacement" strain tensor defined as

$$\epsilon_{mn} = \frac{1}{2}\left(\frac{\partial u_m}{\partial x_n} + \frac{\partial u_n}{\partial x_m}\right) \tag{11}$$

where $\boldsymbol{u} = [u_1, u_2, u_3] = [u_x, u_y, u_z]$ and $x_1 = x$, $x_2 = y$, and $x_3 = z$.

## 2.3 Interfacial Reaction Kinetics

In this section, we write the equations that model the reaction kinetics at the electrode|electrolyte interface. The formulation is largely based on non-equilibrium kinetic models used for LIBs [48, 56, 57]. Subsection 2.3.1 outlines the mechanism of interfacial reaction taking place at the electrolyte|cathode interface, which is also linked to the Li current. Similar expressions are also developed for the anode|electrolyte interface, see subsection 2.3.2.

### 2.3.1 Electrolyte|Cathode Interface

The energy landscape for Li insertion at the electrolyte|cathode interface, *i.e.*, reaction (6), is shown schematically in Figure 1. The left and right convex curves indicate the Gibbs free energy

before and after insertion, respectively. To allow for Li insertion into the cathode, a driving force needs to be applied to offset the system from its thermodynamic equilibrium. This requires the left-hand curve to shift upward to favor the forward reaction in (6). Conversely, during charge the reverse reaction in (6) needs to be favored. Therefore, the left-hand curve has to shift downwards. The magnitude of the shift is given by the electrochemical potential difference [57], *i.e.*,

$$\Delta g_C = \mu_{Li}(E) + \tilde{\mu}_-(C) - \tilde{\mu}_v(E) - \mu_{Li}(C) \tag{12}$$

The $\Delta g_C$ can be divided into two parts [50]:

$$\Delta g_C = \Delta g_{E \to C} - \Delta g_{C \to E} \tag{13}$$

where $\Delta g_{E \to C}$ corresponds to the forward reaction (*i.e.* Li jumps from the electrolyte to the cathode) and $\Delta g_{C \to E}$ to the reverse reaction (*i.e.* Li hops from the cathode to the electrolyte). We note that, as illustrated in Figure 1, $\Delta g_{E \to C}$ and $\Delta g_{C \to E}$ can be written as

$$\Delta g_{E \to C} = -\Delta g_{E \to C}^{\ddagger} + \alpha_C \Delta g_C \tag{14a}$$

$$\Delta g_{C \to E} = -\Delta g_{C \to E}^{\ddagger} - (1 - \alpha_C)\Delta g_C \tag{14b}$$

where $\alpha_C$ is the symmetry coefficient and $\Delta g_{E \to C}^{\ddagger}$ and $\Delta g_{C \to E}^{\ddagger}$ are the reaction barriers of the forward and reverse reactions. It should be noted that, as shown in Figure 1, $\Delta g_{E \to C}^{\ddagger} = \Delta g_{C \to E}^{\ddagger} = \Delta g_C^{\ddagger}$. To simplify the notation especially in later derivations, we will take $\frac{1}{\gamma_C^{\ddagger}} = \exp\left(-\frac{\Delta g_C^{\ddagger}}{k_B T}\right)$, where $\gamma_C^{\ddagger}$ is the activity coefficient of the transition-state. We should note that $\gamma_C^{\ddagger}$ can be determined from transition-state theory [57, 58].

The net Li-ion flux across the electrolyte|cathode interface is given by

$$J_{Li} = J_{E \to C} - J_{C \to E} \tag{15}$$

where $J_{E \to C}$ (or $J_{C \to E}$) is the flux of Li atoms from the electrolyte (cathode) to the cathode (electrolyte). $J_{E \to C}$ and $J_{C \to E}$ can be written as [48, 59]

$$J_{E \to C} = \frac{J_{E \to C}^0}{\gamma_C^{\ddagger}} \exp\left(\frac{\alpha_C \Delta g_C}{eV_{th}}\right) \tag{16a}$$

$$J_{C \to E} = \frac{J_{C \to E}^0}{\gamma_C^{\ddagger}} \exp\left(-\frac{(1 - \alpha_C)\Delta g_C}{eV_{th}}\right) \tag{16b}$$

where $V_{\text{th}} = \frac{k_B T}{e}$ is the thermal voltage, and $J^0_{E \to C}$ and $J^0_{C \to E}$ are two pre-exponential terms that depend on the concentrations of the defects at the interface.

As illustrated in the schematic in the top portion of Figure 1, the insertion of a Li atom into the cathode happens if a lattice Li from the electrolyte side of the interface jumps to a nearby empty Li site in the cathode. We note that the hopping rate of Li from the electrolyte to the cathode is proportional to 1) the Li concentration in the electrolyte; and 2) the availability of empty sites in the cathode. Similarly, the hopping rate of Li from the cathode to the electrolyte is proportional to 1) the Li concentration in the cathode; and 2) the availability of empty sites (*i.e.* Li vacancies) in the electrolyte. Therefore, $J^0_{E \to C}$ and $J^0_{C \to E}$ can be written as

$$J^0_{E \to C} = k_{E \to C} \left( c_v^{\max}(E) - c_v(E) \right) \left( c_{\text{Li}}^{\max}(C) - c_{\text{Li}}(C) \right) \tag{17a}$$

$$J^0_{C \to E} = k_{C \to E} c_{\text{Li}}(C) c_v(E) \tag{17b}$$

where $c_v^{\max}(E)$ and $c_{\text{Li}}^{\max}(C)$ are the maximum number of Li sites that Li ions can occupy in the electrolyte and cathode, respectively. It should be noted that if a Li site is not occupied, a Li vacancy is present in that site. Further, since the total number of Li sites in the electrolyte and cathode are constants, the sum of Li ions and Li vacancies in each domain are also constant, *i.e.*, $c_{\text{Li}}(E) + c_v(E) = c_v^{\max}(E)$ and $c_{\text{Li}}(C) + c_v(C) = c_{\text{Li}}^{\max}(C)$.

By substituting (15) with (16), we obtain that

$$J_{\text{Li}} = \frac{J^0_{E \to C}}{\gamma_C^{\ddagger}} \exp\left( \frac{\alpha_C \Delta g_C}{e V_{\text{th}}} \right) - \frac{J^0_{C \to E}}{\gamma_C^{\ddagger}} \exp\left( -\frac{(1-\alpha_C) \Delta g_C}{e V_{\text{th}}} \right) \tag{18}$$

The latter can further be simplified to

$$J_{\text{Li}} = J^0_C \left( \exp\left( \frac{\alpha_C \Delta g_C^*}{V_{\text{th}}} \right) - \exp\left( -\frac{(1-\alpha_C) \Delta g_C^*}{V_{\text{th}}} \right) \right) \tag{19}$$

where $J^0_C$ and $\Delta g_C^*$ are given by

$$J^0_C = \frac{(J^0_{E \to C})^{\alpha_C} (J^0_{C \to E})^{1-\alpha_C}}{\gamma_C^{\ddagger}} \tag{20a}$$

$$\Delta g_C^* = \mu_{\text{Li}}^*(E) - \tilde{\mu}_-^*(C) + \tilde{\mu}_v^*(E) + V_{\text{oc}} \tag{20b}$$

$V_{\text{oc}}$ in (20b) is the open-circuit voltage, which is a function of the concentration of Li in the cathode [18, 60, 61], *i.e.*,

$$V_{oc} = V_{oc}(c_{Li}(C)) \qquad (21)$$

More details regarding the derivation can be found in the Supplementary Information (SI), section A.

### 2.3.2 Anode|Electrolyte Interface

The energy landscape encountered by the Li atoms while jumping from the anode to the electrolyte, *i.e.*, reaction (7), is the reverse of that schematically shown in Figure 1. To remove Li from the anode and place it on the electrolyte, the following "driving force" is needed:

$$\Delta g_A = \tilde{\mu}_v(E) + \mu_{Li}(A) - \mu_{Li}(E) - \tilde{\mu}_-(A) \qquad (22)$$

As shown for the electrolyte|cathode interface, we can divide $\Delta g_A$ into a forward (anode to electrolyte) and a reverse (electrolyte to anode) component, *i.e.*,

$$\Delta g_A = \Delta g_{A \to E} - \Delta g_{E \to A} \qquad (23)$$

with

$$\Delta g_{A \to E} = -\Delta g^{\ddagger}_{A \to E} + \alpha_A \Delta g_A \qquad (24a)$$

$$\Delta g_{E \to A} = -\Delta g^{\ddagger}_{E \to A} - (1 - \alpha_A)\Delta g_A \qquad (24b)$$

where $\alpha_A$ is the symmetry coefficient and $\Delta g^{\ddagger}_{A \to E}$ and $\Delta g^{\ddagger}_{E \to A}$ are the activation energies for the forward and reverse reactions. As noted already in subsection 2.3.1, those two quantities can be assumed to be identical, *i.e.*, $\Delta g^{\ddagger}_A = \Delta g^{\ddagger}_{A \to E} = \Delta g^{\ddagger}_{E \to A}$.

Therefore, the net Li flux across the anode|electrolyte interface can be written as

$$J_{Li} = J_{A \to E} - J_{E \to A} \qquad (25)$$

Here, the Li metal anode is a reservoir with a constant Li concentration. Hence, the pre-exponential terms of $J_{A \to E}$ and $J_{E \to A}$ depend only on the concentration of Li ions in the electrolyte. Similarly, the flux from the anode to the electrolyte depends only on the concentration of empty sites (*i.e.* the concentration of Li vacancies) in the electrolyte. Therefore, we can write

$$J_{A \to E} = \frac{J^0_{A \to E}}{\gamma^{\ddagger}_A} \exp\left(\frac{\alpha_A \Delta g_A}{eV_{th}}\right) \qquad (26a)$$

$$J_{E \to A} = \frac{J^0_{E \to A}}{\gamma^{\ddagger}_A} \exp\left(-\frac{(1-\alpha_A)\Delta g_A}{eV_{th}}\right) \tag{26b}$$

with

$$J^0_{A \to E} = k_{A \to E} c_v(E) \tag{27a}$$

$$J^0_{E \to A} = k_{E \to A}\left(c^{max}_v(E) - c_v(E)\right) \tag{27b}$$

where $\frac{1}{\gamma^{\ddagger}_A} = \exp\left(-\frac{\Delta g^{\ddagger}_A}{k_B T}\right)$.

Similar to the case of the electrolyte|cathode interface, the reaction kinetics equation for the anode|electrolyte interface can be obtained by substituting (25) with (26) and can be written as

$$J_{Li} = \frac{J^0_{A \to E}}{\gamma^{\ddagger}_A} \exp\left(\frac{\alpha_A \Delta g_A}{eV_{th}}\right) - \frac{J^0_{E \to A}}{\gamma^{\ddagger}_A} \exp\left(-\frac{(1-\alpha_A)\Delta g_A}{eV_{th}}\right) \tag{28}$$

We note that (28) can be further simplified to obtain

$$J_{Li} = J^0_A \left( \exp\left(\frac{\alpha_A \Delta g^*_A}{V_{th}}\right) - \exp\left(-\frac{(1-\alpha_A)\Delta g^*_A}{V_{th}}\right) \right) \tag{29}$$

where $J^0_A$ and $\Delta g^*_A$ are given by

$$J^0_A = \frac{(J^0_{A \to E})^{\alpha_A} (J^0_{E \to A})^{1-\alpha_A}}{\gamma^{\ddagger}_A} \tag{30a}$$

$$\Delta g^*_A = -\mu^*_{Li}(E) + \tilde{\mu}^*_-(A) - \tilde{\mu}^*_v(E) \tag{30b}$$

The derivation of (19) and (29) are given in section A of the SI.

## 2.4 Discussion

The governing equations and constitutive relations of the general modeling framework detailed above are summarized in Table S.1. In the following sections (3 to 5), we will specialize the PNP, (1), the mechanical equilibrium, (8), and the reaction kinetics, (19) and (29), equations to three specific models of the galvanostatic discharge of SSBs. The first model (section 3) is electroneutral. While analogous models has been developed in the literature [18, 24], we show that it can be derived as a particular approximation of our more complete model. The second model (section 4) is non-neutral, and we will discuss in detail how to formulate mathematical deviations from electroneutrality in the SCLs. The third model (section 5) is the ECM model. We will show that the SCL formation leads to stresses at interfaces and significantly modifies the interfacial reaction rates. All models are developed in a 1D geometry with the conventions shown in Figure 2. The anode is assumed to be a 0D reservoir of Li, placed at $x = 0$. The anode|electrolyte interface and electrolyte|cathode interface are at $x = 0$ and $x = L_\text{E}$, respectively ($L_\text{E}$ is the thickness of the electrolyte). The cathode ends at $x = L_\text{E} + L_\text{C}$ ($L_\text{C}$ is the thickness of the cathode).

## 3. Electroneutral Model

Typically, the thickness of the SE is between several hundreds and tens of thousands of nanometers [17, 62, 63]. However, the SCL's thickness is comparable to the lattice parameter [64]. In other words, the electrolyte is much thicker than the SCLs adjacent to each interface. Therefore, it can be safely assumed that the concentration of charged species does not change in the bulk of the electrolyte [65]. It follows that (1a) and (1c) can be omitted, and the bulk of the electrolyte can be treated as an Ohmic resistor characterized by a constant electric field or equivalently a linearly varying electric potential.

As for the bulk of the cathode, we note that common cathode materials can either be metallic or p-type semiconductors. For example, $\text{Li}_y\text{CoO}_2$ is metallic for $y_\text{Li} < 0.75$ [66, 67] and is a p-type semi-conductor for $y_\text{Li} > 0.75$ [66, 67], where $y_\text{Li} = \frac{c_\text{Li}}{c_\text{Li}^\text{max}}$ is Li concentration with respect to its maximum allowed value. To simplify the mathematical exposition, we consider Li ions and electrons to be the two charged species diffusing within the bulk of the cathode. Due to electroneutrality, the local concentration of Li ions is equal to that of electrons, *i.e.*, $c_{\text{Li}^+} = c_-$, implying that the right-hand side of (1c) equals zero. Then, it is sufficient to solve the boundary value problem formulated on equations (1a) and (1b), in order to describe the Li transport [65]. Moreover, the electric field in the cathode is also constant as it is in the electrolyte. The electric potential drop in the cathode can then be modeled as an Ohmic loss. In order to estimate the cell potential for this electroneutral model, we first solve for the transport of Li in the cathode domain. Doing so allows us to determine the Li concentration. Then, we evaluate: 1) the potential drop due to the charge transfer at the two electrode|electrolyte interfaces; 2) the open-circuit voltage, which depends on the concentration of Li in the cathode; and 3) the Ohmic losses in the electrolyte and the cathode.

## 3.1 Li Transportation in the Cathode

Due to electroneutrality and the fact that $c_{Li^+} = c_- = c_{Li}$, one only needs to solve one governing equation that describes the Li transport in the cathode, i.e.,

$$\frac{\partial c_{Li}}{\partial t} + \frac{\partial J_{Li}}{\partial x} = 0 \tag{31}$$

where $J_{Li} = \frac{j_{Li^+}}{e} = -\frac{\sigma_{Li^+}}{e}\frac{\partial \tilde{\mu}^*_{Li^+}}{\partial x}$ is the mass flux of Li. It is shown in section B of the SI that (31) can further be simplified to

$$\frac{\partial c_{Li}}{\partial t} + \frac{\partial}{\partial x}\left(-\tilde{D}_{chem}\frac{\partial c_{Li}}{\partial x}\right) = 0 \tag{32}$$

where $\tilde{D}_{chem}$ is the chemical diffusivity of the cathode and is a function of $c_{Li}$ [48]. (32) is equipped with a constant flux boundary condition at the electrolyte|cathode interface (i.e. $x = L_E$) and a blocking boundary condition at the other end of the cathode (i.e. $x = L_E + L_C$). In other words, we have

$$-\tilde{D}_{chem}\frac{\partial c_{Li}}{\partial x}\bigg|_{x=L_E^+} = -J_{Li} \tag{33a}$$

$$-\tilde{D}_{chem}\frac{\partial c_{Li}}{\partial x}\bigg|_{x=L_E+L_C} = 0 \tag{33b}$$

$J_{Li}$ in (33a) is a constant under galvanostatic discharge. It should also be noted that $J_{Li}$ depends on the charge transfer rates as $j_{Li^+}$ does, see (19). Such a dependence will be further discussed in the following section.

## 3.2 Charge Transfer Kinetics

### 3.2.1 Electrolyte|Cathode Interface

The general charge transfer equations illustrated in subsection 2.3 can be recast in a conventional Butler-Volmer form if we assume that only the electric potential varies as a result of departure from equilibrium [59]. In such a case, the charge transfer current density $j_{Li^+} = eJ_{Li}$ is just a function of the (step) electric potential variation across each interface (see section C.1 of the SI for details). In particular, (19) can be rewritten as

$$j_{\text{Li}^+} = j_C^0 \left( \exp\left(\frac{\alpha_C \eta_C}{V_{\text{th}}}\right) - \exp\left(-\frac{(1-\alpha_C)\eta_C}{V_{\text{th}}}\right) \right) \qquad (34)$$

where $\eta_C = \phi_{E|C}(L_E^-) - \phi_C(L_E^+) + V_{oc}\left(c_{\text{Li}}|_{x=L_E^+}\right)$ is the charge transfer overpotential at the electrolyte|cathode interface. In $\eta_C$, $\phi_{E|C}$ and $\phi_C$ are the electric potentials in the electrolyte and cathode, respectively. For the electroneutral model, the concentration of Li vacancies in the bulk of the electrolyte is a constant, i.e., $c_v(L_E^-) = c_v^s = c_v^0$. We also use the same notation found in the SSB model literature [17-21], by defining $y_{\text{Li}}(t,x) = \frac{c_{\text{Li}}(t,x)}{c_{\text{Li}}^{\max}}$ and $y_{\text{Li}}^s = y_{\text{Li}}(t, L_E^+)$, where $y_{\text{Li}}$ corresponds to the local Li fraction with respect to the maximum Li concentration in $\text{Li}_y\text{CoO}_2$. Therefore, by utilizing the fact that $j_C^0 = eJ_C^0$ and substituting $c_v(E)$ and $c_{\text{Li}}(C)$ in $J_C^0$, see (20), with $c_v^0$ and $c_{\text{Li}}^{\max} y_{\text{Li}}^s$, respectively, we may set the exchange current density $j_C^0$ in (34) as

$$j_C^0 = (k_C^0)_{EN} (y_{\text{Li}}^s)^{1-\alpha_C} (1 - y_{\text{Li}}^s)^{\alpha_C} \qquad (35)$$

where $(k_C^0)_{EN} = \frac{k_{E \to C}^{\alpha_C} k_{C \to E}^{1-\alpha_C} c_{\text{Li}}^{\max} e}{\gamma_C^{\ddagger}} (c_v^0)^{1-\alpha_C} (c_v^{\max} - c_v^0)^{1-\alpha_C}$ is the reaction constant for the electrolyte|cathode interface.

### 3.2.2 Anode|Electrolyte Interface

The charge transfer equation can also be rewritten for the anode|electrolyte interface (29) using the conventional Butler-Volmer formalism (see section C.2 of the SI):

$$j_{\text{Li}^+} = j_A^0 \left( \exp\left(\frac{\alpha_A \eta_A}{V_{\text{th}}}\right) - \exp\left(-\frac{(1-\alpha_A)\eta_A}{V_{\text{th}}}\right) \right) \qquad (36)$$

where $\eta_A = \phi_A(0^-) - \phi_{E|A}(0^+)$ is the charge transfer overpotential across the anode|electrolyte interface. Since the concentration of Li vacancies in the electrolyte is a constant, $c_v(0^+) = c_v^0$. By utilizing the fact that $j_A^0 = eJ_A^0$ and substituting $c_v(E)$ in $J_A^0$ (see (30a)) with $c_v^0$, $j_A^0$ in (36) can be rewritten as

$$j_A^0 = (k_A^0)_{EN} \qquad (37)$$

where $j_A^0$ is a constant with $(k_A^0)_{EN} = \frac{k_{A \to E}^{\alpha_A} k_{E \to A}^{1-\alpha_A} e}{\gamma_A^{\ddagger}} (c_v^0)^{\alpha_A} (c_v^{\max} - c_v^0)^{1-\alpha_A}$.

### 3.3 Cell Voltage

The potential drop across the entire SSB cell is equal to the difference in the reduced electrochemical potential of electrons from the anode to the cathode, *i.e.*,

$$V_{\text{cell}} = \tilde{\mu}_-^*|_{x=(L_E+L_C)^+} - \tilde{\mu}_-^*|_{x=0^-} \qquad (38)$$

This last expression can be rewritten as a function of 1) the open circuit potential $V_{\text{oc}}$; 2) the overpotentials at the two interfaces (*i.e.* $\eta_C$ and $\eta_A$); and 3) the overpotentials due to the Ohmic losses over the electrolyte and cathode (*i.e.* $\eta_{R_E}$ and $\eta_{R_C}$, respectively). In other words, we can set

$$V_{\text{cell}} = V_{\text{oc}}\left(c_{\text{Li}}|_{x=L_E^+}\right) - \eta_C - \eta_A - \eta_{R_E} - \eta_{R_C} \qquad (39)$$

The terms on the right-hand side of the equation above are shown schematically in Figure 2 (a). In particular, $V_{\text{oc}}\left(c_{\text{Li}}|_{x=L_E^+}\right)$ is given by

$$V_{\text{oc}}\left(c_{\text{Li}}|_{x=L_E^+}\right) = -\frac{\mu_{\text{Li}}^0}{e} + V_{\text{th}} \ln\left(\gamma_{\text{Li}} \tilde{c}_{\text{Li}}|_{x=L_E^+}\right) \qquad (40)$$

The latter can be approximated by fitting the experimental open-circuit voltage [60]. Moreover, the Ohmic losses over the electrolyte and cathode, *i.e.*, $\eta_{R_E}$ and $\eta_{R_C}$, can be computed as

$$\eta_{R_E} = \frac{j_{\text{tot}} L_E}{\sigma_E} \qquad (41a)$$

$$\eta_{R_C} = \frac{j_{\text{tot}} L_C}{\sigma_C} \qquad (41b)$$

where $\sigma_E$ and $\sigma_C$ are the electrical conductivities in the electrolyte and cathode, respectively. $j_{\text{tot}}$ is the total discharge current density with $j_{\text{tot}} = j_{\text{Li}^+} + j_-$ in the bulk of the cathode.

In summary, we have used the general framework introduced in section 2 to formulate the electroneutral model. The relevant equations used in this model are summarized in Table S.2. The implementation to simulate the galvanostatic discharge voltage is described in section D.1 of the SI.

## 4. Non-electroneutral Model

The electroneutral model assumes step changes of concentrations and potentials across interfaces. However, this assumption is unphysical. In this section, we apply the general framework introduced in section 2 and model the sharp variations of concentrations and potentials taking place across the interfaces as being connected to deviations from neutrality in the SCLs [21]. Mathematically, accounting for deviations from neutrality in the SCLs amounts to solving a singular perturbation problem by dimensional rescaling, perturbation, and expansion [68].

Using singular perturbations, we can show that, while the SCLs are charge non-neutral, the bulk of the materials remains electroneutral. Therefore, equations (32) and (33) can still be used to model Li transportation in the bulk of the cathode. Furthermore, the bulk of the electrolyte has a constant concentration of Li vacancies and a constant electric potential gradient. It follows that the Ohmic loss equations (41) can still model the potential drop across the bulk of the electrolyte and cathode. However, the electroneutral model cannot be used for the interfaces, where the concentration and electric potential variations are significant. Figure 2 (b) summarizes schematically the non-electroneutral model and shows the SCL domains.

To analyze the SCL, we split the electrolyte|cathode interface into two sub-domains: $(L_\text{E} - \delta x_\text{E|C}, L_\text{E}]$ and $[L_\text{E}, L_\text{E} + \delta x_\text{C})$. Here, the length scales $\delta x_\text{E|C}$ and $\delta x_\text{C}$ are loosely the SCL thicknesses of the electrolyte and cathode sides of the electrolyte|cathode interface, respectively. Physically, these two parameters represent the length scales of variation of the physical properties (e.g. concentrations and electric potentials). Similarly, we can define domain $(-\delta x_\text{A}, 0]$ and $[0, \delta x_\text{E|A})$, which are the two SCL regions of the anode|electrolyte interface. It should be noted that $\delta x_\text{A}$ and $\delta x_\text{E|A}$ have a similar physical meaning as $\delta x_\text{E|C}$ and $\delta x_\text{C}$. While the indices C and A refer to cathode and anode, respectively, E|C and E|A denote the electrolyte side of the SCL facing the cathode and anode, respectively.

Since the scale of the SCL is much smaller than that of the bulk, the analysis is performed using singular perturbations [65, 69]. As a first step, the coordinate system is transformed by defining $X = \frac{x - L_\text{E}}{\lambda_\text{D}}$ for the electrolyte|cathode interface and $X = \frac{x}{\lambda_\text{D}}$ for the anode|electrolyte interface, where $\lambda_\text{D}$ is the Debye length of the corresponding SCL sub-domain[2]. The $x$ to $X$ coordinate is schematically shown in the bottom panel of Figure 2 (b). If we take $\lambda_\text{D} \ll \delta x_\text{C}$, $\lambda_\text{D} \ll \delta x_\text{E|C}$, $\lambda_\text{D} \ll \delta x_\text{A}$, and $\lambda_\text{D} \ll \delta x_\text{E|A}$, it follows that $\frac{\delta x_\text{C}}{\lambda_\text{D}} \to \infty$, $\frac{\delta x_\text{E|C}}{\lambda_\text{D}} \to \infty$, $\frac{\delta x_\text{A}}{\lambda_\text{D}} \to \infty$, and $\frac{\delta x_\text{E|A}}{\lambda_\text{D}} \to \infty$ and that $X \in$ $(-\infty, 0]$ or $X \in [0, \infty)$. In the following subsections, we will introduce the SCL model, which tracks the concentrations of the charged species and the electric potentials in the electrolyte, cathode, and anode. After that, we will discuss how these SCLs are matched across each interface.

---

[2] As shall be seen in later subsections, $\lambda_\text{D}$ is different for electrolyte, cathode, and anode.

## 4.1 Space Charge Layer in the Electrolyte

As mentioned in subsection 2.1, it is assumed that Li vacancy is the only mobile species in the electrolyte. Because of the strong electric fields, the concentration of Li vacancies is expected to vary significantly in the region adjacent to each electrode|electrolyte interface. Further, we note that the concentration of Li vacancies is limited by the total number of available lattice sites. Therefore, we will utilize (5) and write the reduced chemical potential of Li vacancies as

$$\tilde{\mu}_v^* = -\frac{\mu_v^0}{e} - V_{\text{th}} \ln \frac{\tilde{c}_v}{\beta_v - \tilde{c}_v} + \phi \qquad (42)$$

where $\beta_v = \frac{c_v^{\max}}{c_v^0}$ with the Li site concentration $c_v^{\max}$ and $c_v^0$ is the concentration of Li vacancies in the bulk of the electrolyte. The characteristic length scale of the SCL is given by the Debye length $\lambda_{\text{D,E}} = \sqrt{\frac{\varepsilon_0 \varepsilon_{r,\text{E}} k_B T}{e^2 c_v^0}}$, where $\varepsilon_{r,\text{E}}$ is the relative permittivity of the electrolyte [70]. The PNP equations (1) are used for modeling the electrolyte side of the SCL at the electrolyte|cathode interface, i.e., for $x \in (L_{\text{E}} - \delta x_{\text{E}|\text{C}}, L_{\text{E}}]$. In order to conduct the analysis, we non-dimensionalize (1) by taking $X_{\text{E}} = \frac{x}{\lambda_{\text{D,E}}}$, $\tilde{c}_v = \frac{c_v}{c_v^0}$, $\tilde{\phi} = \frac{\phi}{V_{\text{th}}}$, $\tilde{j}_v = \frac{j_v \lambda_{\text{D,E}}}{c_v^0 D_v e}$, and $\tilde{t} = \frac{t}{\lambda_{\text{D,E}}^2 / D_v}$, where $D_v$ is the diffusion coefficient of Li vacancies in the electrolyte. This leads to the following set of equations:

$$\frac{\partial \tilde{c}_v}{\partial \tilde{t}} + \frac{\partial \tilde{j}_v}{\partial X_{\text{E}}} = 0 \qquad (43\text{a})$$

$$\tilde{j}_v = -\tilde{c}_v \frac{\partial}{\partial X_{\text{E}}} \left( \frac{\tilde{\mu}_v^*}{V_{\text{th}}} \right) = -\tilde{c}_v \frac{\partial}{\partial X_{\text{E}}} \left( -\ln \frac{\tilde{c}_v}{\beta_v - \tilde{c}_v} + \tilde{\phi} \right) \qquad (43\text{b})$$

$$-\frac{\partial^2 \tilde{\phi}}{\partial X_{\text{E}}^2} = 1 - \tilde{c}_v \qquad (43\text{c})$$

We first assess the values of various physical parameters. We note that $j_v \sim 0.1 - 10$ mA/cm$^2$, $c_v^0 \sim 10^{27}$/m$^3$, $\lambda_{\text{D,E}} \sim 10^{-10} - 10^{-11}$ m, and $D_v \sim 10^{-12}$ m/s$^2$ [18]. This implies that $\tilde{j}_v \leq 10^{-5}$. In other words, $\tilde{j}_v \sim 0$, while all other terms in (43) are $\sim 1$. Therefore, we can discard (43a) by simply setting $\tilde{j}_v = 0$. This implies that from (43b), the electrochemical potential of vacancies is uniform over the entire SCL, i.e., $\tilde{\mu}_v^*|_{X_E=0^+} = \tilde{\mu}_v^*|_{X_E\to-\infty}$. Therefore, we obtain

$$\tilde{c}_v = \frac{\beta_v e^{\Delta\tilde{\phi}_{E|C}}}{e^{\Delta\tilde{\phi}_{E|C}} + \beta_v - 1} \tag{44}$$

where $\Delta\tilde{\phi}_{E|C} = \tilde{\phi}(X_E) - \tilde{\phi}|_{X_E\to-\infty}$. By substituting (44) into (43b), we can write

$$-\frac{\partial^2 \Delta\tilde{\phi}_{E|C}}{\partial X_E^2} = 1 - \frac{\beta_v e^{\Delta\tilde{\phi}_{E|C}}}{e^{\Delta\tilde{\phi}_{E|C}} + \beta_v - 1} \tag{45}$$

The boundary condition at the electrolyte|cathode interface ($X_E = 0^+$) and the bulk of the electrolyte ($X_E \to -\infty$) for the boundary value problem (45) are as follows

$$\left.\frac{\partial \Delta\tilde{\phi}_{E|C}}{\partial X_E}\right|_{X_E\to-\infty} = 0 \tag{46a}$$

$$\Delta\tilde{\phi}_{E|C}|_{X_E=0} = \Delta\tilde{\phi}_{E|C}^0 \tag{46b}$$

Since the electric fields in the SCL are far stronger than in the bulk of the electrolyte, we can take $\left.\frac{\partial \Delta\tilde{\phi}_{E|C}}{\partial \tilde{x}_E^+}\right|_{X_E\to-\infty} \approx 0$. Moreover, the governing equations (45) can also be utilized for the electrolyte side of the anode|electrolyte interface by merely substituting $\Delta\tilde{\phi}_{E|C}$ with $\Delta\tilde{\phi}_{E|A} = \tilde{\phi}(X_E) - \tilde{\phi}|_{X_E\to\infty}$, where $X_E \in [0, \infty)$.

## 4.2 Space Charge Layer in the Cathode

As mentioned in subsection 3.1, when we simulate the bulk of the cathode, we assume that Li ions and electrons are the two mobile charged species. In this subsection, we take the Li cathode to be a p-type semi-conductor that has both mobile holes and Li ions.

In the bulk of the cathode, i.e., $x \in [L_E + \delta x_C, L_E + L_C]$, $c_{Li^+} = c_- = c_{Li}$ due to electroneutrality. Further, the maximum concentration of mobile electrons equals the maximum concentration of Li ions, i.e., $c_-^{max} = c_{Li^+}^{max}$. From the definition of a hole and electroneutrality, it follows that $c_h = c_-^{max} - c_- = c_{Li^+}^{max} - c_{Li^+}$. Moreover, we assume that the concentration of holes is maximum when $c_- = 0$. Therefore, $c_h^{max} = c_-^{max} = c_{Li^+}^{max}$.

At $x = L_E + \delta x_C$, i.e., the intersection between the cathode-side SCL and the bulk of the cathode, we take $c_{Li^+}(t, L_E + \delta x_C) = c_{Li^+}^0$ and $c_-(t, L_E + \delta x_C) = c_-^0$. Therefore, $c_h(t, L_E + \delta x_C) = c_h^0 = c_{Li^+}^{max} - c_{Li^+}^0$. It should be noted that $c_h^0$ and $c_{Li^+}^0$ are functions of time. They can be approximated by solving the model of the bulk of the cathode described in subsection 3.1.

We can then rewrite Poisson's equation (1c) for the cathode by using the relative permittivity of the cathode $\varepsilon_{r,C}$ and the concentration of holes and Li ions. Using $X_C = \frac{x - L_E}{\lambda_{D,C}}$, $\lambda_{D,C} = \sqrt{\frac{\varepsilon_0 \varepsilon_{r,C} k_B T}{e^2 c_{Li^+}^{max}}}$, $\tilde{c}_h = \frac{c_h}{c_h^0}$, $\tilde{c}_{Li^+} = \frac{c_{Li^+}}{c_{Li^+}^0}$, and $\tilde{\phi} = \frac{\phi}{V_{th}}$ allows us to non-dimensionalize Poisson's equation as

$$-\frac{\partial^2 \tilde{\phi}}{\partial X_C^2} = \left(\frac{\tilde{c}_h}{\beta_h} + \frac{\tilde{c}_{Li^+}}{\beta_{Li^+}} - 1\right) \tag{47}$$

with $\beta_h = \frac{c_h^{max}}{c_h^0}$, $\beta_{Li^+} = \frac{c_{Li^+}^{max}}{c_{Li^+}^0}$ and $c_h^{max} = c_{Li^+}^{max} = c_h^0 + c_{Li^+}^0$. We consider site exclusion for both holes and Li ions in the SCL, see equation (5). Therefore, we can set

$$\tilde{\mu}_h^* = \frac{\mu_h^0}{e} + V_{th} \ln \frac{c_h}{c_h^{max} - c_h} + \phi \tag{48a}$$

$$\tilde{\mu}_{Li^+}^* = \frac{\mu_{Li^+}^0}{e} + V_{th} \ln \frac{c_{Li^+}}{c_{Li^+}^{max} - c_{Li^+}} + \phi \tag{48b}$$

Similar to what was done for the electrolyte side of the SCL (see subsection 4.1), the non-dimensional fluxes of Li ions and holes can be defined as $\tilde{j}_{Li^+} = \frac{j_{Li^+} \lambda_{D,C}}{c_{Li^+}^{max} D_{Li^+} e}$ and $\tilde{j}_h = \frac{j_h \lambda_{D,C}}{c_h^{max} D_h e}$, respectively, where $\tilde{j}_{Li^+}, \tilde{j}_h \sim \tilde{j}_v$. Therefore, we can write $\tilde{j}_{Li^+}, \tilde{j}_h \sim 0$ and discard (1a) [65]. It follows that the reduced electrochemical potentials $\tilde{\mu}_h^*$ and $\tilde{\mu}_{Li^+}^*$ are constant in the SCL and that

$$\tilde{c}_h = \frac{\beta_h e^{-\Delta \tilde{\phi}_C}}{e^{-\Delta \tilde{\phi}_C} + \beta_h - 1} \tag{49a}$$

$$\tilde{c}_{Li^+} = \frac{\beta_{Li^+} e^{-\Delta \tilde{\phi}_C}}{e^{-\Delta \tilde{\phi}_C} + \beta_{Li^+} - 1} \tag{49b}$$

where for notational convenience $\Delta \tilde{\phi}_C = \tilde{\phi}(X_C) - \tilde{\phi}\big|_{X_C \to \infty}$. Substituting (49) into (47) gives

$$-\frac{\partial^2 \Delta\tilde{\phi}_C}{\partial X_C^2} = \frac{e^{-\Delta\tilde{\phi}_C}}{e^{-\Delta\tilde{\phi}_C} + \beta_h - 1} + \frac{e^{-\Delta\tilde{\phi}_C}}{e^{-\Delta\tilde{\phi}_C} + \beta_{Li^+} - 1} - 1 \qquad (50)$$

with the boundary conditions

$$\left.\frac{\partial \Delta\tilde{\phi}_C}{\partial X_C}\right|_{X_C \to \infty} = 0 \qquad (51a)$$

$$\left.\Delta\tilde{\phi}_C\right|_{X_C=0} = \Delta\tilde{\phi}_C^0 \qquad (51b)$$

## 4.3 Space Charge Layer in the Anode

The charge stored in the electrolyte side of the SCL of the anode|electrolyte interface needs to be balanced by the accumulation/depletion of electrons at the anode side. We can adapt Poisson's equation (1c) in this SCL ($x \in (-\delta x_A, 0]$) with the relative permittivity of the anode material $\varepsilon_{r,A}$ and the concentration of electrons $c_-$. By non-dimensionalizing with $\tilde{c}_- = \frac{c_-}{c_-^0}$, $X_A = \frac{x}{\lambda_{D,A}}$, and $\tilde{\phi} = \frac{\phi}{V_{th}}$, where $\lambda_{D,A} = \sqrt{\frac{\varepsilon_0 \varepsilon_{r,A} k_B T}{e^2 c_-^0}}$ is the Debye length of the anode, we obtain

$$-\frac{\partial^2 \tilde{\phi}}{\partial X_A^2} = 1 - \tilde{c}_- \qquad (52)$$

The electrochemical potential of electrons in metals is given by [71-73]

$$\tilde{\mu}_- = \left(\frac{3}{8\pi}\right)^{2/3} \frac{\hbar^2 c_-^{0\,2/3}}{2m_e} \tilde{c}_-^{2/3} - e\phi \qquad (53)$$

where $\hbar$ is Planck's constant and $m_e$ is the mass of the electron. We note that the Fermi energy of electrons $E_F = \left(\frac{3}{8\pi}\right)^{2/3} \frac{\hbar^2 c_-^{0\,2/3}}{2m_e} = 4.7$ eV [74]. Following the same argument as the one used to model the electrolyte and cathode sides of the SCL (see subsection 4.1 and 4.2), we assume that the non-dimensional electron flux is zero and, therefore, we can avoid solving the mass conservation equation (1a) of the PNP equation set. This also implies that the reduced electrochemical potential $\tilde{\mu}_-^* = -\frac{\tilde{\mu}_-}{e}$ is constant within the SCL. Therefore, we can write

$$\tilde{c}_-^{2/3} = \frac{\Delta\tilde{\phi}_A}{\xi} + 1 \qquad (54)$$

where, for convenience, we take $\xi = \left(\frac{3}{8\pi}\right)^{2/3} \frac{\hbar^2 c_0^{2/3}}{2 m_e e V_{\text{th}}}$ and $\Delta\tilde{\phi}_A = \tilde{\phi}(X_A) - \tilde{\phi}\big|_{X_A \to \infty}$. Incidentally, we note that $\frac{\Delta\tilde{\phi}_A}{\xi} + 1 \geq 0$. If (54) is substituted into (52), we obtain

$$-\frac{\partial^2 \Delta\tilde{\phi}_A}{\partial X_A^2} = 1 - \left(\frac{\Delta\tilde{\phi}_A}{\xi} + 1\right)^{\frac{3}{2}} \tag{55}$$

The boundary conditions at the anode|electrolyte interface ($X_A = 0$) and the bulk of the anode ($X_A \to -\infty$) are

$$\frac{\partial \Delta\tilde{\phi}_A}{\partial X_A}\bigg|_{X_A \to -\infty} = 0 \tag{56a}$$

$$\Delta\tilde{\phi}_A\big|_{X_A = 0} = \Delta\tilde{\phi}_A^0 \tag{56b}$$

## 4.4 Matching the Space Charge Layers between Electrodes and the Electrolyte

To ensure physical consistency, dielectric fields at the two sides of the SCL need to be equal as schematically illustrated in the bottom panel of Figure 2 (b). We assume that no surface charge accumulates at the infinitesimally thin surface, where the two phases intersect. The electric displacement field $\mathbf{D} = \varepsilon_0 \varepsilon_r \mathbf{E}$ follows Maxwell 1$^{\text{st}}$ law [50] and it needs to be continuous [75]. Therefore,

$$[\![\mathbf{D} \cdot \mathbf{n}]\!] = 0 \tag{57}$$

where $[\![(\bullet)]\!]$ denotes the jump of $(\bullet)$ at the interface of interest and $\mathbf{n}$ is the unit vector normal to that interface. At the electrolyte|cathode interface, we have

$$[\![\mathbf{D} \cdot \mathbf{n}]\!]_{\text{EC}} = \left(\mathbf{D}_{\text{E}|\text{C}} - \mathbf{D}_{\text{C}}\right) \cdot \mathbf{n} = 0 \tag{58}$$

where $\mathbf{D}_{\text{E}|\text{C}}$ and $\mathbf{D}_{\text{C}}$ are the displacement field at the electrolyte and cathode sides of the interface, respectively. Similarly, at the anode|electrolyte interface, we have

$$[\![\mathbf{D} \cdot \mathbf{n}]\!]_{\text{AE}} = \left(\mathbf{D}_{\text{A}} - \mathbf{D}_{\text{E}|\text{A}}\right) \cdot \mathbf{n} = 0 \tag{59}$$

where $\mathbf{D}_{\text{A}}$ and $\mathbf{D}_{\text{E}|\text{A}}$ are the displacement field in the anode and electrolyte sides of the interface, respectively. Since $\mathbf{E} = -\nabla\phi$ and we select $\mathbf{n}$ to be the unit vector along the *x*-axis (*i.e.* $\mathbf{n} = \mathbf{e}_x$), we can write that for the electrolyte|cathode interface

$$\left.\frac{\partial \Delta \tilde{\phi}_C}{\partial X_C}\right|_{X_C=0} = \frac{\lambda_{D,C}\varepsilon_{r,E}}{\lambda_{D,E}\varepsilon_{r,C}} \left.\frac{\partial \Delta \tilde{\phi}_{E|C}}{\partial X_E}\right|_{X_E=0} \tag{60}$$

where $\lambda_{D,E}$ and $\lambda_{D,C}$ are the Debye length of the electrolyte and cathode, respectively. Also, $\varepsilon_{r,E}$ and $\varepsilon_{r,C}$ are the relative permittivities of the electrolyte and cathode materials, respectively. Similarly, for the anode|electrolyte interface, we can write

$$\left.\frac{\partial \Delta \tilde{\phi}_A}{\partial X_A}\right|_{X_A=0} = \frac{\lambda_{D,A}\varepsilon_{r,E}}{\lambda_{D,E}\varepsilon_{r,A}} \left.\frac{\partial \Delta \tilde{\phi}_{E|A}}{\partial X_E}\right|_{X_E=0} \tag{61}$$

where $\lambda_{D,A}$ is the Debye length of the anode and $\varepsilon_{r,A}$ is the relative permittivity of the anode material. The analytical expressions obtained following (60) and (61) are given in section E.4 of the SI. Another matching condition over each interface is that the electric potential is continuous. It follows that the total electric potential drop over a specific electrode|electrolyte interface is the sum of the electric potential drop over the corresponding electrode and electrolyte sub-domains. Therefore, we can define, for the electrolyte|cathode interface,

$$\Delta \tilde{\phi}_{EC} = \Delta \tilde{\phi}_{E|C}^0 - \Delta \tilde{\phi}_C^0 \tag{62}$$

where $\Delta \tilde{\phi}_{EC} = \frac{\phi_C(X_C \to \infty) - \phi_{E/C}(X_E \to -\infty)}{V_{th}} = \frac{V_{oc}(c_{Li}|_{x=L_E+\delta x_C}) - \eta_C}{V_{th}}$, $\Delta \tilde{\phi}_{E|C}^0 = \Delta \tilde{\phi}_{E|C}|_{X_E=0}$ and $\Delta \tilde{\phi}_C^0 = \Delta \tilde{\phi}_C|_{X_C=0}$. Also, for the anode|electrolyte interface,

$$\Delta \tilde{\phi}_{AE} = \Delta \tilde{\phi}_A^0 - \Delta \tilde{\phi}_{E|A}^0 \tag{63}$$

where $\Delta \tilde{\phi}_{AE} = \frac{\phi_{E|A}(X_E \to \infty) - \phi_A(X_A \to -\infty)}{V_{th}} = -\frac{\eta_A}{V_{th}}$, $\Delta \tilde{\phi}_A^0 = \Delta \tilde{\phi}_A|_{X_A=0}$ and $\Delta \tilde{\phi}_{E|A}^0 = \Delta \tilde{\phi}_{E|A}|_{X_E=0}$. (62) and (63) are schematically illustrated in the bottom panel of Figure 2 (b).

## 4.5 Charge Transfer Kinetics

As a result of the formation of the SCLs, the charge transfer kinetics equation of the electrolyte|cathode interface needs modifying. While the equation is formally similar to the (34) used in the electroneutral model, $\tilde{c}_{v,E}$ is no longer a constant. This implies that the exchange current density (20) becomes

$$j_C^0 = \frac{k_{E \to C}^{\alpha_C} k_{C \to E}^{1-\alpha_C} c_{Li}^{max} e}{\gamma_C^{\ddagger}} (y_{Li}^s)^{1-\alpha_C} (1 - y_{Li}^s)^{\alpha_C} (c_v^s)^{\alpha_C} (c_v^{max} - c_v^s)^{1-\alpha_C} \tag{64}$$

where $y_{Li}^s = y_{Li}(t, L_E^+)$ and $c_v^s = c_v(t, L_E^-)$. To allow a meaningful comparison between the electroneutral and non-electroneutral model, (64) is normalized with respect to the exchange current density computed for the electroneutral model given by (35). Therefore, we can write

$$j_C^0 = (k_C^0)_{EN} (y_{Li}^s)^{1-\alpha_C} (1 - y_{Li}^s)^{\alpha_C} \frac{(\tilde{c}_v^s)^{\alpha_C} (\beta_v - \tilde{c}_v^s)^{1-\alpha_C}}{(\beta_v - 1)^{1-\alpha_C}} \tag{65}$$

where $\tilde{c}_v^s = c_v(t, L_E^-)/c_v^0$, $c_v^0 = c_v(L_E - \delta x_{E|C})$ is the concentration of vacancies in the bulk of the electrolyte. Correspondingly, the charge transfer at the anode|electrolyte interface is similar to that used in the electroneutral model (36) with the modified exchange current density given by

$$j_A^0 = (k_A^0)_{EN} \frac{(\tilde{c}_v^s)^{\alpha_A} (\beta_v - \tilde{c}_v^s)^{1-\alpha_A}}{(\beta_v - 1)^{1-\alpha_A}} \tag{66}$$

where $\tilde{c}_v^s = c_v(t, 0^+)/c_v^0$. The use of (66) implies that the charge transfer at the anode|electrolyte depends on the Li vacancy concentration in the electrolyte side of the SCL.

In summary, we used singular perturbations to simplify the analysis with the relevant equations summarized in Table S.3. The implementation of the non-electroneutral model is reported in section D.2 of the SI.

## 5. Electro-Chemo-Mechanical Model

We develop a third model that couples the electrochemistry of SCL with mechanics. To do this, in addition to the governing equations introduced in sections 3 and 4, we solve (in the domain shown in the upper panel of Figure 2 (b)) the mechanical equilibrium equation (9). The latter mechanical equilibrium can be recast as

$$\frac{\partial}{\partial x}\left(\sigma_{xx} + \frac{\varepsilon_0 \varepsilon_r}{2}\left(\frac{\partial \phi}{\partial x}\right)^2\right) = 0 \tag{67}$$

In our model, the local chemical composition is linked to the concentration of Li vacancies in the electrolyte and Li ions in the cathode. The corresponding concentrations in the expansion-free state of electrolyte and cathode are $c_v' = c_v^0$ and $c_{Li^+}' = c_{Li^+}^{max}$, respectively, see (11) for the notation. As the present model is 1D with coordinate $x$, the displacements along the $x$, $y$ and $z$-directions are independent of $y$ and $z$. Furthermore, displacements along the $y$ and $z$-directions are equal to zero, i.e., $u_y = 0$ and $u_z = 0$. These conditions mean that the physical system is invariant along the $yz$ plane [76]. It follows from (11) that the strains along the $y$ and $z$-directions and the corresponding off-diagonal terms can be taken to be zero, i.e., $\epsilon_{yy} = \epsilon_{zz} = \epsilon_{xy} = \epsilon_{yx} = \epsilon_{xz} = \epsilon_{zx} = 0$. Therefore, the stress tensor is diagonal and its components are

$$\sigma_{xx} = 2G\epsilon_{xx} + \kappa\epsilon_{xx} - \zeta(c_i - c_i') \tag{68a}$$

$$\sigma_{yy} = \kappa\epsilon_{xx} - \zeta(c_i - c_i') \tag{68b}$$

$$\sigma_{zz} = \kappa\epsilon_{xx} - \zeta(c_i - c_i') \tag{68c}$$

where we note that $\sigma_{yy} = \sigma_{zz}$ and the hydrostatic stress $\sigma_h = \frac{\sigma_{xx}+\sigma_{yy}+\sigma_{zz}}{3}$. Moreover, the ECM potential (4) computed with the assumption of site exclusion (*i.e.* using (2) as the chemical potential) is given as follows:

$$\tilde{\mu}_{i,\text{ECM}} = \mu_i^0 + k_B T \ln\left(\frac{c_i}{c_i^{\max} - c_i}\right) + z_i e\phi - \Omega\sigma_h \tag{69}$$

In the following subsections, we will formulate the ECM framework for modeling the bulk of the cathode and the SCLs of the two electrode|electrolyte interfaces shown in Figure 2 (b). The bulk of the electrolyte is not explicitly modeled because the concentration of Li vacancies and the electric field are constant. Therefore, $\sigma_{xx}$, $\sigma_{yy}$, and $\sigma_{zz}$ are also constant as implied by (67) and (68).

## 5.1 The Bulk of the Cathode

In the bulk of the cathode, because of electroneutrality, the electric field $\mathbf{E} = -\nabla\phi$ is a constant. Therefore, the mechanical equilibrium equation (67) can be simplified as

$$\frac{\partial \sigma_{xx}}{\partial x} = 0 \tag{70}$$

This implies that $\sigma_{xx}$ is constant and dependent on the boundary stress. For the boundary condition where a pre-stress $P_x$ is applied to the SSB, see the bottom right panel of Figure 2 (b), we can set

$$\sigma_{xx} = 2G\epsilon_{xx} + \kappa\epsilon_{xx} - \zeta(c_{Li} - c'_{Li}) = P_x \tag{71}$$

for all $x \in [L_E, L_E + L_C]$. It follows from (68b) and (68c) that $\sigma_{yy}$ and $\sigma_{zz}$ can be written as

$$\sigma_{yy} = \sigma_{zz} = P_x - 2G\epsilon_{xx} \tag{72}$$

We should point out that $\sigma_{yy}$ and $\sigma_{zz}$ are functions of the Li concentration (see (68b) and (68c)). Therefore, we also need to solve the Li bulk transportation equation using (31) and the boundary condition (33), but with the ECM potential of Li having the form of

$$\tilde{\mu}_{Li,ECM} = \mu_{Li}^0 + k_BT \ln(\gamma_{Li}\tilde{c}_{Li}) - \Omega\sigma_h \tag{73}$$

By solving (31) together with (71) and (72), we can obtain $\sigma_{yy}$, $\sigma_{zz}$, and $\sigma_h$ in the bulk of the cathode. It should be noted from (71) and (72) that $P_x$ only shifts the magnitude of the stresses in the SSB. However, reality is much more complex as interfaces are neither completely coherent nor 0D, and the pre-stress may help suppress the delamination [77]. While our analysis can be carried out for any $P_x$, we will assume $P_x = 0$ for simplicity.

## 5.2 Space Charge Layer in the Electrolyte

To solve the ECM problem at the SCL, we first recast the mechanical equilibrium equation (67) in $X_E$, the SCL coordinate. To shorten the writing and without loss of generality, we solve the equations with respect to the differences in stress and electric potential between a point in the SCL of the electrolyte|cathode interface and the bulk of the electrolyte. The quantities are formally defined as $\Delta\sigma_{xx,E|C} = \sigma_{xx}(X_E) - \sigma_{xx}|_{X_E \to -\infty}$ and $\Delta\phi_{E|C} = \phi(X_E) - \phi|_{X_E \to -\infty}$. They replace $\phi$ and $\sigma_{xx}$ in (67). After rescaling and non-dimensionalization, i.e., $\Delta\tilde{\sigma}_{xx,E|C} = \frac{\Delta\sigma_{xx,E|C}}{k_BTc_v^0}$, $\Delta\tilde{\phi}_{E|C} = \frac{\Delta\phi_{E|C}}{V_{th}}$, and $X_E = \frac{x-L_E}{\lambda_{D,E}}$, (67) becomes

$$\frac{\partial}{\partial X_E}\left(\Delta\tilde{\sigma}_{xx,E|C} + \frac{1}{2}\left(\frac{\partial \Delta\tilde{\phi}_{E|C}}{\partial X_E}\right)^2\right) = 0 \tag{74}$$

The boundary condition of (74) is such that at the bulk ($X_E \to \infty$),

$$\Delta\tilde{\sigma}_{xx,E|C}\Big|_{X_E \to \infty} = 0 \tag{75}$$

Moreover, using the same argument as the one employed for the non-electroneutral model (see subsection 4.1), the non-dimensional flux of Li vacancies can be approximated to be zero in the SCL coordinate, $X_E$. Therefore, the ECM potential $\tilde{\mu}_{v,\text{ECM}}$ is a constant and the following relation holds for the electrolyte side of the SCL:

$$\ln\left(\frac{\tilde{c}_v(\beta_v - \tilde{c}_v)}{\beta_v - 1}\right) + \Delta\tilde{\phi}_{E|C} + \Omega_E c_v^0 \Delta\tilde{\sigma}_{h,E|C} = 0 \tag{76}$$

where $\Delta\tilde{\sigma}_{h,E|C} = \tilde{\sigma}_h(X_E) - \tilde{\sigma}_h|_{X_E \to -\infty}$ is the hydrostatic stress difference between a point in the electrolyte-side SCL and the bulk of the electrolyte.

In addition to (74) and (76), another governing equation is Poisson's equation, see (43c) in subsection 4.1. As discussed in subsections 4.1 and 4.4, (43c) is subject to the boundary conditions (46) and the matching condition for the two sides of the interfaces (60).[3] Moreover, expressions similar to (74) and (76) can also be derived for the electrolyte side of the anode|electrolyte interface by substituting $\Delta\tilde{\phi}_{E|C}$ with $\Delta\tilde{\phi}_{E|A} = \tilde{\phi}(X_E) - \tilde{\phi}|_{X_E \to \infty}$ and $\Delta\tilde{\sigma}_{mn,E|A} = \tilde{\sigma}_h(X_E) - \tilde{\sigma}_h|_{X_E \to \infty}$.

## 5.3 Space Charge Layer in the Cathode

The mechanical equilibrium equation (67) will also be applied to the cathode and recast in the SCL coordinate $X_C$, which is specific to the cathode. As was done in subsection 5.2, the following differences, $\Delta\sigma_{xx,C} = \sigma_{xx}(X_C) - \sigma_{xx}|_{X_C \to \infty}$ and $\Delta\phi_C = \phi(X_C) - \phi|_{X_C \to \infty}$, are defined. Using the non-dimensionalization $\Delta\tilde{\sigma}_{xx,C} = \frac{\Delta\sigma_{xx,C}}{k_B T c_{\text{Li}^+}^0}$, $\Delta\tilde{\phi}_C = \frac{\Delta\phi_C}{V_{\text{th}}}$, and $X_C = \frac{x - L_E}{\lambda_{D,C}}$, (67) becomes

$$\frac{\partial}{\partial X_C}\left(\frac{\Delta\tilde{\sigma}_{xx,C}}{\beta_{\text{Li}^+}} + \frac{1}{2}\left(\frac{\partial\Delta\tilde{\phi}_C}{\partial X_C}\right)^2\right) = 0 \tag{77}$$

---

[3] Our analysis is conducted using as a variable the stress difference, $\Delta\sigma_{mn}$, instead of the actual stress, $\sigma_{mn}$. Since this is a simple translation of the stress tensor, it follows that $[\![\boldsymbol{\sigma} \cdot \mathbf{n}]\!]_{AE} = (\boldsymbol{\sigma}_A - \boldsymbol{\sigma}_{E|A}) \cdot \mathbf{n} = 0$ and $[\![\boldsymbol{\sigma} \cdot \mathbf{n}]\!]_{EC} = (\boldsymbol{\sigma}_{E|C} - \boldsymbol{\sigma}_C) \cdot \mathbf{n} = 0$ are automatically satisfied at $x = 0$ and $x = L_E$, respectively.

Its boundary condition is

$$\Delta\tilde{\sigma}_{xx,C}\big|_{X_C\to\infty} = 0 \tag{78}$$

Of the two charged species ($h^\bullet$ and $Li^+$) present, only Li ions are assumed to lead to a compositional expansion/contraction of the cathode material. Under the hypothesis that the non-dimensional Li ion flux is negligible (see subsection 4.2), the ECM potential of Li ions $\tilde{\mu}_{Li^+,ECM}$ is constant implying that $\tilde{\mu}_{Li^+,ECM}(X_C) = \tilde{\mu}_{Li^+,ECM}\big|_{X_C\to\infty}$. Therefore, we can write

$$\ln\left(\frac{\tilde{c}_{Li^+}(\beta_{Li^+} - \tilde{c}_{Li^+})}{\beta_{Li^+} - 1}\right) + \Delta\tilde{\phi}_C + \Omega_C c^0_{Li^+}\Delta\tilde{\sigma}_{h,C} = 0 \tag{79}$$

where $\Delta\tilde{\sigma}_{h,C} = \tilde{\sigma}_h(X_C) - \tilde{\sigma}_h\big|_{X_C\to\infty}$ is the difference in hydrostatic stress between a point in the cathode side of the SCL and the bulk of the cathode directly adjacent to the SCL. Another governing law is Poisson's equation (47), which is subject to the boundary condition (51) and the matching condition (60). Also, the concentration of holes continues to be given by (48a).

## 5.4 Space Charge Layer in the Anode

Similar to the discussion in subsections 5.2 and 5.3, we can also recast the mechanical equilibrium equation (67) in the coordinate system specific to the anode's side of the SCL. This can be done by defining $\Delta\sigma_{xx,A} = \sigma_{xx}(X_A) - \sigma_{xx}\big|_{X_A\to-\infty}$ and $\Delta\phi_A = \phi(X_A) - \phi\big|_{X_A\to-\infty}$ and using the non-dimensionalization $\Delta\tilde{\sigma}_{xx,A} = \frac{\sigma_{xx,A}}{k_B T c^0_-}$, $\Delta\tilde{\phi}_A = \frac{\Delta\phi_A}{V_{th}}$, and $X_A = \frac{x}{\lambda_{D,A}}$. It follows that (67) becomes

$$\frac{\partial}{\partial X_A}\left(\Delta\tilde{\sigma}_{xx,A} + \frac{1}{2}\left(\frac{\partial\Delta\tilde{\phi}_A}{\partial X_A}\right)^2\right) = 0 \tag{80}$$

which is subject to the boundary condition

$$\Delta\tilde{\sigma}_{xx,A}\big|_{X_A\to-\infty} = 0 \tag{81}$$

Moreover, electrons are the only charged species in the anode (see subsection 2.4). By assuming that the concentration of electrons does not contribute to the compositional expansion, we can continue to use (55) as the governing equation with the boundary conditions (56) and the matching condition (61).

## 5.5 Charge Transfer Kinetics

The emergence of stresses at the interfaces impacts the chemical potentials. This implies that the charge transfer kinetics is also modified. In fact, the driving force for the reaction, see (12), has two additional components, $\Omega_E \sigma_h$ and $\Omega_C \sigma_h$, compared to the conventionally used electrochemical potentials [48]. The modified electrochemical potentials at the electrolyte and cathode sides of the interface are

$$\tilde{\mu}_v(E) - \mu_{Li}(E) = \mu_v^0 - \mu_{Li}^0 + k_B T \ln \frac{c_v}{c_v^{max} - c_v} - e\phi_E - \Omega_E \sigma_h \tag{82a}$$

$$\mu_{Li}(C) - \tilde{\mu}_-(C) = \mu_{Li}^0 - \mu_-^0 + k_B T \ln \frac{c_{Li}}{c_{Li}^{max} - c_{Li}} + e\phi_C - \Omega_C \sigma_h \tag{82b}$$

We note that (82) has the same form as the ECM potential given in (69). By substituting (82) into (20b), we obtain that the Li current density at the electrolyte|cathode interface can be modeled by

$$j_{Li^+} = j_C^0 \left( \exp\left( \frac{\alpha_C(\eta_C + \sigma_h(\Omega_C + \Omega_E)/e)}{V_{th}} \right) - \exp\left( \frac{-(1-\alpha_C)(\eta_C + \sigma_h(\Omega_C + \Omega_E)/e)}{V_{th}} \right) \right) \tag{83}$$

where $j_C^0$ is given in (65). As for the anode|electrolyte interface, we note that only the chemical potential at the electrolyte side of the interface is modified by the mechanical energy. Therefore, by substituting (82a) into (30b), we have that

$$j_{Li^+} = j_A^0 \left( \exp\left( \frac{\alpha_A(\eta_A - \sigma_h \Omega_E/e)}{V_{th}} \right) - \exp\left( \frac{-(1-\alpha_A)(\eta_A - \sigma_h \Omega_E/e)}{V_{th}} \right) \right) \tag{84}$$

where $j_A^0$ is given in (66).

In this section, we formulated the ECM model. The relevant equations and implementation are summarized in Table S.4 and section D.3 of the SI, respectively.

## 6. Result and Discussion

We modeled an SSB with Li metal as the anode, lithium phosphorus oxynitride (LiPON) as the SE, and $LiCoO_2$ as the cathode. We considered LiPON because its physical properties are well characterized [62, 78, 79], and experimental data is available [18]. We first fitted the electroneutral model against the experimental discharge curves of Fabre *et al.* [18] (subsection 6.1). We then solved the non-electroneutral model for the SCLs formed at the interfaces. In particular, we studied the impact of the SCL on the interfacial charge transfer (subsection 6.2). We finally solved the ECM model and conducted a parametric study so as to identify the factors that control the stress exerted at the SCL of the electrolyte|cathode interface (subsection 6.3). All simulations were performed using COMSOL Multiphysics interfaced with MATLAB.

## 6.1 Electroneutral Model

First, we simulated the galvanostatic discharge of the SSB from an initial state of charge of 0.5 ($Li_yCoO_2$ with $y = 0.5$) with the battery assumed to operate at Li fraction between $y_{Li} = 0.5$ and $y_{Li} = 1$. Following the work of Fabre et al. [18], the chemical diffusivity $\widetilde{D}_{chem}$ used for simulating the Li transport of the cathode is a function of $y_{Li}$ and is given in Figure S.4 of the SI. Figure S.3 shows $y_{Li}$ across the bulk of the cathode as a function of time for $j_{tot}$ = 0.24 mA/cm², 0.35 mA/cm², and 0.48 mA/cm². Upon discharge, the Li concentration inside the cathode increases. The discharge of the battery stops when $y_{Li}(t, L_E^+) = y_{Li}^s = 1$, i.e., the maximum cathode capacity is reached at the electrolyte|cathode interface causing a capacity loss since $y_{Li} < 1$ in the bulk of the cathode, i.e., $x \in [L_E, L_E + L_C]$, see Figure S.3. Capacity loss increases as the current density increases. We stress that, within this model, Li ions can no longer enter the cathode because there are no more vacant sites available for Li insertion.

The Li concentration at the boundary can be used to compute the open circuit potential, $V_{oc}\left(c_{Li}|_{x=L_E^+}\right)$, typically an empirically measured function.[4] Using the cell potential, (39), and following the workflow illustrated in Figure S.1, we can estimate the model parameters, i.e., $(k_C^0)_{EN}$, $(k_A^0)_{EN}$, $\sigma_C$, and $\sigma_E$, by comparing the discharge data with the electroneutral model, see Figure 3 (a). The input and fitted parameters are tabulated in Table 1. The corresponding sum of the squared residuals is reported in Table S.5. Figure 3 (b) shows the overpotentials $\eta_C$, $\eta_A$, $\eta_{R_C}$, and $\eta_{R_E}$, as defined in subsection 3.2 and 3.3, of the individual physical processes as a function of capacity for $j_{tot}$ = 0.24 mA/cm². As illustrated in Figure 3 (b), the overpotential due to Li transport inside the electrolyte, i.e., $\eta_{R_E}$, contributes to the highest potential loss among the four components.

---

[4] The open circuit potential have the empirical expression of $V_{oc}\left(c_{Li}|_{x=L_E^+}\right) = V_{oc}\left(y_{Li}|_{x=L_E^+}\right) = \frac{-219.027+322.003y_{Li}^2-198.242y_{Li}^4+254.911y_{Li}^6-467.807y_{Li}^8+207.168y_{Li}^{10}}{-44.337+36.643y_{Li}^2-3.4302y_{Li}^4+113.081y_{Li}^6-182.567y_{Li}^8+80.3097y_{Li}^{10}}$ as obtained from reference [22].

This is attributed to the relatively low conductivity of LiPON ($\sigma_E \approx 2 \times 10^{-6}$ S/cm) [80]. The second highest overpotential component is $\eta_C$, the one that is due to Li transfer at the electrolyte|cathode interface. Unlike other overpotentials, $\eta_C$ increases during discharge, and is particularly high when the discharge of the SSB nears completion. This can be explained by inspecting the reaction rate model (6) and noting that the rate is proportional to the concentration of empty sites at the boundary of the cathode. As one nears full discharge, all empty sites are filled. This implies that, since the current density is fixed, the charge transfer barrier for Li transport across the electrolyte|cathode interface increases.

## 6.2 Non-electroneutral Model

### 6.2.1 Space Charge Layer Profile

Electroneutrality, as discussed in section 6.1, does not allow modelling of the SCLs and their influence on the interfacial charge transfer. To overcome this limitation, we followed the workflow depicted in Figure S.2 and solved the non-electroneutral model, see section 4. The parameters used in the simulation are reported in Table 2. Since the relative permittivity of metals is extremely large (ideally infinite) [81], we set $\varepsilon_{r,A} = 1000$ for ease of analysis. Following the literature [79, 82], we assume that the concentration of Li vacancies in the bulk of LiPON to be around 4% of the total available Li sites. We also take $\beta_v = c_v^{\max}/c_v^0 = 10$. This value is chosen such that the maximum concentration of Li vacancies is less than that of the lattice Li sites. Otherwise, if the concentration of Li vacancies saturates in the electrolyte, lattice collapse may occur. Alternatively, a more accurate $\beta_v$ value can be estimated using ab-initio simulations, where one can relax the lattice structure of the SE with an increasing vacancy concentration in order to determine whether the lattice is structurally unstable [83].

The electric potential and the distribution of the charged species at the anode|electrolyte interface are shown in Figure 4. Near the anode|electrolyte interface, the deviation from electroneutrality leads to a depletion of electrons in the anode and build-up of Li vacancies in the electrolyte. It is important to stress that unlike other models previously reported in the literature [20, 21], which consider only the SCL in the electrolyte, our model matches the SCL formed in both the anode and electrolyte sides of the interface. The SCL does not change during galvanostatic discharge because the Li concentration in the anode and $-\eta_A$ are both constant. On the other hand, when the

SSB discharges at higher current densities, the depletion of electrons in the anode side and the accumulation of vacancies in the electrolyte side increase due to the larger $-\eta_A$.

We further compare the SCL thickness in the two sides of the anode|electrolyte interface. The SCL thickness in one side of the interface can be defined as the width $\delta x$ of the region that contains 99.9% of the total SCL charge [84], i.e., $\delta x = h\lambda_D$ is such that $\frac{\int_0^h (\tilde{c}_i - 1)dX}{\int_0^\infty (\tilde{c}_i - 1)dX} = 0.999$. Therefore the thickness of the SCL side of the anode is $\delta x_A = 9.66$ nm and that of the electrolyte side is $\delta x_{E|A} = 0.63$ nm. We must also note that the SCL thickness is insensitive to the current density.

If we compare the electric potential drop of the two sides of the SCL, we note that the anode side's drop $\Delta\tilde{\phi}_A^0$ is smaller than that of the electrolyte side's, i.e., $\Delta\tilde{\phi}_{E|A}^0$. For $j_{tot} = 0.24$ mA/cm$^2$, the maximum electric potential gradient in the anode side is $\left.\frac{\partial \Delta\phi_A}{\partial x}\right|_{X_A=0} = -5.68 \times 10^5$ V/m while in the electrolyte side it is $\left.\frac{\partial \Delta\phi_{E|A}}{\partial x}\right|_{X_E=0} = -3.42 \times 10^7$ V/m. In other words, the electric field is weaker for the anode side compared to the electrolyte side. Such a difference can be explained by the matching condition (61). For the anode|electrolyte interface, we noted that the relative permittivity of the anode $\varepsilon_{r,A}$ (=1000) is much larger than that of the electrolyte $\varepsilon_{r,E}$ (=16.6). Then, from (61), $\frac{\lambda_{D,A}\varepsilon_{r,E}}{\lambda_{D,E}\varepsilon_{r,A}} = \left(\frac{c_v^0 \varepsilon_{r,E}}{c_-^0 \varepsilon_{r,A}}\right)^{1/2} \sim 0.01$ and that $-\left.\frac{\partial \Delta\tilde{\phi}_A}{\partial X_A}\right|_{X_A=0} \ll -\left.\frac{\partial \Delta\tilde{\phi}_{E|A}}{\partial X_E}\right|_{X_E=0}$. Physically, a large permittivity implies a large polarization, which reduces the electric field, or equivalently, the potential gradient (since $\mathbf{E} = -\nabla\phi$), in the material [75]. Indeed, as we further increase the relative permittivity of $\varepsilon_{r,A}$ to 10000, the electric field in the anode decreases and $\delta x_A$ increases to 30.5 nm (see Figure S.6 (a) and (c)). On the other hand, the electric field in the electrolyte increases as $\delta x_{E|A}$ remains unchanged (see Figure S.6 (b) and (d)).

For the electrolyte|cathode interface, the potential drop over the SCL, i.e., $\Delta\phi_{EC} = V_{oc}(c_{Li}|_{x=L_E+\delta x_C}) - \eta_C$, can be up to 4.3 V. As shown in Figure 5 (a), (c), and (d), the large $\Delta\phi_{EC}$ leads to the accumulation of Li vacancies in the electrolyte and both Li ions and holes in the cathode. The accumulation of charge is so strong that the species concentrations reach the maximum values allowed by the assumed thermodynamics, i.e., $c_v^{max}$, $c_h^{max}$, and $c_{Li^+}^{max}$. Also, as shown in Figure 5 (b) and (e), $\Delta\tilde{\phi}_{E|C}^0 = \Delta\tilde{\phi}_{E|C}(X_E = 0)$ and $\Delta\tilde{\phi}_C^0 = \Delta\tilde{\phi}_C(X_C = 0)$ are of

comparable magnitude because of the similar relative permittivity of electrolyte and cathode. For $j_{tot} = 0.24 \text{ mA/cm}^2$, the maximum electric fields for the electrolyte and cathode sides of the SCL are $\frac{\partial \Delta \phi_{E|C}}{\partial x}\big|_{X_E=0} = 1.11 \times 10^{10}$ V/m and $\frac{\partial \Delta \phi_C}{\partial x}\big|_{X_C=0} = 1.24 \times 10^{10}$ V/m, respectively, when the bulk Li concentration is $y_{Li}^0 = y_{Li}(t, L_E + \delta x_C) = 0.5$. Both values are several orders of magnitude greater than those encountered in the bulk and at the other interface.

Using the same criterion as the one employed for the anode|electrolyte interface, we compute the thickness of the electrolyte side of the SCL to be $\delta x_{E|C} = 0.71$ nm and that of the cathode side of the SCL to be $\delta x_C = 0.49$ nm if $y_{Li}^0 = 0.5$. Both thicknesses have the same order of magnitude as those previously computed in the literature [21, 37]. Moreover, as shown in Figure 5 (a), (c), and (d), the SCL thickness decreases for the electrolyte side and increases for the cathode side during discharge. Such a change is unlike the one encountered at the SCL of the anode|electrolyte interface and is due to the decrease of $V_{oc}(c_{Li}|_{x=L_E+\delta x_C})$ and the increase of $\eta_C$ during discharge. While $\eta_C$ is a function of the current density of the Li ions, and hence, the total current density $j_{tot}$ (see (34)), the concentration of the species and the electric potential in the SCL does not change significantly with $j_{tot}$ (in the range from 0.24 mA/cm$^2$ to 0.48 mA/cm$^2$). This is because, within the $j_{tot}$ range studied, $\eta_C \sim 0.01$ V and the value of $\eta_C$ is much smaller than that of $V_{oc}(c_{Li}|_{x=L_E+\delta x_C})$.

Moreover, the SCL in the two sub-domains of the electrolyte|cathode interface depends on $\beta_v = \frac{c_v^{max}}{c_v^0}$. As shown in Figures S.7 and S.8, the lower the $\beta_v$ is, the thicker the SCL in the electrolyte side and the larger $\Delta \tilde{\phi}_{E|C}^0$. As it is well known, the SCL in the electrolyte side forms to compensate for the charges accumulated in the cathode side [85]. Such a charge compensation in the electrolyte is limited by the availability of vacant sites in the electrolyte. As $\beta_v$ reduces, the charge density in the electrolyte side of the SCL decreases. A thicker electrolyte-side SCL is then needed to compensate for the charge in the cathode. Moreover, as reflected in (S.23), the electric field $\frac{\partial \Delta \tilde{\phi}_{E|C}}{\partial X_E}$ for the same potential difference $\Delta \tilde{\phi}_{E|C}$ reduces as $\beta_v$ decreases. Therefore, a larger $\Delta \tilde{\phi}_{E|C}^0$ is needed to match conditions (60) and (62).

One should note that our model did not consider the presence of a Stern layer, which is widely accepted in modeling the SCL of liquid electrolyte|electrode interfaces [36]. Such a layer forms as the solvent molecules of the liquid electrolyte reorient and prevent the mobile ions from reaching the electrode. In turn, this mechanism leads to a region in the vicinity of the electrode|electrolyte interface [86, 87] that is depleted of mobile ions. For the SE, a steric effect is unlikely. Instead, at least for perfectly matching interfaces, the only factor that limits the concentration of charged species is the site exclusion effect, which is embedded in the assumed electrochemical potential, see (42) and (48). Two weaknesses of the present model need to be noted. First, interfaces are hardly perfect, and the contact between the materials may lead to physical gaps and the emergence of secondary phases. Second, as a result of the high concentrations and the potential changes in polarizability within the SCL, one may need to apply a model for the relative permittivity that is concentration-dependent [88].

## 6.2.2 Impact of Space Charge Layer on the Charge Transfer Kinetics

One important conclusion that can be drawn from the above discussion is that the species concentrations in the SCL differ significantly from those in the bulk. We note that such concentration changes may affect the transfer of charge and ions across them. In this subsection, we investigate the impact of the species' concentrations at interfaces on 1) exchange current densities given in (65) and (66) and 2) corresponding overpotentials $\eta_C$ and $\eta_A$.

One should note that it is not trivial to solve $\eta_C$. This is because the concentration of Li vacancies in the electrolyte side and the concentration of Li ions in the cathode side reach their maximum thermodynamically allowed values. Such an accumulation of charged species at $X_E = 0$ and $X_C = 0$ leads to a negligible exchange current density and an infinite $\eta_C$ as the pre-exponential term (65) is zero. We address this issue by utilizing, instead of the exact concentration of the charged species at $X_E = 0$ and $X_C = 0$, their average over the lattices planes constituting the SCL of the electrolyte|cathode interface, as illustrated in Figure S.9. In other words, we estimate (65) as:

$$j_C^0 \approx (k_C^0)_{EN} \bar{y}_{Li}^{1-\alpha_C} (1 - \bar{y}_{Li})^{\alpha_C} \frac{\bar{c}_v^{\alpha_C} (\beta_v - \bar{c}_v)^{1-\alpha_C}}{(\beta_v - 1)^{1-\alpha_C}} \tag{85}$$

with

$$\bar{y}_{\text{Li}} = \overline{\frac{c_{\text{Li}^+}}{c_{\text{Li}^+}^{\max}}} \cong \frac{\int_0^{h_C} c_{\text{Li}^+}/c_{\text{Li}^+}^{\max}\, dX_C}{h_C} \tag{86a}$$

$$\bar{\tilde{c}}_v \cong \frac{\int_{-h_{E|C}}^{0} \tilde{c}_v\, dX_E}{h_{E|C}} \tag{86b}$$

where $h_{E|C}$ and $h_C$ are the non-dimensional thicknesses of the electrolyte and the cathode, respectively, that are considered for solving the charge transfer equation. We assume $h_{E|C}$ and $h_C$ to be twice the lattice parameters, which are approximately equal to the SCL thickness, of the corresponding domain (see Figure S.9). Since the SCL thicknesses of the two sides of the interface are of the order of the lattice parameter [64, 89], the region in the electrolyte where $\tilde{c}_v = \beta_v$ and the region in the cathode where $c_{\text{Li}^+} = c_{\text{Li}^+}^{\max}$ are penetrable. In other words, as long as there exists Li ions within $[-h_{E/C}, 0]$ and vacant site within $[0, h_C]$, charge transfer can happen.

The lattice parameter of LiPON, which we denote as $\Delta l_E$, is about 6.92 Å [90], whereas that of LiCoO$_2$, which we indicate as $\Delta l_C$, is about 2.83Å [91].[5] Therefore, we set $h_{E|C} = \frac{2\Delta l_E}{\lambda_{D,E}} = 7.86$ and $h_C = \frac{2\Delta l_C}{\lambda_{D,C}} = 21.3$. The experimental discharge curves are then fitted with respect to $(k_C^0)_{\text{EN}}$, $(k_A^0)_{\text{EN}}$, $\sigma_C$, and $\sigma_E$. Figure 3 (c) shows the discharge potential fitted using the non-electroneutral model. The estimated physical parameters are reported in Table 1. The corresponding sum of square residuals is given in Table S.5. As shown in Figure 3 (d), the overpotential contributed by the charge transfer at the anode|electrolyte interface $\eta_A$ remains similar to that of the electroneutral model. This implies that the SCL does not alter the charge transfer kinetics at the anode|electrolyte interface significantly. The small impact of the SCL on the kinetics can be attributed to the small energy barrier of Li transfer at this interface. In contrast, the charge transfer overpotential at the electrolyte|cathode interface $\eta_C$ is significantly larger if the SCL is included in the model. The

---

[5] For LiPON, we consider the Li$_4$P$_2$O$_7$ polymorph, whose lattice parameter is a $=$ 8.56Å, b $=$ 7.11Å, and c $=$ 5.19Å, and we take the average of a, b, and c for our calculation. For LiCoO$_2$, a $=$ b $=$ 2.83Å, and c $=$ 14.1Å. Since Li can only diffuse along the ab-plane, we directly take a $=$ 2.83Å for our calculation.

increase in $\eta_C$ implies that, under constant-current discharge, an additional energy barrier is needed for the interfacial charge transfer process to occur. Such an extra energy barrier compensates for the fewer Li ions in the electrolyte and Li vacant sites in the cathode that decrease the exchange current density. Our results thus suggest that including the SCL in the model affects the computed charge transfer kinetic overpotentials, particularly, at the electrolyte|cathode interface.

## 6.3 Electro-Chemo-Mechanical Model

The results obtained for the non-electroneutral model (section 6.2) indicate that the large electric potential difference over the electrolyte|cathode interface causes a strong accumulation of Li ions and holes on the cathode side of the electrolyte|cathode interface and of Li vacancies in the electrolyte side of the same interface. The large electric fields accompanied by strong deviation from the bulk concentrations result in net body forces that in turn lead to stress accumulation [55]. In order to gain a better understanding of the generated stress, we extend our analysis to include the mechanics model developed in section 5. We focus only on the electrolyte|cathode interface since the average electric fields at the electrolyte|cathode interface are several orders of magnitude higher than those experienced at the anode|electrolyte interface, see section 6.2. We also compute the stress in the bulk of the cathode as it is necessary for investigating its impact on the charge transfer at the electrolyte|cathode interface.

The parameters used in the ECM model are reported in Table 3. It is worth noting that the partial molar volume of the cathode, $\Omega_C$, is taken to be negative because the $Li_yCoO_2$ contracts as the Li concentration increases [92]. As for the LiPON electrolyte, it is typical to assume the partial molar volume to be zero [41]. To generalize our analysis, we assume that $\Omega_E = -10^{-7}$ m³/mol, such that $\Omega_E$ is of the same magnitude as that of the $Li_yCoO_2$ cathode. Also, the negative sign implies that the insertion of Li vacancies causes contraction of the electrolyte.

### 6.3.1 Mechanical Stresses in the Bulk of the Cathode

We first solve for the stress in the bulk of the cathode. Figure 6 shows the in-plane and hydrostatic stresses, *i.e.*, $\sigma_{yy}$ and $\sigma_h$, respectively. $\sigma_{xx}$ equals zero because of the boundary condition (71) with $P_x = 0$. Initially, there is a compressive hydrostatic stress in the bulk of the cathode. The hydrostatic stress is compressive because the cathode material expands as Li is removed from it during charging. Upon discharge with a $j_{\text{tot}} = 0.24 \text{ mA/cm}^2$, Li ions insert into the cathode, causing a reduction in the magnitude of the compressive stress. The compressive stress follows the same trend as the Li concentration as computed with the electroneutral model (see Figure S.5). At $x = L_E$, the cathode becomes expansion free when it is fully occupied, *i.e.*, when $y_{\text{Li}}^S = 1$.

### *6.3.2 Mechanical Stresses in the Space Charge Layers (without Considering their Impact on the Interfacial Charge Transfer)*

We first investigate how stress is generated in the SCL by assuming that interfacial charge transfer does not depend on the stress. In other words, we use the current density equation (34) with the exchange current density (65) to compute $\Delta\tilde{\phi}_{\text{EC}}$. Figure 7 shows that significant stresses (in the order of GPa's) are exerted in the SCL along both the out-of-plane (perpendicular to the interface) and in-plane (parallel to the interface) directions. The compressive nature of the stress is a result of counter SCL charges attraction over the interface. The stresses are larger across than along the interface. The out-of-plane stresses for both sides of the interface and the in-plane stress for the electrolyte side decrease during discharge due to the smaller electric potential drop and smaller accumulation of charged species.

In contrast, the in-plane stress in the cathode side shows a more complicated feature. Particularly, at $y_{\text{Li}}^0 = 0.5$, the stress at $X_C < 5$ is negative, whereas the stress at $X_C > 5$ is positive relative to the bulk of cathode. Such a feature can be attributed to a competition between the charges attraction over the interface and the volumetric change due to the Li accumulation. As mentioned above, the charges attraction over the interface can induce a negative stress next to the interface. On the other hand, a higher Li concentration in the SCL leads to stronger lattice contraction at the SCL compared to the bulk. Such a difference in lattice contraction can induce a positive stress in the cathode side of the SCL. At $y_{\text{Li}}^0 = 0.5$ and $X_C < 5$, the attraction by the counter SCL dominates leading to a local compressive stress. As the effect of charges attraction reduces for larger $X_C$, the effect of local compositional shrinkage then dominates leading to a tensile stress at $X_C > 5$. As the SSB discharge, the Li concentration difference between the SCL and the bulk diminish, the compositional shrinkage relative to the bulk reduces. Therefore, the positive stress feature disappear and only negative stress dominated for a larger $y_{Li}^0$. The negative stress magnitude at $X_C = 0$ also increases as a result.

If we inspect the two components of the ECM potential, *i.e.*, (76) and (79), we find that $|\Omega c_i^0 \Delta \tilde{\sigma}_h| \sim 0.1$ and $\left| \ln\left(\frac{\tilde{c}_i(\beta_i - \tilde{c}_i)}{\beta_i - 1}\right) + \Delta \tilde{\phi} \right| \sim 10^2$. In other words, $|\Omega c_i^0 \Delta \tilde{\sigma}_h| \ll \left| \ln\left(\frac{\tilde{c}_i(\beta_i - \tilde{c}_i)}{\beta_i - 1}\right) + \Delta \tilde{\phi} \right|$ for the values of $\Omega_E$ and $\Omega_C$ considered. This implies that, *de facto*, the mechanical energy terms $\Omega c_i^0 \Delta \tilde{\sigma}_h$ in (76) and (79) can be neglected. Consequently, the distributions of charged species and electric potential in the SCL are identical to those computed for the non-electroneutral model, see Figure 5.

To further understand which specific factors affect the stress generation the most, we assess how the stress generated varies following a modification in the partial molar volumes (*i.e.* $\Omega_E$ and $\Omega_C$) and permittivities (*i.e.* $\varepsilon_{r,E}$ and $\varepsilon_{r,C}$). Specifically, Figure 8 reports the stresses at $X_E = 0$ and $X_C = 0$, *i.e.*, $\Delta\sigma_{mn,E|C}^0 = \Delta\sigma_{mn,E|C}(X_E = 0)$ and $\Delta\sigma_{mn,C}^0 = \Delta\sigma_{mn,C}(X_C = 0)$, respectively as a function of $\Omega_E$ and $\Omega_C$. As shown in Figure 8, the stresses along the in-plane direction reduces as the magnitude of $\Omega_E$ and $\Omega_C$ increase, suggesting that they are most sensitive to volumetric change due to the accumulation of charged species. In contrast, the out-of-plane stresses are relatively insensitive to $\Omega_E$ and $\Omega_C$.

We also studied the impact of permittivities on the stress exerted in the SCL by accounting for the influence of $\frac{\lambda_{D,C} \varepsilon_{r,E}}{\lambda_{D,E} \varepsilon_{r,C}} = \left(\frac{c_v^0 \varepsilon_{r,E}}{c_{Li}^{max} \varepsilon_{r,C}}\right)^{1/2}$ on $\Delta\sigma_{mn,C}^0$ and $\Delta\sigma_{mn,E|C}^0$, where $\frac{\lambda_{D,C} \varepsilon_{r,E}}{\lambda_{D,E} \varepsilon_{r,C}}$ originates from the SCL matching condition (60). Consistent with the discussion, the SCL of the anode|electrolyte interfaces (see section 6.2.1), $\frac{\lambda_{D,C} \varepsilon_{r,E}}{\lambda_{D,E} \varepsilon_{r,C}}$ affects $\Delta\tilde{\phi}_C^0$ and $\Delta\tilde{\phi}_{E|C}^0$. As illustrated in Figure S.10 (a), increasing $\frac{\lambda_{D,C} \varepsilon_{r,E}}{\lambda_{D,E} \varepsilon_{r,C}}$, reduces $\Delta\tilde{\phi}_{E|C}^0$ and increases $\Delta\tilde{\phi}_C^0$. This is because a larger $\frac{\lambda_{D,C} \varepsilon_{r,E}}{\lambda_{D,E} \varepsilon_{r,C}}$ implies a larger permittivity in the electrolyte side relative to the cathode side. A large permittivity reduces the electric fields and increases the SCL thickness. This leads to a smaller potential drop for the electrolyte side relative to that of the cathode side. As the potential drop of the electrolyte side decreases, the stresses along both the out-of-plane and in-plane direction decrease as shown in Figure S.10 (b). This further validates that all components of the stress tensors are dominated by the potential drop over the SCL of the electrolyte|cathode interface.

### 6.3.3 Impact of the Mechanics on the Interfacial Charge Transfer

In subsection 6.3.2 the interfacial charge transfer is assumed to be independent of the stresses emerging in the two sides of the interface. Here we relax this hypothesis. The stress-modified charge transfer kinetic rate (83) depends on the interfacial hydrostatic stress $\sigma_h(X_C = 0) = \sigma_h(X_C \to \infty) + \Delta\sigma_h(X_C = 0)$. Using the $\sigma_h(X_C \to \infty)$ in the bulk of the cathode (see subsection 6.3.1), we can iteratively estimate the $\eta_C$ and $\Delta\sigma_h(X_C = 0)$ at the interface by solving the stress-modified charge transfer equation (83) and the governing equations of the cathode side of the SCL, i.e., (47), (77), and (79). Figure 9 compares the charge transfer overpotential $\eta_C$ as estimated from 1) the stress-modified model, i.e., (83), and 2) the electroneutral model, i.e., (34). The overpotential computed with the stress-modified model and the electroneutral model are denoted as $(\eta_C)_{ECM}$ and $(\eta_C)_{EN}$, respectively. For both cases, the reaction constant $(k_C^0)_{EN}$ is considered as identical and is equal to 4.40 A/m².

As shown in Figure 9, $(\eta_C)_{ECM} - (\eta_C)_{EN} < 0$ and is of the order of several tens of millivolts for the $\Omega_E$ and $\Omega_C$ considered. This implies that the mechanical stresses induced at the interface reduces the charge transfer barrier. The computed $(\eta_C)_{ECM} - (\eta_C)_{EN}$ value is of the same order of magnitude to that of $\eta_{C,EN}$. For the $\Omega_E$ and $\Omega_C$ taken in subsection 6.3.1 (i.e. $\Omega_E = -10^{-7}$ m³/mol and $\Omega_C = -7.28 \times 10^{-7}$ m³/mol) and $j_{tot} = 0.24$ mA/cm², $(\eta_C)_{ECM} - (\eta_C)_{EN} = -47.8$ mV at $y_{Li}^0 = 0.5$ and $(\eta_C)_{ECM} - (\eta_C)_{EN} = -37.6$ mV at $y_{Li}^0 = 0.98$. In other words: $(\eta_C)_{ECM} = -33.9$mV and $(\eta_C)_{EN} = 13.9$ mV at $y_{Li}^0 = 0.5$. $(\eta_C)_{ECM} = 10.3$mV and $(\eta_C)_{EN} = 47.9$ mV at $y_{Li}^0 = 0.98$. This clearly shows that the compressive stress at the electrolyte|cathode interface has an a positive impact on the interfacial charge transfer, especially when the partial molar volumes of the electrolyte and cathode are significant. Therefore, one should not overlook the stress exerted at the SCL when studying the interfacial reaction rates as it could impact the charge transfer overpotential.

It should be noted that applying the stress-modified charge transfer reaction equation (83) has an impact on the stress generated. Figure 10 shows $\Delta\sigma_{xx}^0$ and $\Delta\sigma_{yy}^0 = \Delta\sigma_{zz}^0$ computed as a function of $\Omega_E$ and $\Omega_C$ for both the electrolyte and cathode side of the interface. The corresponding hydrostatic stresses $\Delta\sigma_{h,E|C}^0$ and $\Delta\sigma_{h,C}^0$ are given in Figure S.11 of the SI. As shown in Figure 10 (a) and (b), the stress in the out-of-plane direction reduces monotonically as $\Omega_E$ and $\Omega_C$ increase.

This monotonicity is analogous to that encountered for $\Delta\eta$, as given in Figure 9. As previously suggested in subsection 6.3.2, the out-of-plane stress is closely linked to the interfacial overpotential. Conversely, changing the partial molar volumes $\Omega_E$ and $\Omega_C$ modifies the in-plane stress significantly. Again, this finding points to one of the conclusions drawn in subsection 6.3.2: the accumulation of the charged species has a significant impact on the in-plane stresses.

## 6.4 Remarks

We need to emphasize that our analysis in section 6.2 and 6.3 focused mainly on the formation of the SCLs, the associated stresses, and the impact of such ECM coupling on the charge transfer kinetics. Other potential interfacial phenomena are yet to be addressed, including 1) the electrochemical decomposition of the interlayer region; 2) the impact of imperfect interfacial contacts on the SCL, charge transfer kinetics and mechanics; 3) emerging nonlinearities in the deformations; and 4) mechanical failure.

Several references suggest that the chemical composition of the electrode|electrolyte interface is different from that of the bulk [93-96]. For example, in-situ and ex-situ characterizations of the interfacial regions indicate that some Co from the $LiCoO_2$ side may diffuse towards LiPON [93, 94]. This leads to the formation of an interlayer of a few microns thick, which is much larger than the SCL as predicted by our model [95, 96]. One may need to apply a phase-field model to investigate the phase evolution and the corresponding reactions at the interface [97].

Furthermore, as mentioned in the introduction, the utilization of rigid and hard ceramic SEs may lead to microscopic gaps at the electrode|electrolyte interfaces [30, 31]. Such complex microstructures may, in turn, alter the SCL and the charge transfer kinetic rates described in this work. As a follow up to the current work, we suggest taking the complex interfacial microstructure into account. In particular, one can utilize homogenization and volume averaging methods to upscale the above-proposed equations from the microscopic to the macroscopic (cell-level) scale to resolve the interfacial microstructure [48, 98].

Also, plastic deformations may take place at the interfaces, especially in the case of significant compositional expansions [99]. The mechanical failure of SSBs is another exciting area that needs further work. As mentioned in the introduction, Bucci *et al.* [41, 100] have studied the formation and propagation of micro-cracks in the SSB by focusing on the failure occurring at a negative electrode, which consists of randomly distributed active Si particles embedded in a rigid SE matrix. Unlike our model, the influence of the SCL was neglected since the model is electroneutral. Due to the significant stresses arising in the SCL, interfacial failure studies should consider the impact of the SCL.

## 7. Conclusions

This work develops a general electro-chemo-mechanical model for SSBs that is consistent with the laws of mechanics and thermodynamics. The general framework is used to formulate three sub-models, *i.e.*, the electroneutral model, the non-electroneutral model, and the ECM model. In particular, the electroneutral model allows the simulation of the discharge potential together with the overpotentials of the various components of the SSB. Besides, the non-electroneutral model illustrates the characteristics of the SCLs in the electrode|electrolyte interfaces. In particular, we considered the effect of charged species accumulation on the charge transfer kinetics. The result showed that the deviation of electroneutrality could lead to a notable increase in charge transfer overpotential, particularly at the electrolyte|cathode interface.

Moreover, we investigated the coupling between electrochemistry and mechanics at the SCL of the electrolyte|cathode interface using our ECM model. Our model predicted considerable stress exerted at the SCL of the electrolyte|cathode interface. Our parametric study suggested that the out-of-plane stress is dominated by the potential drop across the SCL. In contrast, the compositional expansion associated with charge accumulation at the SCL is also an important controlling factor for the in-plane stress. In addition, stresses associated with the SCLs of the electrolyte|cathode interface lead to the decrease in the interfacial charge transfer overpotential, particularly when the partial molar volume of the electrolyte and cathode materials are significant.

This work highlights the importance of considering the SCLs when studying the charge transfer kinetics and the mechanical stress emerging at the interfaces of SSBs. The model provides the essential starting point for investigating other critical phenomena observed in SSBs, such as interfacial delamination and reaction at the interlayer. The continuum level study of these physical phenomena will likely facilitate the development of SSBs as the next-generation energy storage devices.

## Acknowledgments

The authors acknowledge the support from the Research Grants Council of Hong Kong (projects 16227016, and 16204517), the Hong Kong Innovation and Technology Fund (ITS/292/18FP), and the Guangzhou Science and Technology Program (No. 201807010074). T.H. Wan thanks the support from the Hong Kong PhD Fellowship Scheme.

# List of Symbols

| Symbols | Description | Unit |
|---|---|---|
| $\boldsymbol{j}_i$ | Current density of the mobile charged species $i$, with $\boldsymbol{j}_i = z_i e \boldsymbol{J}_i$ | A/m$^2$ |
| $j_{\text{tot}}$ | Total discharge current density | A/m$^2$ |
| $\boldsymbol{J}_i$ | Flux of the mobile charged species $i$, with $\boldsymbol{J}_i = \boldsymbol{j}_i / z_i e$ | 1/m$^2$s |
| $D_i$ | Diffusion coefficient of species $i$ | m$^2$/s |
| $\widetilde{D}_{\text{chem}}$ | Chemical diffusion coefficient of the cathode | m$^2$/s |
| $c_i$ | Concentration of species $i$ | 1/m$^3$ |
| $c'_i$ | Concentration of species $i$ in the stress free state | 1/m$^3$ |
| $c_i^0$ | Standard concentration of species $i$ | 1/m$^3$ |
| $c_i^{\max}$ | Maximum concentration of species $i$ available in the lattice | 1/m$^3$ |
| $\tilde{c}_i$ | Normalized concentration of species $i$, $\tilde{c}_i = c_i / c_i^0$ | - |
| $\tilde{c}_v^s$ | $\tilde{c}_v^s = c_v(t, L_E^-)/c_v^0$ for the electrolyte\|cathode interface, $\tilde{c}_v^s = c_v(t, 0^+)/c_v^0$ for the anode\|electrolyte interface | - |
| $\bar{c}_v$ | Average $\tilde{c}_v$ over a non-dimentional thickness $h_{E\|C}$ | - |
| $y_{\text{Li}}$ | $c_{\text{Li}}/c_{\text{Li}}^{\max}$ | - |
| $y_{\text{Li}}^s$ | $y_{\text{Li}}^s = y_{\text{Li}}(t, L_E^+)$ | - |
| $y_{\text{Li}}^0$ | $y_{\text{Li}}^0 = y_{\text{Li}}(t, L_E + \delta x_C)$ | - |
| $\bar{y}_{\text{Li}}$ | Average $y_{\text{Li}}$ over a non-dimentional thickness $h_C$ | - |
| $\varepsilon_{r,M}$ | Relative permittivity ( M = E, C, or A for electrolyte, cathode, or anode, respectively) | - |
| $\sigma_i$ | Conductivity of species $i$ | S/m$^2$ |
| $\sigma_M$ | Electrical conductivity ( M = E or C for electrolyte or cathode, respectively) | S/m$^2$ |
| $\mu_i^0$ | Standard potential of species $i$ | eV |
| $\mu_i$ | Chemical potential of species $i$ | eV |
| $\tilde{\mu}_i$ | Electrochemical potential of species $i$ | eV |
| $\tilde{\mu}_{i,\text{ECM}}$ | Electro-chemo-mechanical potential of species $i$ | eV |
| $\mu_i^*$ | Reduced chemical potential of species $i$, $\mu_i^* = \mu_i / z_i e$ | V |

| Symbol | Description | Units |
|---|---|---|
| $\tilde{\mu}_i^*$ | Reduced electrochemical potential of species $i$, $\tilde{\mu}_i^* = \tilde{\mu}_i/z_i e$ | V |
| $\tilde{\mu}_{i,\text{ECM}}^*$ | Reduced electro-chemo-mechanical potential of species $i$, $\tilde{\mu}_{i,\text{ECM}}^* = \tilde{\mu}_{i,\text{ECM}}/z_i e$ | V |
| $\gamma_i$ | Activity coefficient of species $i$ | - |
| $\beta_i$ | $c_i^{\max}/c_i^0$ | - |
| $\Delta g_\text{M}$ | Driving force of the charge transfer reaction ($\text{M} = \text{C}$ or $\text{A}$ for electrolyte\|cathode interface or anode\|electrolyte interface, respectively) | eV |
| $\Delta g_{\text{M}\to\text{N}}$ | Driving force for the interfacial charge transfer reaction to occur from domain M to N ($\text{M}, \text{N} = \text{E}, \text{C}$ or $\text{A}$ for the electrolyte, cathode or anode, respectively) | eV |
| $\Delta g_{\text{M}\to\text{N}}^\ddagger$ | Activation energy for the interfacial charge transfer reaction to occur from domain M to N ($\text{M}, \text{N} = \text{E}, \text{C}$ or $\text{A}$ for the electrolyte, cathode or anode, respectively) | eV |
| $\gamma_\text{M}^\ddagger$ | Activity coefficient of the transition-state of the charge transfer reaction ($\text{M} = \text{C}$, or $\text{A}$ for electrolyte\|cathode interface or anode\|electrolyte interface, respectively) | – |
| $\alpha_\text{M}$ | Symmetry coefficient for charge transfer reaction ($\text{M} = \text{C}$, or $\text{A}$ for electrolyte\|cathode interface or anode\|electrolyte interface, respectively) | – |
| $J_{\text{M}\to\text{N}}$ | Flux of Li from domain M to N ($\text{M}, \text{N} = \text{E}, \text{C}$ or $\text{A}$ for the electrolyte, cathode or anode, respectively) | $1/\text{m}^2\text{s}$ |
| $J_{\text{M}\to\text{N}}^0$ | Pre-exponential term of $J_{\text{M}\to\text{N}}$ | $1/\text{m}^2\text{s}$ |
| $J_\text{M}^0$ | Apparent exchange current density ($\text{M} = \text{C}$ or $\text{A}$ for the electrolyte\|cathode interface or anode\|electrolyte interface, respectively) | $1/\text{m}^2\text{s}$ |
| $k_{\text{M}\to\text{N}}$ | Reaction constant of $J_{\text{M}\to\text{N}}$ | $\text{m}^4/\text{s}$ or $\text{m/s}$ |

| Symbol | Description | Units |
|---|---|---|
| $(k_M^0)_{EN}$ | Interfacial reaction constant for the electroneutral model (M = C or A for electrolyte\|cathode interface or anode\|electrolyte interface, respectively) | A/m² |
| $L_M$ | Thickness (M = E or C for the electrolyte or cathode, respectively) | m |
| $\Delta l_M$ | Lattice parameter ( M = E or C for the electrolyte or cathode, respectively) | m |
| $\delta x_{E\|M}$ | Thickness of the electrolyte-side SCL ( M = C or A for the electrolyte\|cathode interface or the anode\|electrolyte interface, respectively) | m |
| $\delta x_M$ | Thickness of the electrode-side SCL (M = C or A for the cathode side of the electrolyte\|cathode interface or the anode side of the anode\|electrolyte interface, respectively) | m |
| $\lambda_{D,M}$ | Debye length (M = E, C, or A for the electrolyte, cathode, or anode) | m |
| $X_M$ | $x$-coordinate normalized with respect to $\lambda_{D,M}$ | - |
| $h_{E\|C}$ | Non-dimensional thickness for computing $\bar{c}_v$ | - |
| $h_C$ | Non-dimensional thickness for computing $\bar{y}_{Li}$ | - |
| $V_{th}$ | Thermal voltage, $V_{th} = k_B T/e$ | V |
| $V_{oc}$ | Open circuit voltage | V |
| $\eta_C$ | Charge transfer overpotential at the electrolyte\|cathode interface. $(\eta_C)_{ECM}$ and $(\eta_C)_{EN}$ are the values computed with the ECM model and electroneutral model, respectively. | V |
| $\eta_A$ | Charge transfer overpotential at the anode\|electrolyte interface | V |
| $\eta_{R_C}$ | Ohmic losses in the cathode | V |
| $\eta_{R_E}$ | Ohmic losses in the electrolyte | V |
| $\phi$ | Electric potential | V |
| $\tilde{\phi}$ | Non-dimensionalized electric potential, $\tilde{\phi} = \phi/V_{th}$ | - |
| $\Delta\tilde{\phi}_{E\|M}$ | Normalized electric potential of the electrolyte side of the interface ( M = C or A for the electrolyte\|cathode interface or the anode\|electrolyte interface, respectively) | - |
| $\Delta\tilde{\phi}_M$ | Normalized electric potential of the electrode side of the interface (M = C or A for the cathode side of the electrolyte\|cathode interface or the anode side of the anode\|electrolyte interface, respectively) | - |
| $\xi$ | Defined as $\left(\frac{3}{8\pi}\right)^{2/3}\frac{\hbar^2 c_0^{2/3}}{2m_e eV_{th}}$ | - |
| $\mathbf{D}_{E\|M}$ | Electric displacement field of the electrolyte side of the interface ( M = C or A for the electrolyte\|cathode interface or the anode\|electrolyte interface, respectively) | C/m² |
| $\mathbf{D}_M$ | Electric displacement field of the electrode side of the interface (M = C or A for the cathode side of the electrolyte\|cathode interface or the anode side of the anode\|electrolyte interface, respectively) | C/m² |
| $\mathbf{E}$ | Electric field | V/m |
| $\sigma_{mn}$ | Stress tensor, $\sigma_{mn} = 2G\epsilon_{mn} + \left(\kappa\epsilon_{kk} - \zeta(c_i - c_i')\right)\delta_{mn}$, where $G = \frac{Y}{2(1+v)}$, $\kappa = \frac{2vG}{(1-2v)}$, and $\zeta = \frac{\Omega(3\kappa+2G)}{3}$ | Pa |
| $\sigma_h$ | Hydrostatic stress, $\sigma_h = \frac{\sigma_{xx}+\sigma_{yy}+\sigma_{zz}}{3}$ | Pa |
| $\Delta\tilde{\sigma}_{mn,E\|M}$ | Normalized stress tensor of the electrolyte side of the interface (M = C or A for the electrolyte\|cathode interface or the anode\|electrolyte interface, respectively) | - |
| $\Delta\tilde{\sigma}_{mn,M}$ | Normalized stress tensor of the electrode side of the interface (M = C or A for the cathode side of the electrolyte\|cathode interface or the anode side of the anode\|electrolyte interface, respectively) | - |
| $\Delta\tilde{\sigma}_{h,E\|M}$ | Normalized hydrostatic stress of the electrolyte side of the interface ( M = C or A for the electrolyte\|cathode interface or the anode\|electrolyte interface, respectively) | - |
| $\Delta\tilde{\sigma}_{h,M}$ | Normalized hydrostatic stress of the electrode side of the interface (M = C or A for the cathode side of the electrolyte\|cathode interface or the anode side of the anode\|electrolyte interface, respectively) | - |
| $\mathbf{u}$ | Displacement vector, $\mathbf{u} = [u_1, u_2, u_3] = [u_x, u_y, u_z]$ | m |
| $\epsilon_{mn}$ | Strain tensor, $\epsilon_{mn} = \frac{1}{2}\left(\frac{\partial u_m}{\partial x_n} + \frac{\partial u_n}{\partial x_m}\right)$, $x_1 = x$, $x_2 = y$, and $x_3 = z$ | - |
| $P_x$ | Pre-stress applied to the SSB | Pa |
| $Y_M$ | Young's modulus ( M = E or C for the electrolyte or cathode, respectively ) | Pa |
| $\nu_M$ | Poisson's ratio ( M = E or C for the electrolyte or cathode, respectively ) | - |
| $\Omega_M$ | Partial molar volume ( M = E or C for the electrolyte or cathode, respectively ) | m³/mol |

[37] N.J.J. de Klerk, M. Wagemaker, Space-Charge Layers in All-Solid-State Batteries; Important or Negligible?, ACS Applied Energy Materials, 1 (2018) 5609-5618.

[38] Y. Gao, M. Cho, M. Zhou, Mechanical Reliability of Alloy-Based Electrode Materials for Rechargeable Li-Ion Batteries, Journal of Mechanical Science and Technology, 27 (2013) 1205-1224.

[39] S. Bishop, D. Marrocchelli, C. Chatzichristodoulou, N. Perry, M.B. Mogensen, H. Tuller, E. Wachsman, Chemical Expansion: Implications for Electrochemical Energy Storage and Conversion Devices, Annual Review of Materials Research, 44 (2014) 205-239.

[40] G. Bucci, T. Swamy, S. Bishop, B.W. Sheldon, Y.-M. Chiang, W.C. Carter, The Effect of Stress on Battery-Electrode Capacity, Journal of The Electrochemical Society, 164 (2017) A645-A654.

[41] G. Bucci, T. Swamy, Y.-M. Chiang, W.C. Carter, Modeling of Internal Mechanical Failure of All-Solid-State Batteries during Electrochemical Cycling, and Implications for Battery Design, Journal of Materials Chemistry A, 5 (2017) 19422-19430.

[42] J. Janek, W.G. Zeier, A Solid Future for Battery Development, Nature Energy, 1 (2016) 16141.

[43] Y. Zhu, X. He, Y. Mo, Origin of Outstanding Stability in the Lithium Solid Electrolyte Materials: Insights from Thermodynamic Analyses Based on First-Principles Calculations, ACS Applied Materials & Interfaces, 7 (2015) 23685-23693.

[44] P. Bai, J. Guo, M. Wang, A. Kushima, L. Su, J. Li, F.R. Brushett, M.Z. Bazant, Interactions between Lithium Growths and Nanoporous Ceramic Separators, Joule, 2 (2018) 2434-2449.

[45] F. Ciucci, Y. Hao, D.G. Goodwin, Impedance Spectra of Mixed Conductors: a 2D Study of Ceria, Physical Chemistry Chemical Physics, 11 (2009) 11243-11257.

[46] F. Ciucci, W.C. Chueh, D.G. Goodwin, S.M. Haile, Surface Reaction and Transport in Mixed Conductors with Electrochemically-Active Surfaces: a 2-D Numerical Study of Ceria, Physical Chemistry Chemical Physics, 13 (2011) 2121-2135.

[47] T.H. Wan, F. Ciucci, Continuum Level Transport and Electro-Chemo-Mechanics Coupling—Solid Oxide Fuel Cells and Lithium Ion Batteries, in: S.R. Bishop, N.H. Perry, D. Marrocchelli, B.W. Sheldon (Eds.) Electro-Chemo-Mechanics of Solids, Springer International Publishing, Cham, 2017, pp. 161-189.

[48] W. Lai, F. Ciucci, Mathematical Modeling of Porous Battery Electrodes—Revisit of Newman's Model, Electrochimica Acta, 56 (2011) 4369-4377.
52

|  | Electroneutral Model | Non-electroneutral Model |
|---|---|---|
| Input Parameters | | |
| $L_C$ | 4.42 μm | 4.42 μm |
| $L_E$ | 5.75 μm | 5.75 μm |
| $c_{Li^+}^{max}$ | $3.01 \times 10^{28}/m^3$ | $3.01 \times 10^{28}/m^3$ |
| $\alpha_C$ | 0.5 | 0.5 |
| $\alpha_A$ | 0.5 | 0.5 |
| $h_C$ | / | 21.3 |
| $h_{E|C}$ | / | 7.86 |
| Fitted Parameters | | |
| $(k_C^0)_{EN}$ | 4.40 A/m² | 1.97 A/m² |
| $(k_A^0)_{EN}$ | 15.0 A/m² | 15.5 A/m² |
| $\sigma_C$ | 31 S/cm | 31 S/cm |
| $\sigma_E$ | $1.20 \times 10^{-6}$ S/cm | $1.37 \times 10^{-6}$ S/cm |

Table 1. The input model parameters and the parameters that were obtained by fitting the galvanostatic discharge curves with the electroneutral and non-electroneutral model.

| Parameter | Value | Reference |
|---|---|---|
| $\beta_v$ | 10 | / |
| $\varepsilon_{r,E}$ | 16.6 | [101] |
| $\varepsilon_{r,C}$ | 14.95 | [102] |
| $\varepsilon_{r,A}$ | 1000 | / |
| $c_v^0$ | $3.04 \times 10^{27}/m^3$ | [79] |
| $c_{Li^+}^{max}$ | $3.01 \times 10^{28}/m^3$ | [61] |
| $c_-^0$ | $4.63 \times 10^{28}/m^3$ | [74] |
| $\lambda_{D,E}$ | $8.81 \times 10^{-11} m$ | / |
| $\lambda_{D,C}$ | $2.65 \times 10^{-11} m$ | / |
| $\lambda_{D,A}$ | $1.74 \times 10^{-10} m$ | / |

Table 2. The parameters used for computing the SCLs.

| Parameters | LiPON | LiCoO$_2$ |
|---|---|---|
| Y | 77GPa [103] | 191GPa [104] |
| ν | 0.25 [103] | 0.25 [104] |
| Ω | $-1 \times 10^{-7}$ m$^3$/mol (w.r.t. Li vacancies) | $-7.28 \times 10^{-7}$ m$^3$/mol [92] (w.r.t. Li ions) |

Table 3. The parameters used in the ECM model.

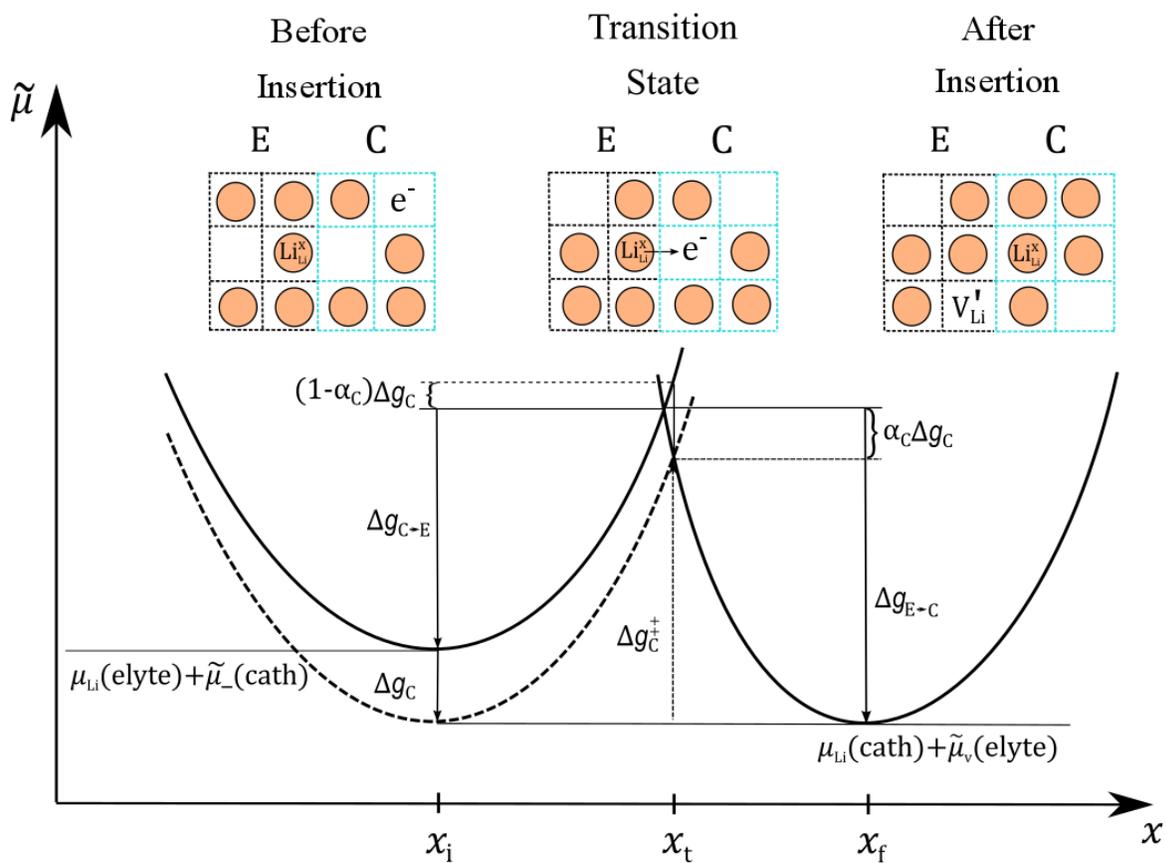

Figure 1. Energy landscape for the Li transfer across the electrolyte|cathode interface. The dashed line represents the energy landscape at equilibrium. The schematics at the top illustrate the initial, transition, and final states.

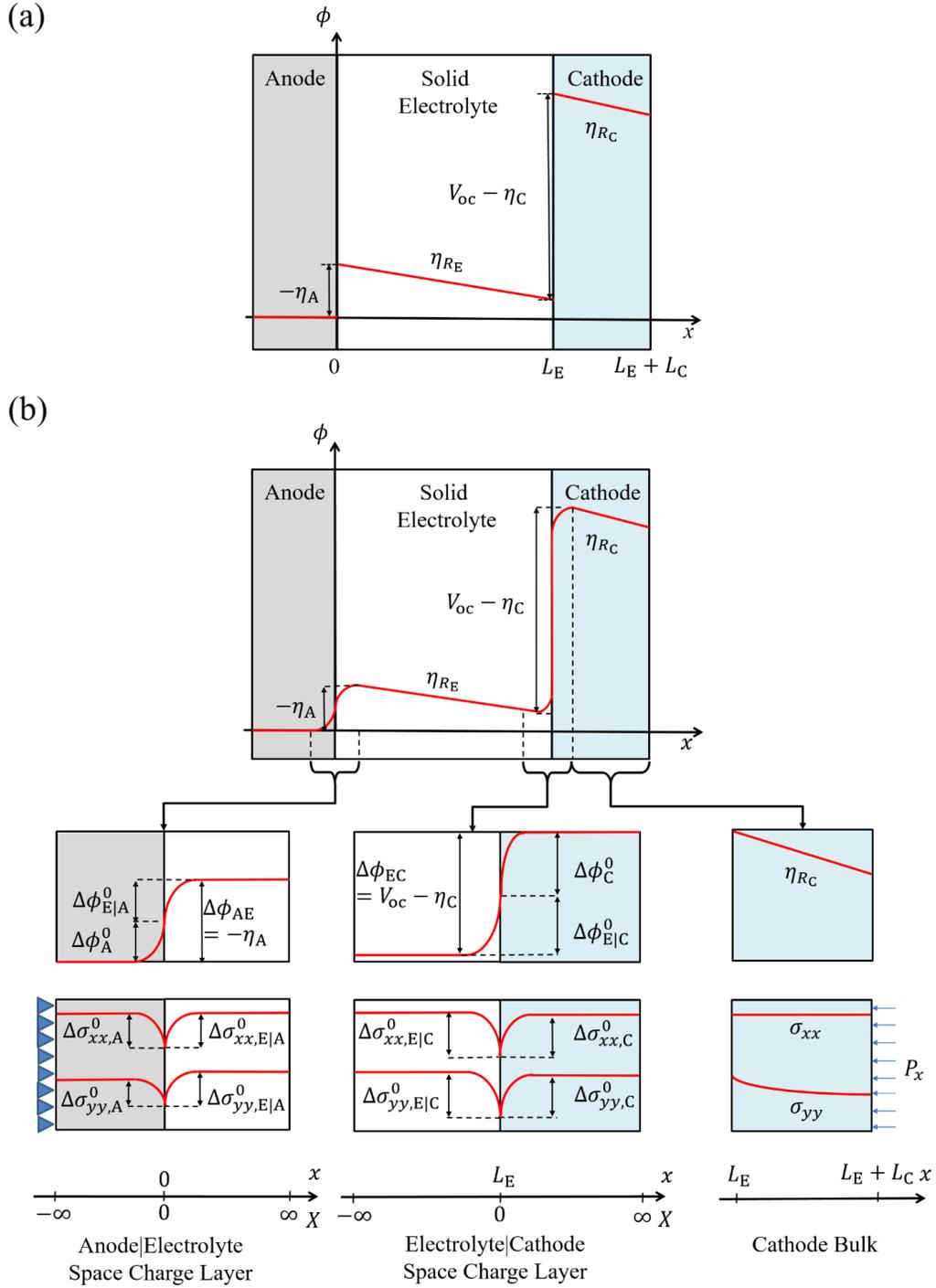

Figure 2. (a) Schematic depiction of the domain and electric potential in the electroneutral. (b) Schematic depiction of the domain, electric potential, and the stresses of the non-electroneutral model and the ECM model. The bottom diagram zooms over the SCLs emerging at the anode|electrolyte and electrolyte|cathode interfaces.

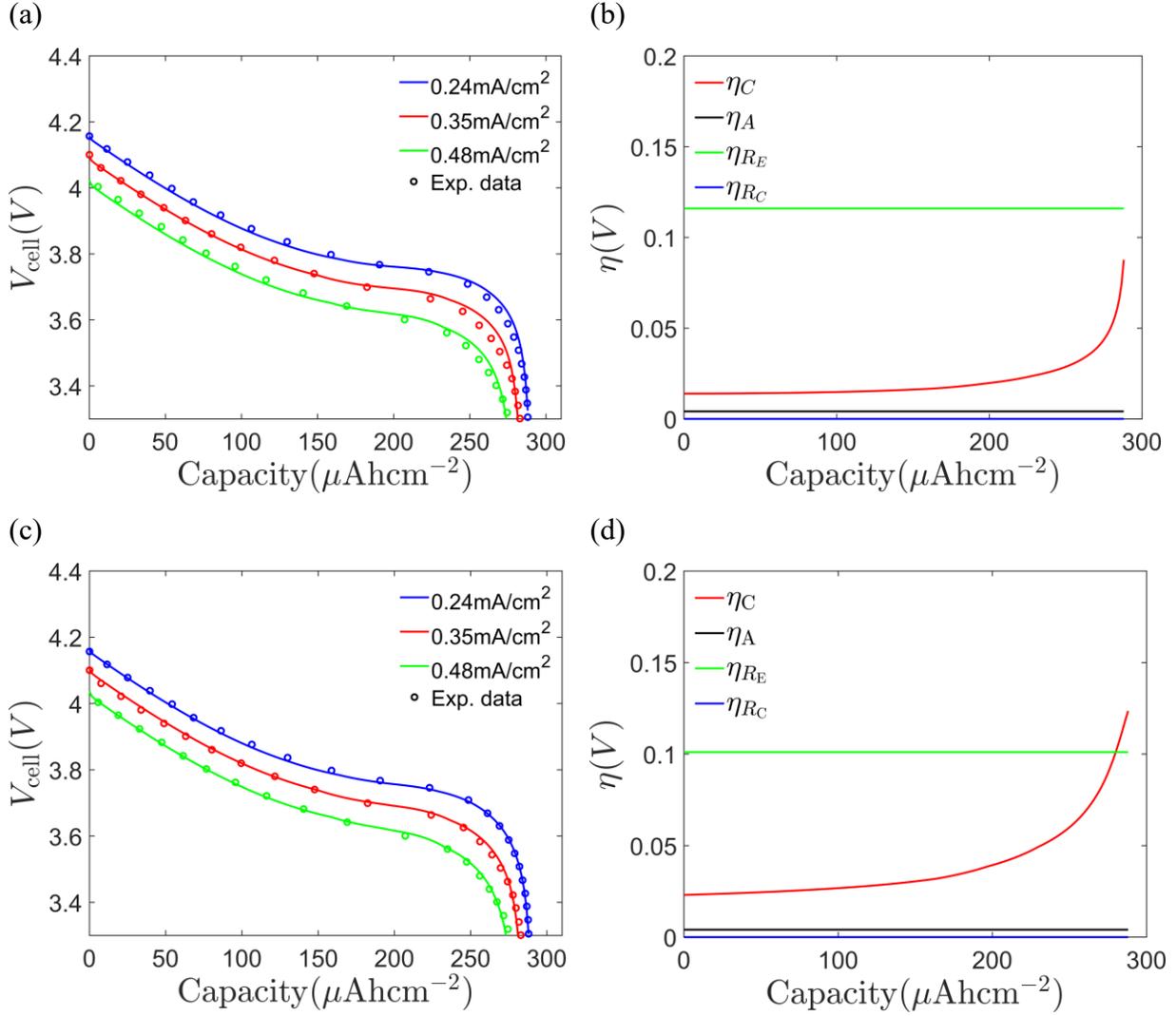

Figure 3. Panels (a) and (c) show the discharge voltage curve computed with the electroneutral and non-electroneutral model, respectively. The dotted and solid lines correspond to experimental data [18] and model output. Panels (b) and (d) show the overpotentials contributions at $j_{\text{tot}} = 0.24$ mA/cm² computed with the electroneutral and non-electroneutral model, respectively.

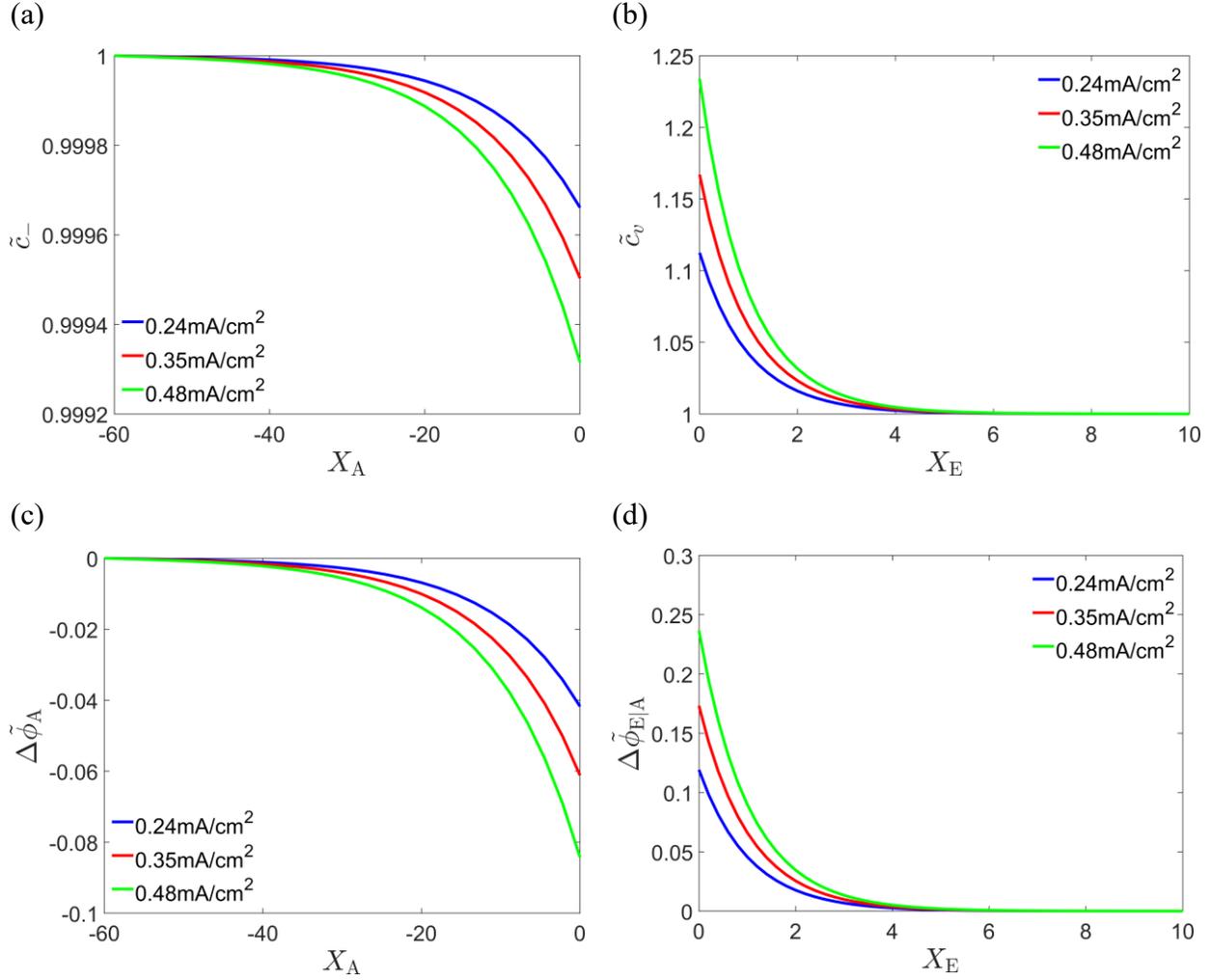

Figure 4. The SCL of the anode|electrolyte interface computed with the non-electroneutral model at various current densities. Panels (a) and (c) show the normalized concentration of electrons and electric potential, respectively, in the anode side of the SCL. Panels (b) and (d) show the normalized concentration of Li vacancies and electric potential, respectively, in the electrolyte side of the SCL.

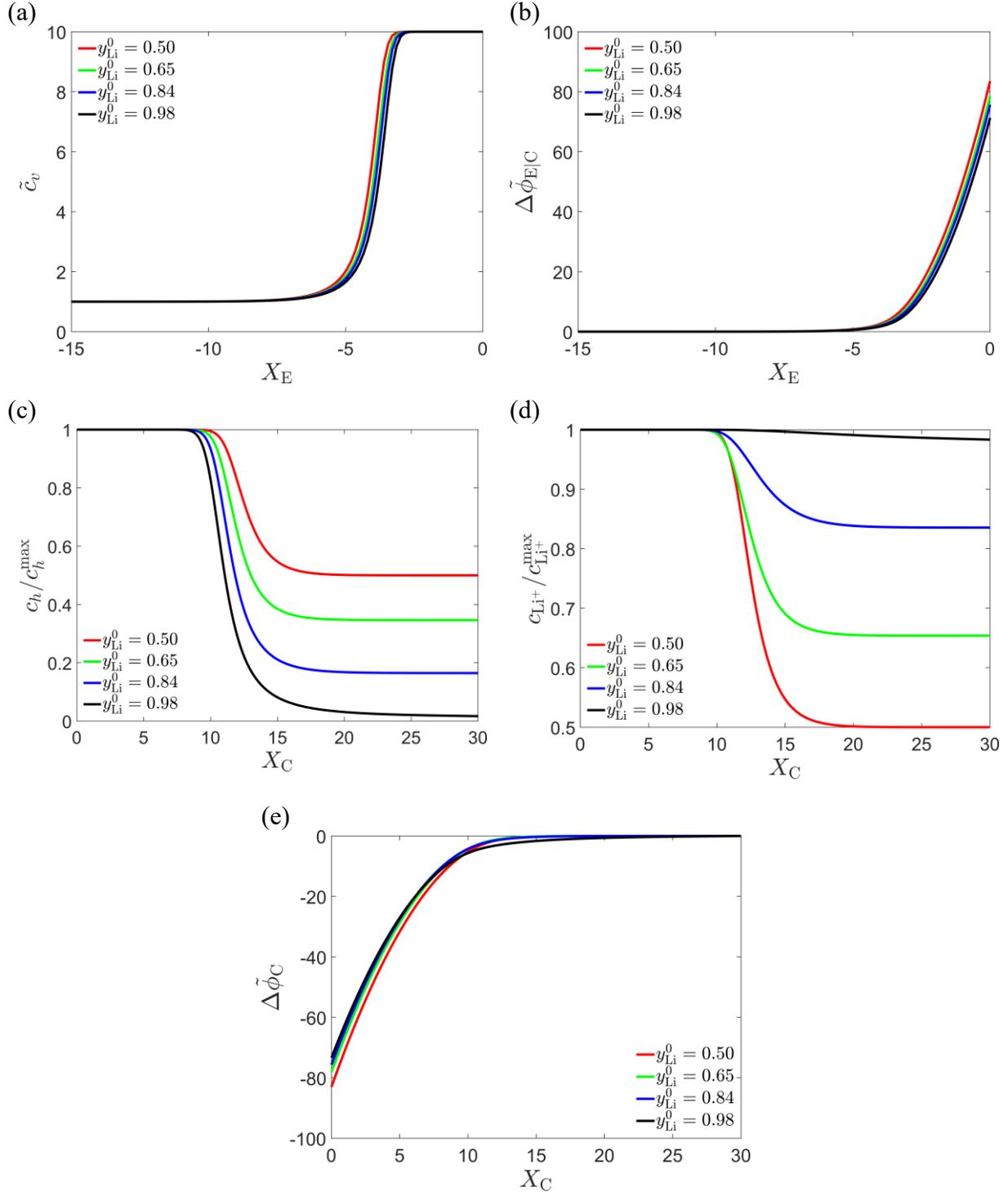

Figure 5. The SCL of the electrolyte|cathode interface calculated with the non-electroneutral model at $j_{\text{tot}} = 0.24$ mA/cm$^2$. Panels (a) and (b) show the normalized concentration of Li vacancies and electric potential in the electrolyte side of the SCL, respectively. Panel (c), (d), and (e) show the normalized holes and Li concentrations and electric potential in the cathode side of the SCL, respectively.

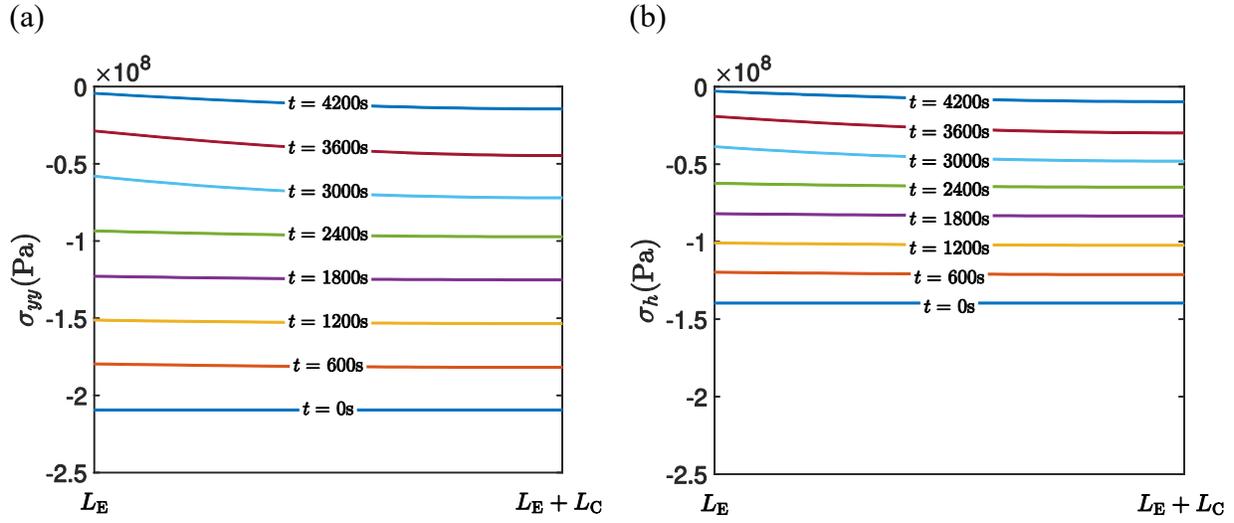

Figure 6. Evolution of the (a) in-plane and (b) hydrostatic stresses in the bulk of the cathode computed at $j_{\text{tot}} = 0.24$ mA/cm$^2$.

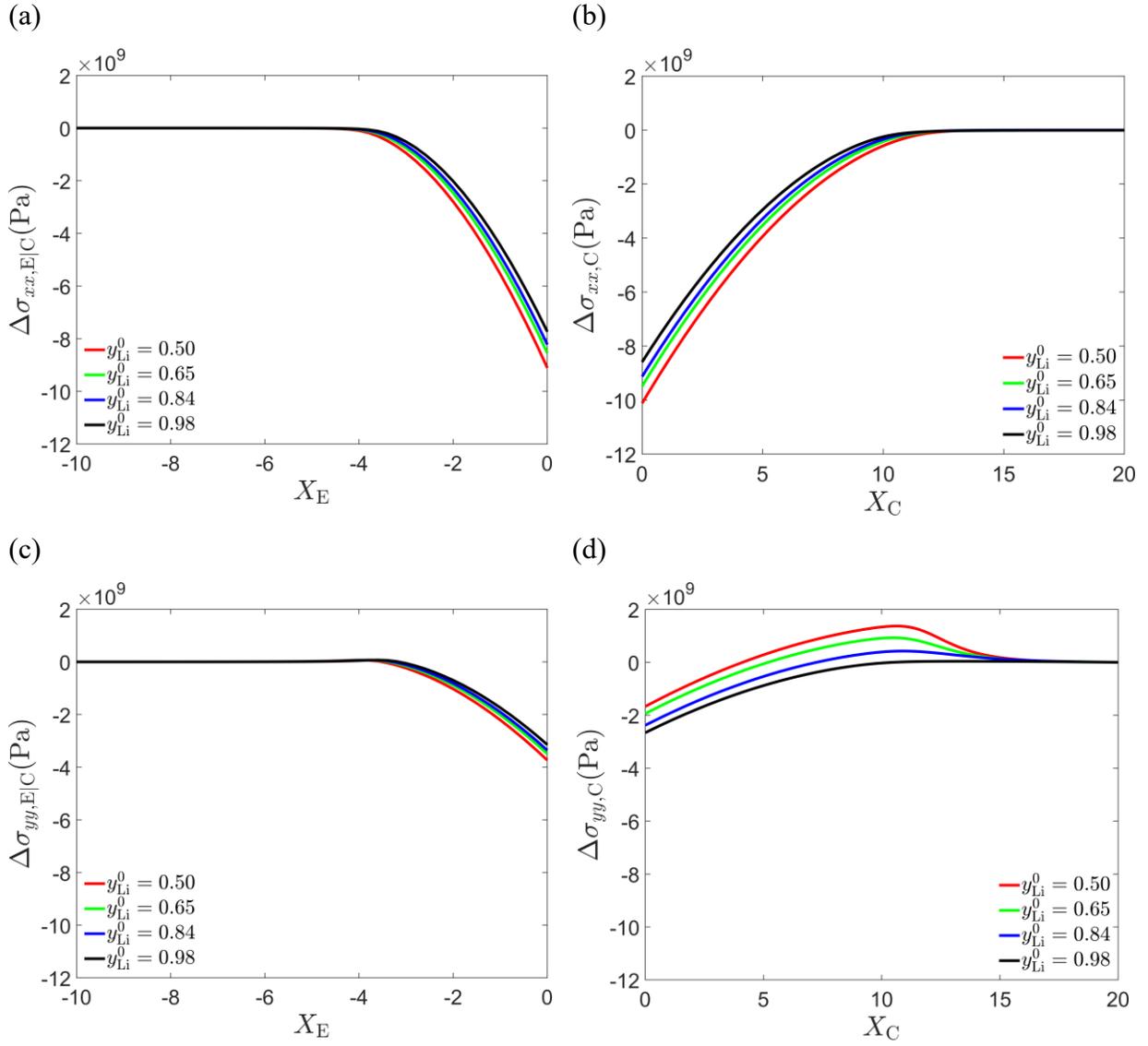

Figure 7. Stress computed in the electrolyte|cathode SCL. Panels (a) and (b) show the out-of-plane stress in the electrolyte and cathode. Panels (c) and (d) show the in-plane stress for electrolyte and cathode.

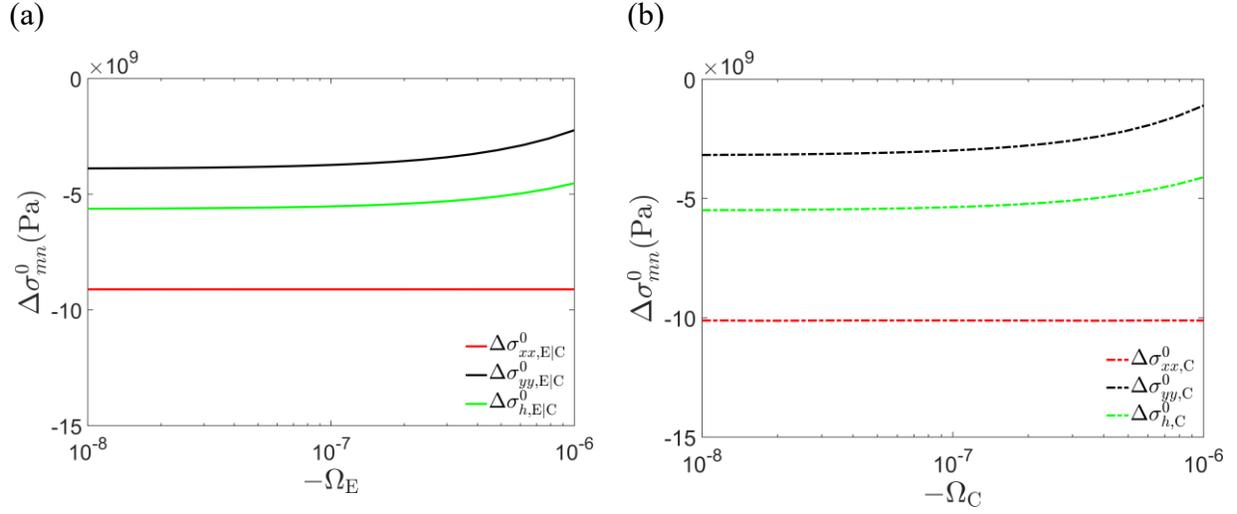

Figure 8. Stresses exerted at the electrolyte|cathode interface ($X_E = 0$ and $X_C = 0$) as a function of $\Omega_E$ and $\Omega_C$. The $\Delta\sigma^0_{mn,E|C} = \Delta\sigma_{mn,E|C}(X_E = 0)$ versus the partial molar volume $\Omega_E$, panel (a). The $\Delta\sigma^0_{mn,C} = \Delta\sigma_{mn,C}(X_C = 0)$ versus the partial molar volume $\Omega_C$, panel (b).

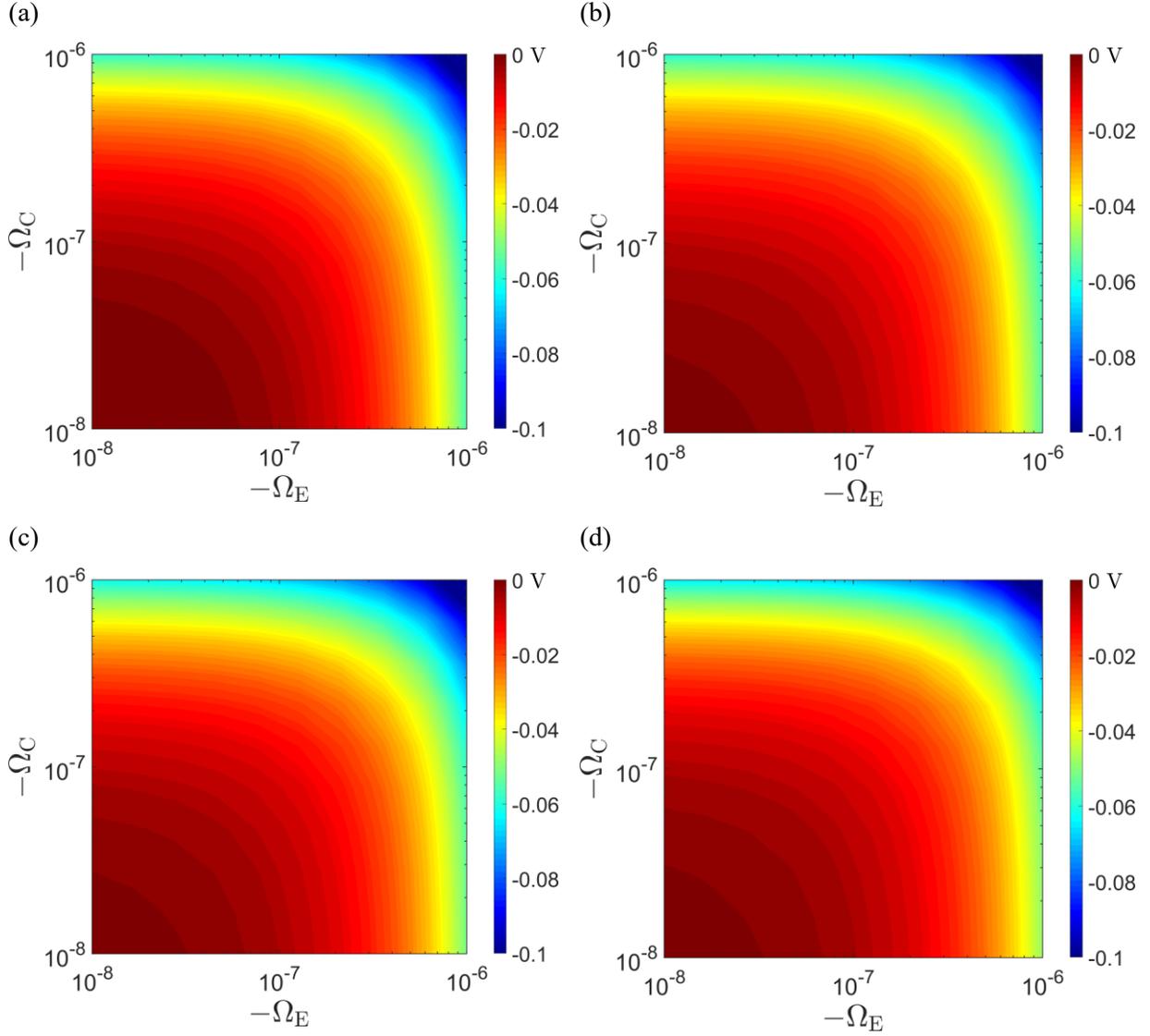

Figure 9. The difference in charge transfer overpotential $(\eta_C)_{ECM} - (\eta_C)_{EN}$ for various $\Omega_E$ and $\Omega_C$ and $j_{tot} = 0.24$ mA/cm$^2$ at (a) $y_{Li}^0 = 0.5$, (b) $y_{Li}^0 = 0.65$, (c) $y_{Li}^0 = 0.84$, and (d) $y_{Li}^0 = 0.98$. $(\eta_C)_{ECM}$ and $(\eta_C)_{EN}$ refer to $\eta_C$ computed from the charge transfer equation with and without the stress modification, respectively.

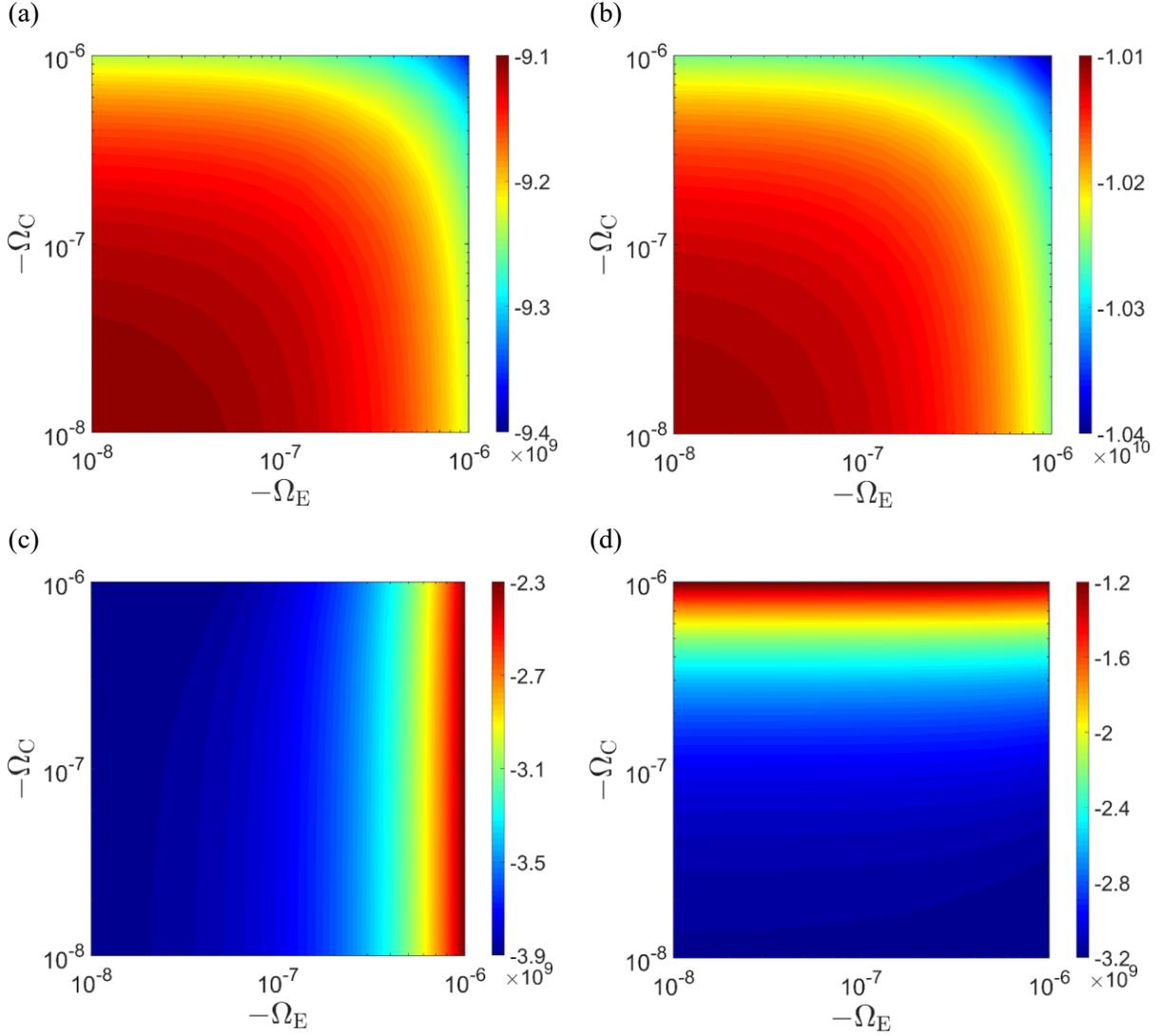

Figure 10. Stresses exerted at the electrolyte|cathode interface ($X_E = 0$ and $X_C = 0$) as a function of $\Omega_E$ and $\Omega_C$. Panels (a) and (c) show $\Delta\sigma^0_{xx,E|C}$ and $\Delta\sigma^0_{yy,E|C}$ in the electrolyte side of the electrolyte|cathode interface, respectively. Panels (b) and (d) present $\Delta\sigma^0_{xx,C}$ and $\Delta\sigma^0_{yy,C}$ in the cathode side of the same interface, respectively. All units are in Pa, $y^0_{Li} = 0.5$, and $j_{tot} = 0.24$ mA/cm$^2$.

# Supplementary Information

## Section A Rearrangement of the Kinetic Equations

As mentioned in the main text, the reaction kinetics equations at the two interfaces, *i.e.*, (18) and (28), can further be simplified. In particular, the Gibbs free energy shifts, *i.e.*, (12) and (22), can be rewritten as a function of the reduced electrochemical potentials of the various species taking part in the reactions, *i.e.*,

$$\Delta g_C^* = \frac{\Delta g_C}{e} = \mu_{Li}^*(E) - \tilde{\mu}_-^*(C) + \tilde{\mu}_v^*(E) - \mu_{Li}^*(C) \qquad \text{(S.1a)}$$

$$\Delta g_A^* = \frac{\Delta g_A}{e} = -\tilde{\mu}_v^*(E) + \mu_{Li}^*(A) - \mu_{Li}^*(E) + \tilde{\mu}_-^*(A) \qquad \text{(S.1b)}$$

At equilibrium, $\mu_{Li}^{*,eq}(E)$ and $\tilde{\mu}_v^{*,eq}(E)$ are constant throughout the electrolyte. This implies that $J_{Li} = 0$ and

$$\Delta g_C^{*,eq} = \mu_{Li}^{*,eq}(E) - \tilde{\mu}_-^{*,eq}(C) + \tilde{\mu}_v^{*,eq}(E) - \mu_{Li}^{*,eq}(C) = 0 \qquad \text{(S.2a)}$$

$$\Delta g_A^{*,eq} = \tilde{\mu}_v^{*,eq}(E) - \mu_{Li}^{*,eq}(A) + \mu_{Li}^{*,eq}(E) - \tilde{\mu}_-^{*,eq}(A) = 0 \qquad \text{(S.2b)}$$

By putting (S.2a) into (16), we obtain that

$$J_{E \to C} = \frac{J_{E \to C}^0}{\gamma_C^\ddagger} \qquad \text{(S.3a)}$$

$$J_{C \to E} = \frac{J_{C \to E}^0}{\gamma_C^\ddagger} \qquad \text{(S.3b)}$$

Similarly, by inserting (S.2b) into (26), we have

$$J_{E \to A} = \frac{J_{E \to A}^0}{\gamma_A^\ddagger} \qquad \text{(S.4a)}$$

$$J_{A \to E} = \frac{J_{A \to E}^0}{\gamma_A^\ddagger} \qquad \text{(S.4b)}$$

By inserting (S.3) and (S.4) into (15) and (25), respectively, and utilizing the fact that $J_{Li} = 0$ at equilibrium, we obtain that

$$J_{E \to C}^0 = J_{C \to E}^0 \qquad \text{(S.5a)}$$

$$J_{A \to E}^0 = J_{E \to A}^0 \qquad \text{(S.5b)}$$

We shall assume that (S.5) continues to hold even if the rate of interfacial charge transfer is not zero. Hence, by inserting (S.5a) and (S.1a) into (18), the total Li ion flux across the electrolyte|cathode interface can be written as

$$J_{\text{Li}} = J_C^0 \left( \exp\left(\frac{\alpha_C \Delta g_C^*}{V_{\text{th}}}\right) - \exp\left(-\frac{(1-\alpha_C)\Delta g_C^*}{V_{\text{th}}}\right) \right) \tag{S.6}$$

where the pre-exponential term $J_C^0$ is given by

$$J_C^0 = \frac{J_{E \to C}^0}{\gamma_C^\ddagger} = \frac{J_{C \to E}^0}{\gamma_C^\ddagger} = \frac{(J_{E \to C}^0)^{\alpha_C}(J_{C \to E}^0)^{1-\alpha_C}}{\gamma_C^\ddagger} \tag{S.7}$$

Moreover, inserting (S.1b) and (S.5b) into (29), we obtain

$$J_{\text{Li}} = J_A^0 \left( \exp\left(\frac{\alpha_A \Delta g_A^*}{V_{\text{th}}}\right) - \exp\left(-\frac{(1-\alpha_A)\Delta g_A^*}{V_{\text{th}}}\right) \right) \tag{S.8}$$

with the pre-exponential term $J_A^0$ given by

$$J_A^0 = \frac{J_{A \to E}^0}{\gamma_A^\ddagger} = \frac{J_{E \to A}^0}{\gamma_A^\ddagger} = \frac{(J_{A \to E}^0)^{\alpha_A}(J_{E \to A}^0)^{1-\alpha_A}}{\gamma_A^\ddagger} \tag{S.9}$$

In addition, we can write, using (S.2a) and (S.2b), that

$$V_{\text{oc}} = \tilde{\mu}_-^{*,\text{eq}}(C) - \tilde{\mu}_-^{*,\text{eq}}(A) = -\mu_{\text{Li}}^{*,\text{eq}}(C) + \mu_{\text{Li}}^{*,\text{eq}}(A) \tag{S.10}$$

where $V_{\text{oc}}$ is the open-circuit voltage. The anode is a Li reservoir. Therefore we can use it as a reference and write that $\mu_{\text{Li}}^{*,\text{eq}}(A) = 0$, obtaining

$$V_{\text{oc}} = -\mu_{\text{Li}}^{*\text{eq}}(C) \tag{S.11}$$

$V_{\text{oc}}$ is a function of the concentration of Li in the cathode [1-3], *i.e.*,

$$V_{\text{oc}} = V_{\text{oc}}(c_{\text{Li}}(C)) \tag{S.12}$$

The $\mu_{\text{Li}}^*(C)$ depends on the concentration of Li only. Therefore, under the assumption of local equilibrium we can take $\mu_{\text{Li}}^*(C) = \mu_{\text{Li}}^{*,\text{eq}}(C)$. Then, by inserting (S.11) into (S.1a), $\Delta g_C^*$ becomes

$$\Delta g_C^* = \mu_{\text{Li}}^*(E) - \tilde{\mu}_-^*(C) + \tilde{\mu}_v^*(E) + V_{\text{oc}} \tag{S.13}$$

Moreover, if we reference the equilibrium chemical potential of Li against Li and follow a similar argument to the one above, we obtain that $\mu_{\text{Li}}^*(A) = 0$, (S.1b) becomes

$$\Delta g_A^* = -\mu_{Li}^*(E) + \tilde{\mu}_-^*(A) - \tilde{\mu}_v^*(E) \tag{S.14}$$

## Section B Governing Equation of the Li Transportation in the Cathode

From (1a), the continuity equation for the Li ions and electrons in the cathode are given by

$$\frac{\partial c_{Li^+}}{\partial t} + \frac{1}{e}\frac{\partial j_{Li^+}}{\partial x} = 0 \tag{S.15a}$$

$$\frac{\partial c_-}{\partial t} - \frac{1}{e}\frac{\partial j_-}{\partial x} = 0 \tag{S.15b}$$

Because of electroneutrality, $c_{Li^+} = c_-$. Then, by subtracting (S.15a) with (S.15b), we obtain

$$\frac{\partial}{\partial x}(j_{Li^+} + j_-) = 0 \tag{S.16}$$

This implies that the total current density $j_{tot} = j_{Li^+} + j_-$ is a constant. Also, from (1b), the fluxes of Li ions and electrons are given by

$$j_{Li^+} = -\sigma_{Li^+}\frac{\partial \tilde{\mu}_{Li^+}^*}{\partial x} \tag{S.17a}$$

$$j_- = -\sigma_-\frac{\partial \tilde{\mu}_-^*}{\partial x} \tag{S.17b}$$

where $\tilde{\mu}_{Li^+}^*$ and $\tilde{\mu}_-^*$ are the reduced electrochemical potentials of Li ions and electrons, respectively. By adding (S.17a) to the product of (S.17b) and $-\sigma_{Li^+}/\sigma_-$, we obtain

$$j_{Li^+} - \frac{\sigma_{Li^+}}{\sigma_-}j_- = -\sigma_{Li^+}\frac{\partial}{\partial x}(\tilde{\mu}_{Li^+}^* - \tilde{\mu}_-^*) \tag{S.18}$$

Because of electroneutrality, the total charge of the cathode is zero and the reaction $Li^+ + e^- \rightleftarrows Li$ is in local thermodynamic equilibrium, implying that

$$\tilde{\mu}_{Li^+}^* - \tilde{\mu}_-^* - \mu_{Li}^* = 0 \tag{S.19}$$

By substituting (S.19) into (S.18) and utilizing the fact that $j_- = j_{Li^+} - j_{tot}$, we obtain

$$j_{Li^+} = -\frac{\sigma_-\sigma_{Li^+}}{\sigma_- + \sigma_{Li^+}}\frac{\partial \mu_{Li}^*}{\partial x} + \frac{\sigma_{Li^+}}{\sigma_- + \sigma_{Li^+}}j_{tot} = -\sigma_{chem}\left(\frac{\partial \mu_{Li}^*}{\partial x} - \frac{j_{tot}}{\sigma_-}\right) \tag{S.20}$$

where $\sigma_{chem}$ is the chemical conductivity defined as

$$\sigma_{\text{chem}} = \frac{\sigma_-\sigma_{\text{Li}^+}}{\sigma_- + \sigma_{\text{Li}^+}} = \frac{D_- D_{\text{Li}^+}}{D_- + D_{\text{Li}^+}} \frac{e^2 c_{\text{Li}^+}}{k_B T} = D_{\text{chem}} \frac{e^2 c_{\text{Li}^+}}{k_B T} \qquad (S.21)$$

where $D_{\text{chem}}$ is the overall diffusivity. One should note that $c_{\text{Li}^+} = c_{\text{Li}}$ in the bulk of the cathode and the ionic current density $j_{\text{Li}^+}$ equals the Li molecular flux $J_{\text{Li}}$ times the electron charge, i.e., $j_{\text{Li}^+} = e J_{\text{Li}}$. Therefore, we can rewrite (S.15a) as

$$\frac{\partial c_{\text{Li}}}{\partial t} + \frac{\partial J_{\text{Li}}}{\partial x} = 0 \qquad (S.22)$$

Using (S.20), (S.21) and the definition $\mu_{\text{Li}}^* = \frac{\mu_{\text{Li}}^0}{e} + \frac{k_B T}{e} \ln\left(\gamma_{\text{Li}} \frac{c_{\text{Li}}}{c_{\text{Li}}^0}\right)$, (S.22) can be rewritten as

$$\frac{\partial c_{\text{Li}}}{\partial t} + \frac{\partial}{\partial x}\left(-\widetilde{D}_{\text{chem}} \frac{\partial c_{\text{Li}}}{\partial x} + \frac{\partial}{\partial x}\left(\frac{\sigma_{\text{chem}}}{\sigma_-} j_{\text{tot}}\right)\right) = 0 \qquad (S.23)$$

where $\widetilde{D}_{\text{chem}} = D_{\text{chem}}\left(1 + \frac{\partial \ln \gamma_{\text{Li}}}{\partial \ln c_{\text{Li}}}\right)$ is the chemical diffusivity. Furthermore, if we use (S.16) and assume that $\frac{\sigma_{\text{chem}}}{\sigma_-}$ is a constant within the cathode, we can simplify (S.23) to

$$\frac{\partial c_{\text{Li}}}{\partial t} + \frac{\partial}{\partial x}\left(-\widetilde{D}_{\text{chem}} \frac{\partial c_{\text{Li}}}{\partial x}\right) = 0 \qquad (S.24)$$

**Section C Reformulation of the Charge Transfer Equations**

*C.1 Electrolyte|Cathode Interface*

In this section, we derive the conventional Butler-Volmer framework used in the electroneutral model from the general framework in section 2.3 of the main text. From (26b), we have

$$\Delta g_C^* = \mu_{\text{Li}}^*(E) - \widetilde{\mu}_-^*(C) + \widetilde{\mu}_v^*(E) + V_{\text{oc}} \qquad (S.25)$$

Upon expanding (S.25) with the definition of reduced electrochemical potential, i.e., $\widetilde{\mu}_i^* = (\mu_i + z_i e \phi)/z_i e$, we have that

$$\mu_{\text{Li}}^*(E) - \widetilde{\mu}_-^*(C) + \widetilde{\mu}_v^*(E) + V_{\text{oc}} = \Delta \mu_{\text{EC}}^* + \phi_{\text{E|C}}(L_E^-) - \phi_C(L_E^+) + V_{\text{oc}}\left(c_{\text{Li}}|_{x=L_E^+}\right) \qquad (S.26)$$

where $\Delta \mu_{\text{EC}}^* = \mu_{\text{Li}}^*(E) - \mu_-^*(C) + \mu_v^*(E)$. Then, we can rewrite (25) to:

$$J_{Li}$$
$$= J_C^0 \left( \exp\left(\frac{\alpha_C \Delta\mu_{EC}^*}{V_{th}}\right) \exp\left(\frac{\alpha_C \left(\phi_{E|C}(L_E^-) - \phi_C(L_E^+) + V_{oc}\left(c_{Li}|_{x=L_E^+}\right)\right)}{V_{th}}\right) \right.$$
$$\left. - \exp\left(\frac{-(1-\alpha_C)\Delta\mu_{EC}^*}{V_{th}}\right) \exp\left(-\frac{(1-\alpha_C)\left(\phi_{E|C}(L_E^-) - \phi_C(L_E^+) + V_{oc}\left(c_{Li}|_{x=L_E^+}\right)\right)}{V_{th}}\right) \right) \quad (S.27)$$

If we assume that the system offset from equilibrium is only because of the electric potential variation, $\Delta\mu_{EC}^*$ in (S.27) equals zero [4]. Also, upon considering that $j_{Li^+} = eJ_{Li}$, we have

$$j_{Li^+} = j_C^0 \left( \exp\left(\frac{\alpha_C \eta_C}{V_{th}}\right) - \exp\left(-\frac{(1-\alpha_C)\eta_C}{V_{th}}\right) \right) \quad (S.28)$$

where $j_C^0 = eJ_C^0$ and $\eta_C = \phi_{E|C}(L_E^-) - \phi_C(L_E^+) + V_{oc}\left(c_{Li}|_{x=L_E^+}\right)$ is the charge transfer overpotential at the electrolyte|cathode interface. (S.28) is the Butler-Volmer formulation typical of the charge transfer kinetics [2].

## C.2 Anode|Electrolyte Interface

From (30b), we have

$$\Delta g_A^* = -\mu_{Li}^*(E) + \tilde{\mu}_-^*(A) - \tilde{\mu}_v^*(E) \quad (S.29)$$

Upon expanding (S.29) with the definition of reduced electrochemical potential, *i.e.*, $\tilde{\mu}_i^* = (\mu_i + z_i e\phi)/z_i e$, we obtain

$$-\mu_{Li}^*(E) + \tilde{\mu}_-^*(A) - \tilde{\mu}_v^*(E) = \phi_A(0^-) - \phi_{E|A}(0^+) + \Delta\mu_{AE}^* \quad (S.30)$$

where $\Delta\mu_{AE}^* = -\mu_{Li}^*(E) + \mu_-^*(A) - \mu_v^*(E)$. Then, we can rewrite (29) to give

$$J_{Li} = J_A^0 \left( \exp\left(\frac{\alpha_A \Delta\mu_{AE}^*}{V_{th}}\right) \exp\left(\frac{\alpha_A \left(\phi_A(0^-) - \phi_{E|A}(0^+)\right)}{V_{th}}\right) \right.$$
$$\left. - \exp\left(-\frac{(1-\alpha_A)\Delta\mu_{AE}^*}{V_{th}}\right) \exp\left(-\frac{(1-\alpha_A)\left(\phi_A(0^-) - \phi_{E|A}(0^+)\right)}{V_{th}}\right) \right) \quad (S.31)$$

If we assume that only electric potential variation contributes to the departure of equilibrium, $\Delta\mu_{AE}^* = 0$. Also, since $j_{Li^+} = eJ_{Li}$, we can rewrite (S.31) into

$$j_{\text{Li}^+} = j_A^0 \left( \exp\left(\frac{\alpha_A \eta_A}{V_{\text{th}}}\right) - \exp\left(-\frac{(1-\alpha_A)\eta_A}{V_{\text{th}}}\right) \right) \tag{S.32}$$

where $\eta_A = \phi_A(0^-) - \phi_{E|A}(0^+)$. (S.32) is in good agreement with the typical charge transfer kinetics at the anode|electrolyte interface using the Butler-Volmer formulation [2].

## Section D Implementation

### D.1 Electroneutral Model

To obtain the discharge voltage from the electroneutral model, one first solves the cathode charge transport equation (32) equipped with the boundary conditions (33). Using $c_{\text{Li}}(t,x)$, the open circuit voltage $V_{\text{oc}}\left(c_{\text{Li}}|_{x=L_E^+}\right)$ can then be obtained as a function of time from (40). After that, one can solve the charge transfer equations (34) and (36). These equations model the reaction at the two interfaces and are used to obtain the cathodic and anodic overpotentials, *i.e.*, $\eta_C$ and $\eta_A$, respectively. The Ohmic losses in the electrolyte and the cathode, *i.e.*, $\eta_{R_E}$ and $\eta_{R_C}$, can be computed with (41) from the dimensions of the SSB and the conductivities of the electrolyte and cathode materials. Using (39), the discharge voltage can finally be obtained and compared to the experimental data.

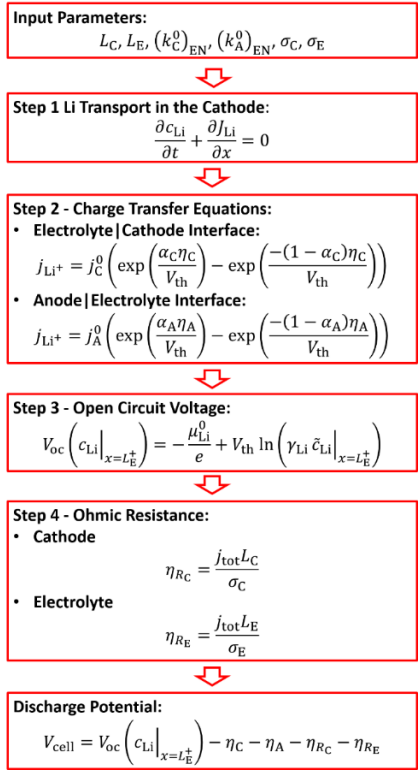

Figure S.1 The workflow of the electroneutral model.

## D.2 Non-Electroneutral Model

To solve the non-electroneutral model, we first resolve the Li transportation in the bulk of the cathode using the same governing equation, *i.e.*, (32), as the electroneutral model. After that, $\eta_C$ and $\eta_A$ are evaluated by solving the charge transfer equations (34) and (36) with the modified exchange current density (65) and (66), respectively. However, as mentioned in subsection 4.5, the modified exchange current densities (65) and (66) depend on the species concentrations in the SCLs, which, in turn, depend on the potential difference across the interfaces. Therefore, we need to use an iterative method to solve for $\eta_C$ and $\eta_A$, together with the concentrations and electric potential distributions in the SCLs.

Trial values of $\eta_C$ and $\eta_A$ are first used to compute $\Delta\tilde{\phi}_{EC} = \dfrac{V_{oc}(c_{Li}|_{x=L_E+\delta x_C}) - \eta_C}{V_{th}}$ and $\Delta\tilde{\phi}_{AE} = -\dfrac{\eta_A}{V_{th}}$. Secondly, given $\Delta\tilde{\phi}_{EC}$, the potential differences in the electrolyte and cathode side of the electrolyte|cathode interface, *i.e.*, $\Delta\tilde{\phi}^0_{E|C}$ and $\Delta\tilde{\phi}^0_{C}$, respectively, are computed by solving a system of equations comprising (60) and (62). Similarly, given $\Delta\tilde{\phi}_{AE}$, the potential differences in the

anode and electrolyte side of the anode|electrolyte interface, *i.e.*, $\Delta\tilde{\phi}_A^0$ and $\Delta\tilde{\phi}_{E|A}^0$, respectively, are evaluated by solving the system of equations consisting of (61) and (63). Thirdly, the electric potential distributions of the SCLs are computed by solving (45), (50), and (55) with the boundary conditions of (46b), (51b), and (56b), respectively. Fourthly, the concentrations of the charged species in the SCLs are obtained using (44), (49), and (54). Fifthly, the computed species concentration are input back to the charge transfer equations (34) and (36) in order to solve for the iterated $\eta_C$ and $\eta_A$ values. Finally, the process is repeated until the $\eta_C$ and $\eta_A$ iteration residuals are less than $10^{-10}$V.

After estimating $\eta_C$ and $\eta_A$ from the non-electroneutral model, we can obtain the electric potential and species concentration in the SCLs. Also, we can compute the overall discharge potential with (39), where $V_{oc}$, $\eta_{R_C}$, and $\eta_{R_E}$ are estimated from the respective equations of the electroneutral model (see subsection D.1).

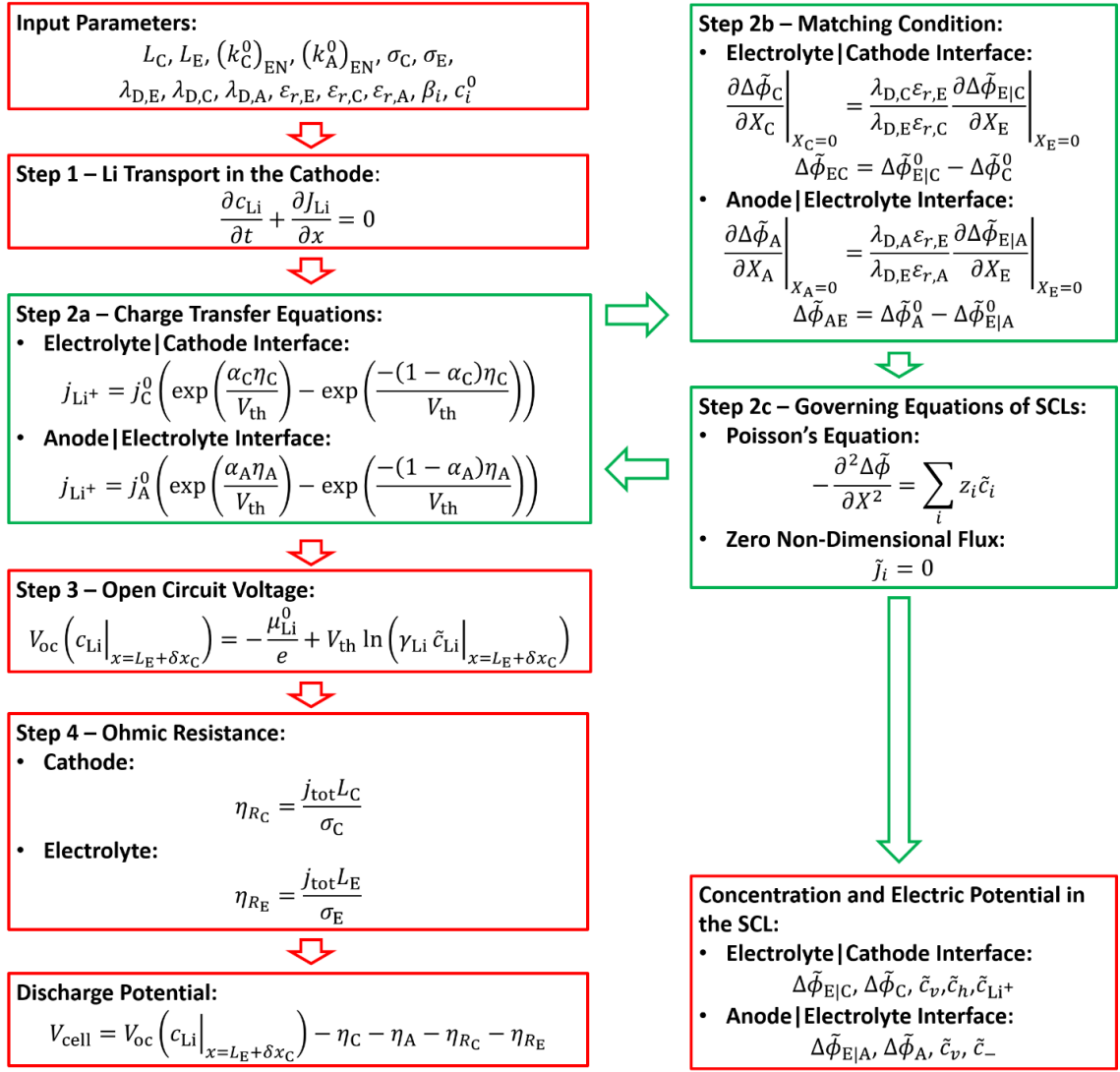

Figure S.2 The workflow of the non-electroneutral model. The green boxes and arrows denote the iteration process of solving the charge transfer equation and the SCL profile.

## D.3 Electro-Chemo-Mechanical Model

First, the transportation equation (31) and the mechanical equilibrium equation (67) are solved for the bulk of the cathode. This leads to computing the Li concentration, $c_{\text{Li}}$, and the mechanical stress, $\sigma_{mn}$, in the bulk of the cathode. Subsequently, the stress-modified charge transfer equation (*i.e.* (83) and (84)) and the governing equations of the SCLs are solved together using a fixed-point iteration.

Firstly, similar to the non-electroneutral model (see subsection D.2), trial $\eta_{\text{C}}$ and $\eta_{\text{A}}$ values are used to estimate the potential difference $\Delta \tilde{\phi}_{\text{EC}}$ and $\Delta \tilde{\phi}_{\text{AE}}$, respectively. Secondly, $\Delta \tilde{\phi}^0_{\text{E|C}}$ and $\Delta \tilde{\phi}^0_{\text{C}}$ are computed by solving a system of equation consisting of (60) and (62). Similarly, $\Delta \tilde{\phi}^0_{\text{E|A}}$ and $\Delta \tilde{\phi}^0_{\text{A}}$ are evaluated by the system of equation comprising (61) and (63). Thirdly, Poisson's equations, *i.e.*, (43c), (47), and (55), the mechanical equation, *i.e.*, (74), (77), and (80), and the equations of constant electrochemical potential, *i.e.*, (76), (79), and (54), are used to compute the concentration of the charged species and the stress differences in the electrolyte, cathode, and anode, respectively. Fourthly, the computed $\tilde{c}_v$, $\tilde{c}_{\text{Li}^+}$, and $\Delta \tilde{\sigma}_{h,\text{C}}$ from the SCL and $\sigma_h$ from the bulk of the cathode are then input into (83) to obtain an iterated $\eta_{\text{C}}$ value. Similarly, the computed $\tilde{c}_v$ and $\Delta \tilde{\sigma}_{h,\text{A}}$ are input into (84) to compute an iterated $\eta_{\text{A}}$ value. Finally, the iteration is repeated with the newly iterated $\eta_{\text{C}}$ and $\eta_{\text{A}}$ values until the iteration residual is less than $10^{-10}$V. The result is an estimation to $\eta_{\text{C}}$ and $\eta_{\text{A}}$, which can then be input to (39), together with $V_{\text{oc}}$, $\eta_{R_{\text{C}}}$, and $\eta_{R_{\text{E}}}$ as computed from the previously developed model (see subsection D.1) to give the stress-modified discharge potential.

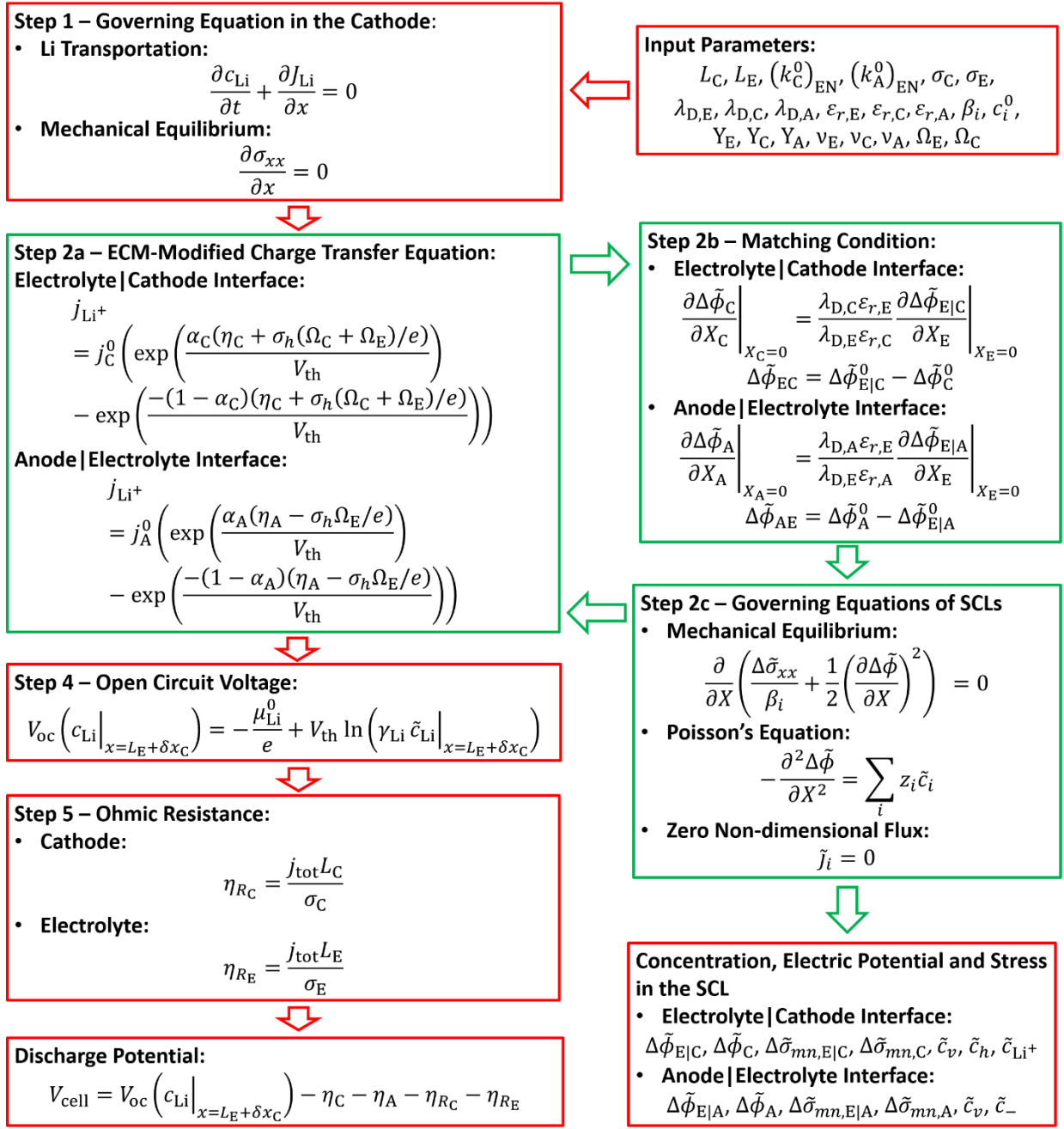

Figure S.3 The workflow of the ECM model. The green boxes and arrows denote the iteration process of solving the stress-modified charge transfer equation and the SCL profile of the electrolyte|cathode interface

# Section E Reformulating the Boundary Condition for Solving the Poisson's Equation in the Space Charge Layer

## E.1 Space Charge Layer in the Electrolyte

We are going to use Poisson's equation to reformulate the boundary condition. First, we rewrite (45):

$$-\frac{\partial^2 \Delta\tilde{\phi}_{E|C}}{\partial X_E^2} = -\frac{(\beta_v - 1)(e^{\Delta\tilde{\phi}_{E|C}} - 1)}{e^{\Delta\tilde{\phi}_{E|C}} + \beta_v - 1} \tag{S.33}$$

Multiplying both sides of (S.33) by $\frac{\partial \Delta\tilde{\phi}_{E|C}}{\partial X_E}$ gives

$$-\frac{1}{2}\frac{\partial}{\partial X_E}\left(\frac{\partial \Delta\tilde{\phi}_{E|C}}{\partial X_E}\right)^2 = -\frac{(\beta_v - 1)(e^{\Delta\tilde{\phi}_{E|C}} - 1)}{e^{\Delta\tilde{\phi}_{E|C}} + \beta_v - 1}\frac{\partial \Delta\tilde{\phi}_{E|C}}{\partial X_E} \tag{S.34}$$

We can then integrate (S.34) from $-\infty$ to $X_E$, and apply the boundary conditions (46). This integration gives

$$-\frac{1}{2}\left[\left(\frac{\partial \Delta\tilde{\phi}_{E|C}}{\partial X_E}\right)^2 - \left(\frac{\partial \Delta\tilde{\phi}_{E|C}}{\partial X_E}\bigg|_{X_E=-\infty}\right)^2\right] = -\beta_v \ln\left(\frac{e^{\Delta\tilde{\phi}_{E|C}} + \beta_v - 1}{\beta_v}\right) + \Delta\tilde{\phi}_{E|C} \tag{S.35}$$

and, therefore,

$$\frac{\partial \Delta\tilde{\phi}_{E|C}}{\partial X_E} = \pm\sqrt{2\left(\beta_v \ln\left(\frac{e^{\Delta\tilde{\phi}_{E|C}} + \beta_v - 1}{\beta_v}\right) - \Delta\tilde{\phi}_{E|C}\right)} \tag{S.36}$$

Since, $\frac{\partial \Delta\tilde{\phi}_{E|C}}{\partial X_E} > 0$ when $\Delta\tilde{\phi}_{E|C} > 0$, and $\frac{\partial \Delta\tilde{\phi}_{E|C}}{\partial X_E} < 0$ when $\Delta\tilde{\phi}_{E|C} < 0$, it follows that

$$\frac{\partial \Delta\tilde{\phi}_{E|C}}{\partial X_E} = \text{sign}(\Delta\tilde{\phi}_{E|C})\sqrt{2\left(\beta_v \ln\left(\frac{e^{\Delta\tilde{\phi}_{E|C}} + \beta_v - 1}{\beta_v}\right) - \Delta\tilde{\phi}_{E|C}\right)} \tag{S.37}$$

By plugging in the boundary condition (48b), we obtain

$$\frac{\partial \Delta\tilde{\phi}_{E|C}}{\partial X_E}\bigg|_{X_E=0} = \text{sign}(\Delta\tilde{\phi}_{E/C}^0)\sqrt{2\left(\beta_v \ln\left(\frac{e^{\Delta\tilde{\phi}_{E/C}^0} + \beta_v - 1}{\beta_v}\right) - \Delta\tilde{\phi}_{E/C}^0\right)} \tag{S.38}$$

A similar result as (S.23) and (S.24) can also be obtained for the electrolyte side of the anode|electrolyte interface by merely substituting $\Delta\tilde{\phi}_{E|C}$ with $\Delta\tilde{\phi}_{E|A} = \tilde{\phi}(X_E) - \tilde{\phi}\big|_{X_E \to \infty}$, where $X_E \in [0, \infty)$.

## E.2 Space Charge Layer in the Cathode

From (50), we have

$$-\frac{\partial^2 \Delta\tilde{\phi}_C}{\partial X_C^2} = \frac{e^{-\Delta\tilde{\phi}_C}}{e^{-\Delta\tilde{\phi}_C} + \beta_h - 1} + \frac{e^{-\Delta\tilde{\phi}_C}}{e^{-\Delta\tilde{\phi}_C} + \beta_{Li^+} - 1} - 1 \quad (S.39)$$

If we then multiply both sides of (S.39) by $\frac{\partial \Delta\tilde{\phi}_C}{\partial \tilde{x}_C^+}$ and rearrange the terms, we obtain

$$-\frac{1}{2}\frac{\partial}{\partial X_C}\left(\frac{\partial \Delta\tilde{\phi}_C}{\partial X_C}\right)^2 = \left(\frac{e^{-\Delta\tilde{\phi}_C}}{e^{-\Delta\tilde{\phi}_C} + \beta_h - 1} + \frac{e^{-\Delta\tilde{\phi}_C}}{e^{-\Delta\tilde{\phi}_C} + \beta_{Li^+} - 1} - 1\right)\frac{\partial \Delta\tilde{\phi}_C}{\partial X_C} \quad (S.40)$$

After integrating (S.40) from $\tilde{x}_C^+$ to $\infty$, and applying the boundary condition (51), we can write

$$\frac{\partial \Delta\tilde{\phi}_C}{\partial X_C} = \pm\sqrt{2\left(\ln\frac{(\beta_h - 1)e^{\Delta\tilde{\phi}_C} + 1}{\beta_h} + \ln\frac{(\beta_{Li^+} - 1)e^{\Delta\tilde{\phi}_C} + 1}{\beta_{Li^+}} - \Delta\tilde{\phi}_C\right)} \quad (S.41)$$

Since $\frac{\partial \Delta\tilde{\phi}_C}{\partial X_C} > 0$ when $\Delta\tilde{\phi}_C < 0$, and $\frac{\partial \Delta\tilde{\phi}_C}{\partial X_C} < 0$ when $\Delta\tilde{\phi}_C > 0$, it follows that

$$\frac{\partial \Delta\tilde{\phi}_C}{\partial X_C} = -\text{sign}(\Delta\tilde{\phi}_C)\sqrt{2\left(\ln\frac{(\beta_h - 1)e^{\Delta\tilde{\phi}_C} + 1}{\beta_h} + \ln\frac{(\beta_{Li^+} - 1)e^{\Delta\tilde{\phi}_C} + 1}{\beta_{Li^+}} - \Delta\tilde{\phi}_C\right)} \quad (S.42)$$

By plugging in the boundary condition (51b), we obtain

$$\frac{\partial \Delta\tilde{\phi}_C}{\partial X_C}\bigg|_{X_C=0}$$
$$= -\text{sign}(\Delta\tilde{\phi}_C^0)\sqrt{2\left(\ln\frac{(\beta_h - 1)e^{\Delta\tilde{\phi}_C^0} + 1}{\beta_h} + \ln\frac{(\beta_{Li^+} - 1)e^{\Delta\tilde{\phi}_C^0} + 1}{\beta_{Li^+}} - \Delta\tilde{\phi}_C^0\right)} \quad (S.43)$$

## E.3 Space Charge Layer in the Anode

From (55), we have

$$\frac{\partial^2 \Delta\tilde{\phi}_A}{\partial X_A^2} = \left(\frac{\Delta\tilde{\phi}_A}{\xi} + 1\right)^{\frac{3}{2}} - 1 \tag{S.44}$$

If we multiply both sides of (S.44) with $\frac{\partial \Delta\tilde{\phi}_A}{\partial X_A}$ and rearrange the terms, we obtain

$$\frac{1}{2}\frac{\partial}{\partial X_A}\left(\frac{\partial \Delta\tilde{\phi}_A}{\partial X_A}\right)^2 = \frac{\partial}{\partial X_A}\left(\frac{2\xi}{5}\left(\frac{\Delta\tilde{\phi}_A}{\xi} + 1\right)^{\frac{5}{2}} - \Delta\tilde{\phi}_A\right) \tag{S.45}$$

We can then integrate both sides from $-\infty$ to $X_A$, and applying the boundary condition of (56). This procedure gives

$$\frac{\partial \Delta\tilde{\phi}_A}{\partial X_A} = \pm\sqrt{\frac{4\xi}{5}\left(\left(\frac{\Delta\tilde{\phi}_A}{\xi} + 1\right)^{\frac{5}{2}} - 1\right) - 2\Delta\tilde{\phi}_A} \tag{S.46}$$

Intuitively, $\frac{\partial \Delta\tilde{\phi}_A}{\partial X_A} > 0$ when $\Delta\tilde{\phi}_A > 0$, and $\frac{\partial \Delta\tilde{\phi}_A}{\partial X_A} < 0$ when $\Delta\tilde{\phi}_A < 0$. Therefore, we can write

$$\frac{\partial \Delta\tilde{\phi}_A}{\partial X_A} = \text{sign}(\Delta\tilde{\phi}_A)\sqrt{\frac{4\xi}{5}\left(\left(\frac{\Delta\tilde{\phi}_A}{\xi} + 1\right)^{\frac{5}{2}} - 1\right) - 2\Delta\tilde{\phi}_A} \tag{S.47}$$

By plugging in the boundary condition (56b), we obtain

$$\left.\frac{\partial \Delta\tilde{\phi}_A}{\partial X_A}\right|_{X_A=0} = \text{sign}(\Delta\tilde{\phi}_A^0)\sqrt{\frac{4\xi}{5}\left(\left(\frac{\Delta\tilde{\phi}_A^0}{\xi} + 1\right)^{\frac{5}{2}} - 1\right) - 2\Delta\tilde{\phi}_A^0} \tag{S.48}$$

## *E.4 Analytical Expression for the Matching Condition between the Electrolyte and the Electrode*

From (60), we have

$$\left.\frac{\partial \Delta\tilde{\phi}_C}{\partial X_C}\right|_{X_C=0} = \frac{\lambda_{D,C}\varepsilon_{r,E}}{\lambda_{D,E}\varepsilon_{r,C}}\left.\frac{\partial \Delta\tilde{\phi}_{E|C}}{\partial X_E}\right|_{X_E=0} \tag{S.49}$$

where $\lambda_{D,E}$ and $\lambda_{D,C}$ are the Debye length of the electrolyte and cathode, respectively. Also, $\varepsilon_{r,E}$ and $\varepsilon_{r,C}$ are the relative permittivities of the electrolyte and cathode materials, respectively. By plugging (S.38) and (S.43) into (S.49), it can be deduced that for the electrolyte|cathode interface the following relation holds:

$$\sqrt{\ln\frac{(\beta_h - 1)e^{\Delta\tilde{\phi}_C^0} + 1}{\beta_h} + \ln\frac{(\beta_{Li^+} - 1)e^{\Delta\tilde{\phi}_C^0} + 1}{\beta_{Li^+}} - \Delta\tilde{\phi}_C^0}$$
$$= \frac{\lambda_{D,C}\varepsilon_{r,E}}{\lambda_{D,E}\varepsilon_{r,C}}\sqrt{\beta_v \ln\left(\frac{e^{\Delta\tilde{\phi}_{E|C}^0} + \beta_v - 1}{\beta_v}\right) - \Delta\tilde{\phi}_{E|C}^0} \quad (S.50)$$

Similarly, from

$$\left.\frac{\partial\Delta\tilde{\phi}_A}{\partial X_A}\right|_{X_A=0} = \frac{\lambda_{D,A}\varepsilon_{r,E}}{\lambda_{D,E}\varepsilon_{r,A}}\left.\frac{\partial\Delta\tilde{\phi}_{E|A}}{\partial X_E}\right|_{X_E=0} \quad (S.51)$$

where $\lambda_{D,A}$ is the Debye length of the anode and $\varepsilon_{r,A}$ is the relative permittivity of the anode material. By plugging (S.38) and (S.48) into (S.51) the following relation holds:

$$\sqrt{\frac{2\xi}{5}\left(\left|\frac{\Delta\tilde{\phi}_A^0}{\xi} + 1\right|^{\frac{5}{2}} - 1\right) - \Delta\tilde{\phi}_A^0} = \frac{\lambda_{D,A}\varepsilon_{r,E}}{\lambda_{D,E}\varepsilon_{r,A}}\sqrt{\beta_v \ln\left(\frac{e^{\Delta\tilde{\phi}_{E|A}^0} + \beta_v - 1}{\beta_v}\right) - \Delta\tilde{\phi}_{E|A}^0} \quad (S.52)$$

# Section F Supplementary Tables

| | Transport of the Charged Species | Equation No. |
|---|---|---|
| Poisson-Nernst-Planck Equations | $$z_i e \frac{\partial c_i}{\partial t} + \nabla \cdot \boldsymbol{j}_i = 0$$ $$\boldsymbol{j}_i = -\sigma_i \nabla \tilde{\mu}_i^* = -\sigma_i \nabla(\mu_i^* + \phi)$$ $$-\varepsilon_r \varepsilon_0 \nabla^2 \phi = \sum_i z_i e c_i$$ | (1) |
| | Mechanics | |
| Equilibrium Equation | $\text{div } \boldsymbol{\sigma} = -(\varepsilon_r \varepsilon_0 \nabla^2 \phi)\nabla \phi$ | (9) |
| Stress Tensor | $\sigma_{mn} = 2G\epsilon_{mn} + (\kappa \epsilon_{kk} - \zeta c_i^*)\delta_{mn}$ | (10) |
| Strain Tensor | $\epsilon_{mn} = \frac{1}{2}\left(\frac{\partial u_m}{\partial x_n} + \frac{\partial u_n}{\partial x_m}\right)$ | (11) |
| | Interfacial Reaction Kinetics | |
| Electrolyte\|Cathode Interface | $$J_{Li} = J_C^0 \left(\exp\left(\frac{\alpha_C \Delta g_C^*}{V_{th}}\right) - \exp\left(-\frac{(1-\alpha_C)\Delta g_C^*}{V_{th}}\right)\right)$$ $$J_C^0 = \frac{(J_{E\to C}^0)^{\alpha_C}(J_{C\to E}^0)^{1-\alpha_C}}{\gamma_C^\ddagger}$$ $$\Delta g_C^* = \mu_{Li}^*(E) - \tilde{\mu}_-^*(C) + \tilde{\mu}_v^*(E) + V_{oc}$$ | (19) and (20) |
| Anode\|Electrolyte Interface | $$J_{Li} = J_A^0 \left(\exp\left(\frac{\alpha_A \Delta g_A^*}{V_{th}}\right) - \exp\left(-\frac{(1-\alpha_A)\Delta g_A^*}{V_{th}}\right)\right)$$ $$J_A^0 = \frac{(J_{A\to E}^0)^{\alpha_A}(J_{E\to A}^0)^{1-\alpha_A}}{\gamma_A^\ddagger}$$ $$\Delta g_A^* = -\mu_{Li}^*(E) + \tilde{\mu}_-^*(A) - \tilde{\mu}_v^*(E)$$ | (29) and (30) |

Table S.1 Summary of the equations introduced in section 2 (general model)

| | Transport of Li | Equation No. |
|---|---|---|
| Cathode | $\dfrac{\partial c_{Li}}{\partial t} + \dfrac{\partial}{\partial x}\left(-\tilde{D}_{chem}\dfrac{\partial c_{Li}}{\partial x}\right) = 0$ | (32) |
| | Charge Transfer Kinetics | |
| Electrolyte\|Cathode Interface | $j_{Li^+} = j_C^0\left(\exp\left(\dfrac{\alpha_C \eta_C}{V_{th}}\right) - \exp\left(-\dfrac{(1-\alpha_C)\eta_C}{V_{th}}\right)\right)$ <br><br> $j_C^0 = (k_C^0)_{EN}(y_{Li}^s)^{1-\alpha_C}(1-y_{Li}^s)^{\alpha_C}$ | (34) and (35) |
| Anode\|Electrolyte Interface | $j_{Li^+} = j_A^0\left(\exp\left(\dfrac{\alpha_A \eta_A}{V_{th}}\right) - \exp\left(-\dfrac{(1-\alpha_A)\eta_A}{V_{th}}\right)\right)$ <br><br> $j_A^0 = (k_A^0)_{EN}$ | (36) and (37) |
| | Cell Voltage | |
| Overall | $V_{cell} = V_{oc}(c_{Li}|_{x=L_E+\delta x_C}) - \eta_C - \eta_A - \eta_{R_C} - \eta_{R_E}$ | (39) |
| Open Circuit | $V_{oc}(c_{Li}|_{x=L_E+\delta x_C}) = -\dfrac{\mu_{Li}^0}{e} + V_{th}\ln(\gamma_{Li}\tilde{c}_{Li}|_{x=L_E+\delta x_C})$ | (40) |
| Ohmic Loss in the Electrolyte and Cathode | $\eta_{R_E} = \dfrac{j_{tot}L_E}{\sigma_E}$ <br><br> $\eta_{R_C} = \dfrac{j_{tot}L_C}{\sigma_C}$ | (41) |

Table S.2 Summary of the equations introduced in section 3 (electroneutral model)

| | Governing Equations for the Space Charge Layers | Equation No. |
|---|---|---|
| Electrolyte | $\tilde{c}_v = \dfrac{\beta_v e^{\Delta\tilde{\phi}_E}}{e^{\Delta\tilde{\phi}_E} + \beta_v - 1}$ $\quad$ $-\dfrac{\partial^2 \Delta\tilde{\phi}_E}{\partial X_E^2} = 1 - \dfrac{\beta_v e^{\Delta\tilde{\phi}_E}}{e^{\Delta\tilde{\phi}_E} + \beta_v - 1}$ | (44) and (45) |
| Cathode | $\tilde{c}_h = \dfrac{\beta_h e^{-\Delta\tilde{\phi}_C}}{e^{-\Delta\tilde{\phi}_C} + \beta_h - 1}$ $\tilde{c}_{Li^+} = \dfrac{\beta_{Li^+} e^{-\Delta\tilde{\phi}_C}}{e^{-\Delta\tilde{\phi}_C} + \beta_{Li^+} - 1}$ $\quad$ $-\dfrac{\partial^2 \Delta\tilde{\phi}_C}{\partial X_C^2} = \dfrac{e^{-\Delta\tilde{\phi}_C}}{e^{-\Delta\tilde{\phi}_C} + \beta_h - 1} + \dfrac{e^{-\Delta\tilde{\phi}_C}}{e^{-\Delta\tilde{\phi}_C} + \beta_{Li^+} - 1} - 1$ | (49) and (50) |
| Anode | $\tilde{c}_-^{2/3} = \dfrac{\Delta\tilde{\phi}_A}{\xi} + 1$ $\quad$ $-\dfrac{\partial^2 \Delta\tilde{\phi}_A}{\partial X_A^2} = 1 - \left|\dfrac{\Delta\tilde{\phi}_A}{\xi} + 1\right|^{\frac{3}{2}}$ | (54) and (55) |
| | Matching Conditions for the Space Charge Layers | |
| Electrolyte\|Cathode Interface | $\left.\dfrac{\partial \Delta\tilde{\phi}_C}{\partial X_C}\right|_{X_C=0} = \dfrac{\lambda_{D,C}\varepsilon_{r,E}}{\lambda_{D,E}\varepsilon_{r,C}} \left.\dfrac{\partial \Delta\tilde{\phi}_{E|C}}{\partial X_E}\right|_{X_E=0}$ $\quad$ $\Delta\tilde{\phi}_{EC} = \Delta\tilde{\phi}_{E|C}^0 - \Delta\tilde{\phi}_C^0$ | (60) and (62) |
| Anode\|Electrolyte Interface | $\left.\dfrac{\partial \Delta\tilde{\phi}_A}{\partial X_A}\right|_{X_A=0} = \dfrac{\lambda_{D,A}\varepsilon_{r,E}}{\lambda_{D,E}\varepsilon_{r,A}} \left.\dfrac{\partial \Delta\tilde{\phi}_{E|A}}{\partial X_E}\right|_{X_E=0}$ $\quad$ $\Delta\tilde{\phi}_{AE} = \Delta\tilde{\phi}_A^0 - \Delta\tilde{\phi}_{E|A}^0$ | (61) and (63) |
| | Charge Transfer Kinetics | |
| Electrolyte\|Cathode Interface | $j_{Li^+} = j_C^0 \left(\exp\left(\dfrac{\alpha_C \eta_C}{V_{th}}\right) - \exp\left(-\dfrac{(1-\alpha_C)\eta_C}{V_{th}}\right)\right)$ $j_C^0 = (k_C^0)_{EN} (y_{Li}^s)^{1-\alpha_C} (1 - y_{Li}^s)^{\alpha_C} \dfrac{(\tilde{c}_v^s)^{\alpha_C}(\beta_v - \tilde{c}_v^s)^{1-\alpha_C}}{(\beta_v - 1)^{1-\alpha_C}}$ | (35) and (65) |
| Anode\|Electrolyte Interface | $j_{Li^+} = j_A^0 \left(\exp\left(\dfrac{\alpha_A \eta_A}{V_{th}}\right) - \exp\left(-\dfrac{(1-\alpha_A)\eta_A}{V_{th}}\right)\right)$ $j_A^0 = (k_A^0)_{EN} \dfrac{(\tilde{c}_v^s)^{\alpha_A}(\beta_v - \tilde{c}_v^s)^{1-\alpha_A}}{(\beta_v - 1)^{1-\alpha_A}}$ | (37) and (66) |

Table S.3 Summary of the equations introduced in section 4 (non-electroneutral model)

| | General Governing Equation and Constitutive Relation | Equation No. |
|---|---|---|
| Equilibrium Equation | $\frac{\partial}{\partial x}\left(\sigma_{xx} - \frac{\varepsilon_0 \varepsilon_r}{2}\left(\frac{\partial \phi}{\partial x}\right)^2\right) = 0$ | (67) |
| Stress Tensor | $\sigma_{xx} = 2G\epsilon_{xx} + \kappa\epsilon_{xx} - \zeta c_i^*$<br>$\sigma_{yy} = \kappa\epsilon_{xx} - \zeta c_i^*$<br>$\sigma_{zz} = \kappa\epsilon_{xx} - \zeta c_i^*$ | (68) |
| | Governing Equation for the Bulk of the Cathode | |
| Bulk of the Cathode | $\sigma_{xx} = P_x$ | (71) |
| | Governing Equations for the Space Charge Layers | |
| Electrolyte | $\frac{\partial}{\partial X_E}\left(\Delta\tilde{\sigma}_{xx,E} + \frac{1}{2}\left(\frac{\partial \Delta\tilde{\phi}_E}{\partial X_E}\right)^2\right) = 0$<br>$\ln\left(\frac{\tilde{c}_v(\beta_v - \tilde{c}_v)}{\beta_v - 1}\right) + \Delta\tilde{\phi}_E + \Omega_E c_v^0 \Delta\tilde{\sigma}_{h,E} = 0$ | (74) and (76) |
| Cathode | $\frac{\partial}{\partial X_C}\left(\frac{\Delta\tilde{\sigma}_{xx,C}}{\beta_{Li^+}} + \frac{1}{2}\left(\frac{\partial \Delta\tilde{\phi}_C}{\partial X_C}\right)^2\right) = 0$<br>$\ln\left(\frac{\tilde{c}_{Li^+}(\beta_{Li^+} - \tilde{c}_{Li^+})}{\beta_{Li^+} - 1}\right) + \Delta\tilde{\phi}_C + \Omega_C c_{Li^+}^0 \Delta\tilde{\sigma}_{h,C} = 0$ | (77) and (79) |
| Anode | $\frac{\partial}{\partial X_A}\left(\Delta\tilde{\sigma}_{xx,A} + \frac{1}{2}\left(\frac{\partial \Delta\tilde{\phi}_A}{\partial X_A}\right)^2\right) = 0$ | (80) |
| | Charge Transfer Kinetics | |
| Electrolyte\|Cathode Interface | $j_{Li^+} = j_C^0 \left(\exp\left(\frac{\alpha_C(\eta_C + \sigma_h(\Omega_C + \Omega_E)/e)}{V_{th}}\right)\right.$<br>$\left. - \exp\left(-\frac{(1-\alpha_C)(\eta_C + \sigma_h(\Omega_C + \Omega_E)/e)}{V_{th}}\right)\right)$<br>$j_C^0 = (k_C^0)_{EN}(y_{Li}^s)^{1-\alpha_C}(1-y_{Li}^s)^{\alpha_C}\frac{(\tilde{c}_v^s)^{\alpha_C}(\beta_v - \tilde{c}_v^s)^{1-\alpha_C}}{(\beta_v - 1)^{1-\alpha_C}}$ | (65) and (83) |
| Anode\|Electrolyte Interface | $j_{Li^+} = j_A^0 \left(\exp\left(\frac{\alpha_A(\eta_A - \sigma_h\Omega_E/e)}{V_{th}}\right)\right.$<br>$\left. - \exp\left(\frac{-(1-\alpha_A)(\eta_A - \sigma_h\Omega_E/e)}{V_{th}}\right)\right)$<br>$j_A^0 = (k_A^0)_{EN}\frac{(\tilde{c}_v^s)^{\alpha_A}(\beta_v - \tilde{c}_v^s)^{1-\alpha_A}}{(\beta_v - 1)^{1-\alpha_A}}$ | (66) and (84) |

Table S.4 Summary of the equations introduced in section 5 (electro-chemo-mechanical model)

| Items | Electroneutral Model | Non-Electroneutral Model |
|---|---|---|
| | Sum of square residual (unit $V^2$) | |
| 0.24 mA/cm² | $1.82 \times 10^{-4}$ | $4.47 \times 10^{-5}$ |
| 0.35 mA/cm² | $2.95 \times 10^{-4}$ | $1.46 \times 10^{-4}$ |
| 0.48 mA/cm² | $2.32 \times 10^{-4}$ | $5.76 \times 10^{-5}$ |
| Total | $7.10 \times 10^{-4}$ | $2.48 \times 10^{-4}$ |

Table S.5 The sum of square residuals of the electroneutral and non-electroneutral models.

**Section G Supplementary Figures**

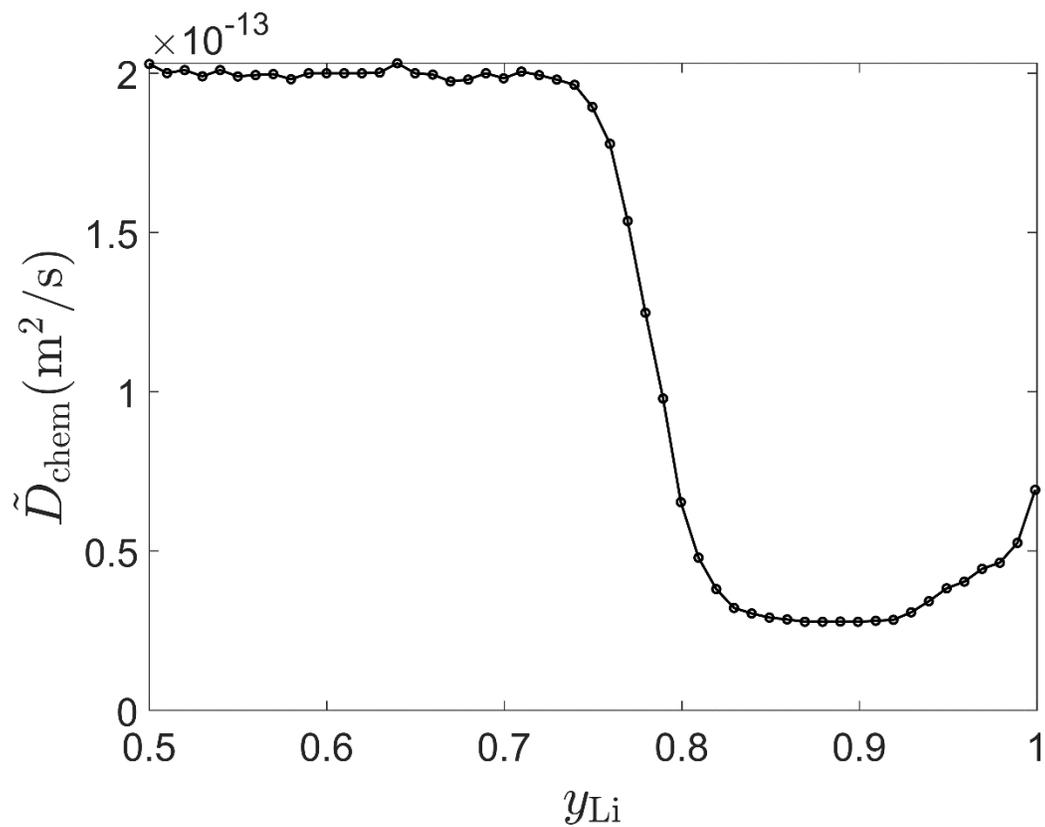

Figure S.4 The variation of chemical diffusivity $\widetilde{D}_{chem}$ with respect to the state of charge $y_{Li}$ for the $Li_yCoO_2$ cathode as obtained from Fabre et al. [2]

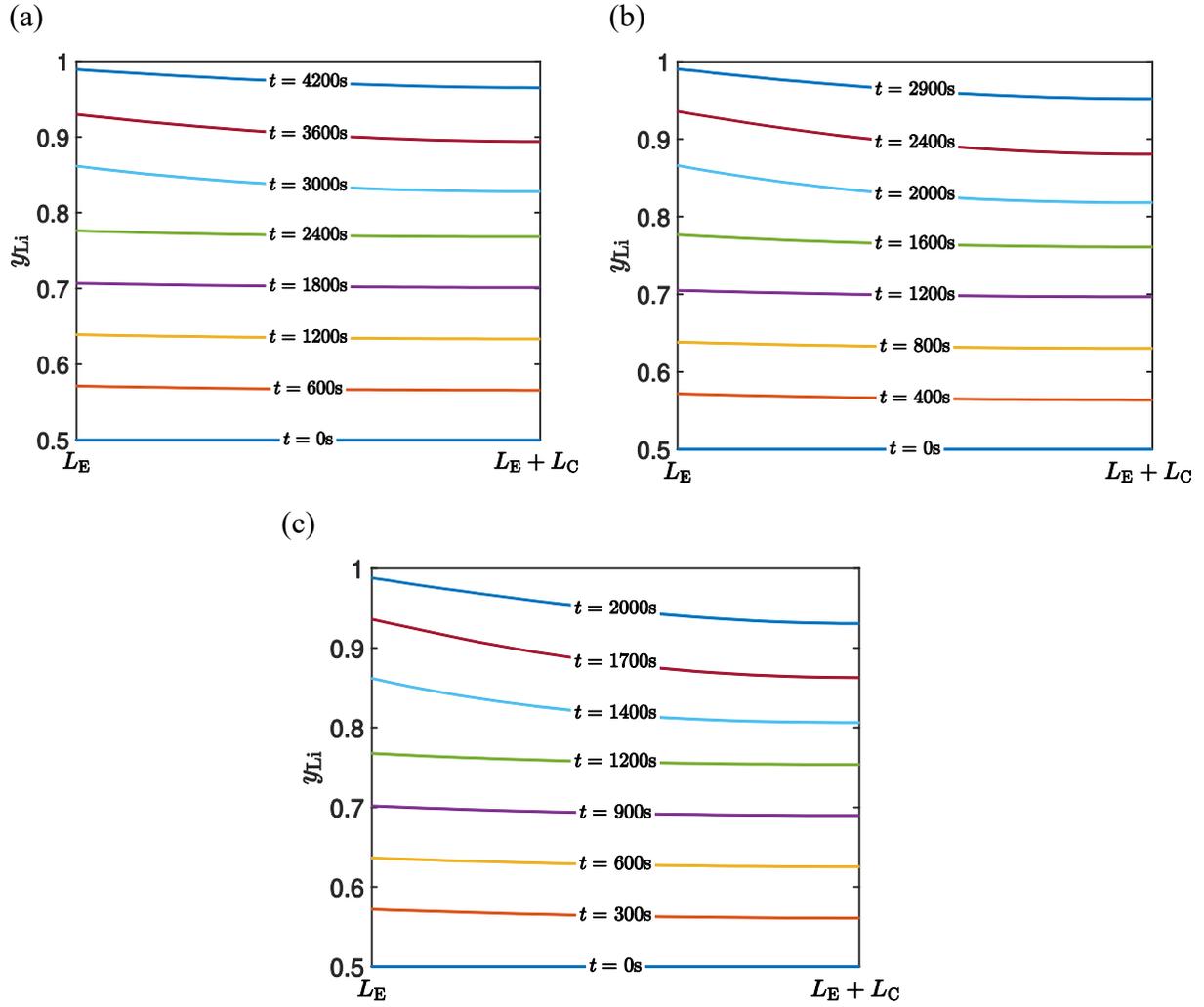

Figure S.5 The Li concentration profile in the cathode computed with the electroneutral model for $j_{\text{tot}} = 0.24$ mA/cm², panel (a), 0.35 mA/cm², panel (b), and 0.48 mA/cm², panel (c).

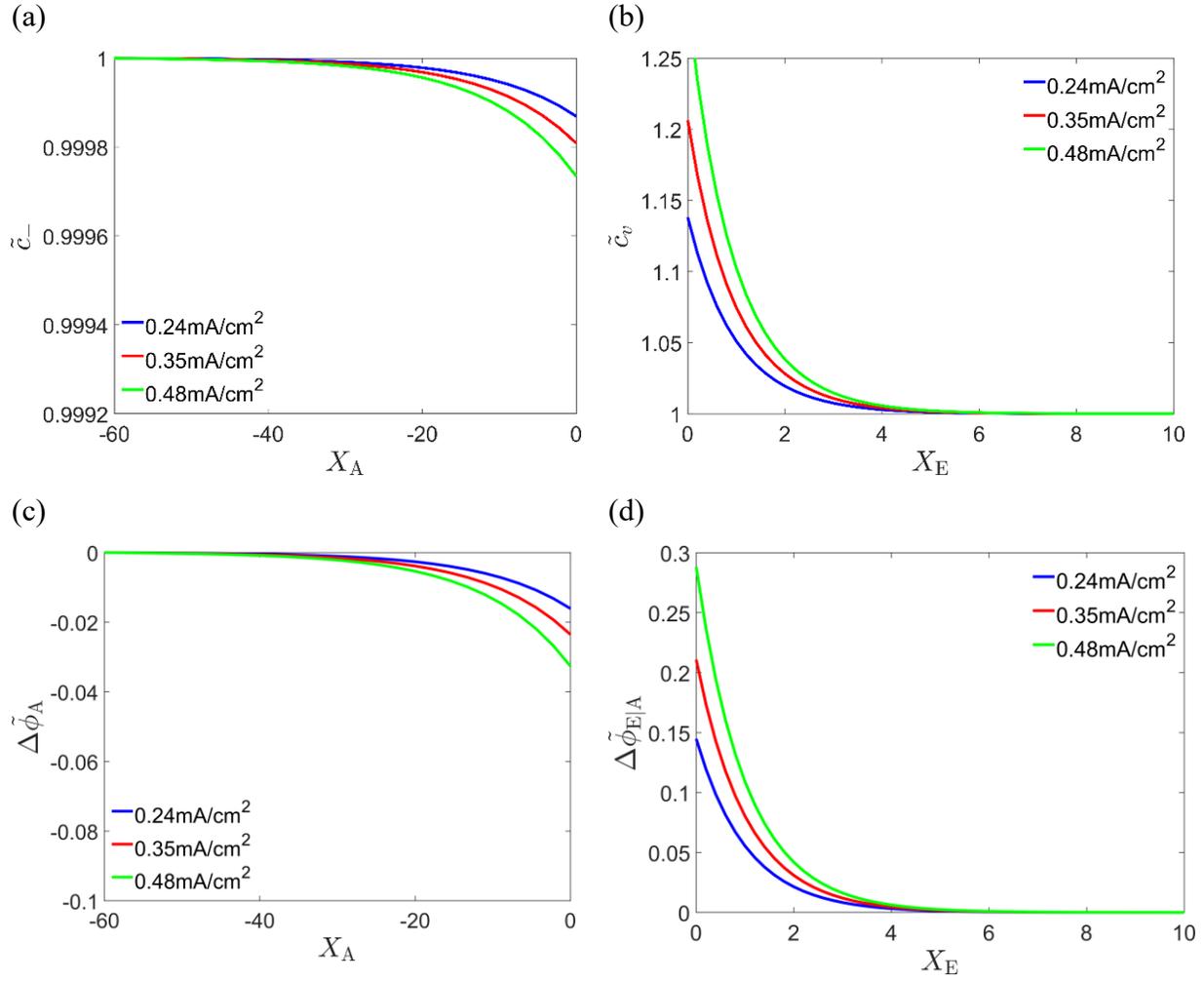

Figure S.6 The SCL at the anode|electrolyte interface computed with the non-electroneutral model at various current densities and $\varepsilon_{r,A} = 10000$. Panels (a) and (c) show the normalized concentration of electrons and electric potential, respectively, in the anode side of the SCL. Panels (b) and (d) show the normalized concentration of Li vacancies and electric potential, respectively, in the electrolyte side of the SCL.

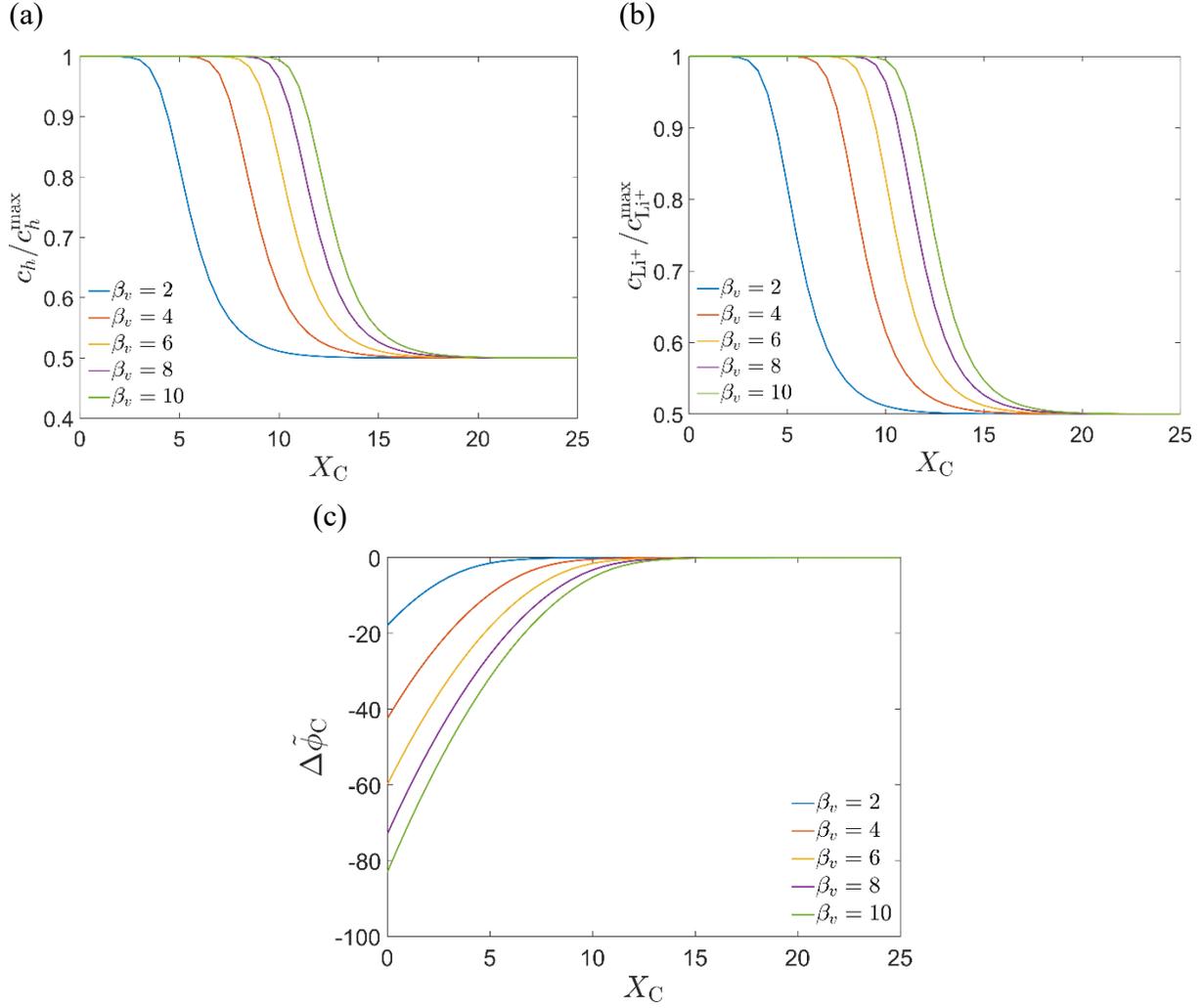

Figure S.7 Influence of $\beta_v$ on the concentration of holes and Li ions and the electric potential at the cathode side of the electrolyte|cathode SCL for $y_{\text{Li}}^0 = 0.5$. Panels (a) and (b) show the normalized holes and lithium ion concentrations, respectively. Panel (c) shows the normalized electrical potential.

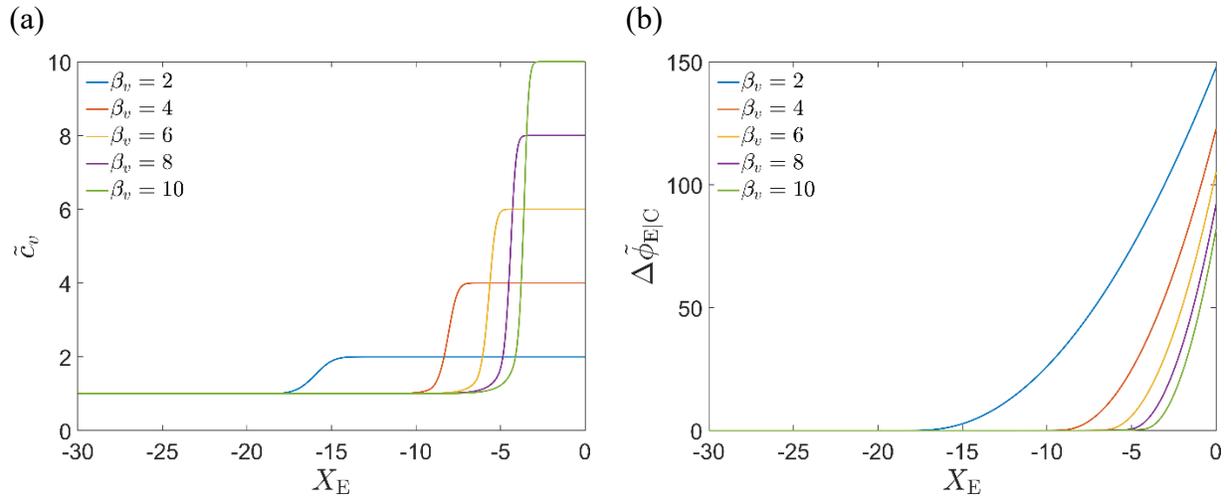

Figure S.8 Influence of $\beta_v$ on the concentration of holes and Li ions and the electric potential in the cathode side of the electrolyte|cathode SCL. Panel (a) shows the normalized lithium vacancies concentration. Panel (b) shows the normalized electrical potential.

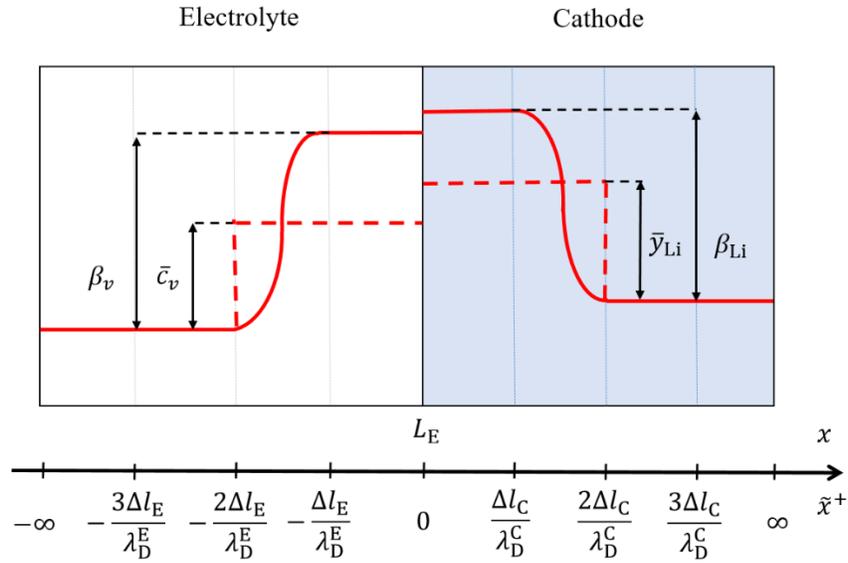

Figure S.9  Illustration of averaging the concentration of the charged species in the SCL of the electrolyte|cathode interface. The red solid lines denote the concentration profile of Li vacancies and Li in the electrolyte and cathode, respectively. The red dash lines denote the averaged concentrations applied for solving the charge transfer equation. The lighter vertical lines denote the lattice plane of each domain.

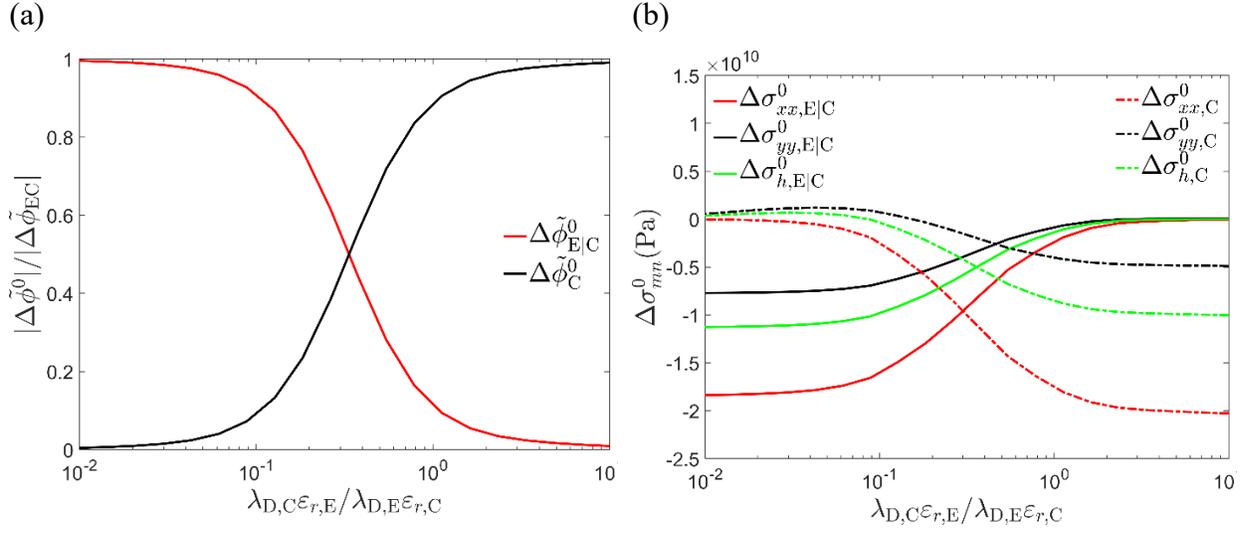

Figure S.10 The electrical potential and stresses exerted at the electrolyte|cathode interface ($X_E = 0$ and $X_C = 0$) as a function of $\lambda_{D,C}\varepsilon_{r,E}/\lambda_{D,E}\varepsilon_{r,C}$. Panel (a) shows the $\Delta\phi^0_{E|C}$ and $\Delta\phi^0_C$ of the electrolyte|cathode SCL for various $\lambda_{D,C}\varepsilon_{r,E}/\lambda_{D,E}\varepsilon_{r,C}$. Panel (b) shows the stress difference $\Delta\sigma^0_{mn,E|C}$ and $\Delta\sigma^0_{mn,C}$ of the electrolyte|cathode SCL for various $\lambda_{D,C}\varepsilon_{r,E}/\lambda_{D,E}\varepsilon_{r,C}$. The solid and dashed lines refer to the $\Delta\sigma$ at the electrolyte and cathode side of the electrolyte|cathode interface, respectively.

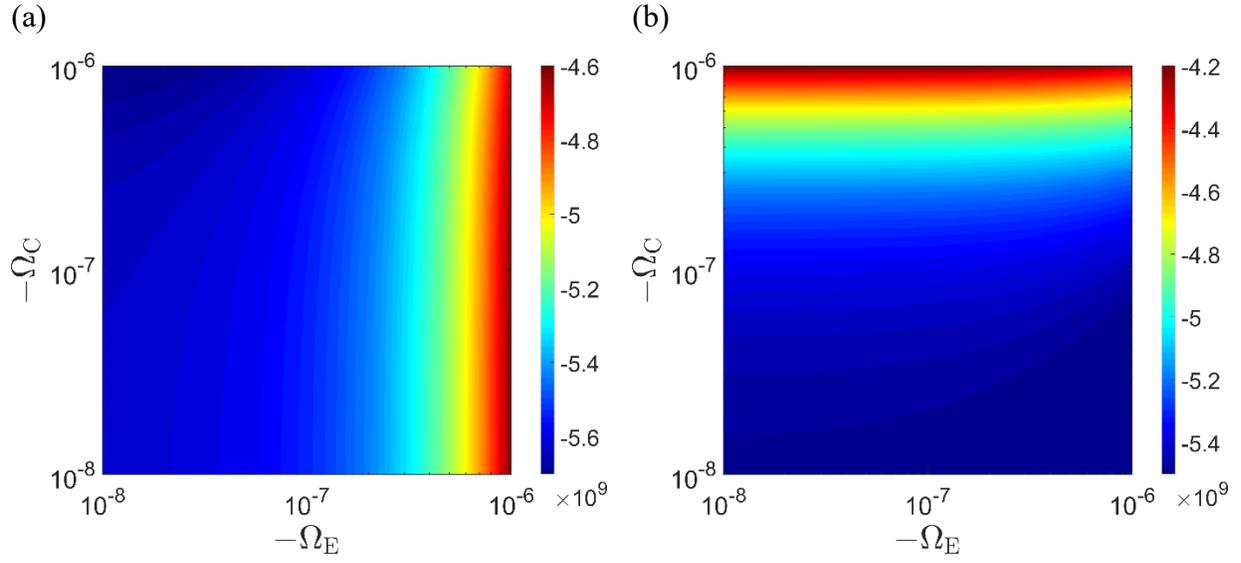

Figure S.11. Hydrostatic stresses exerted at $X_E = 0$ and $X_C = 0$ (electrolyte|cathode interface) as a function of $\Omega_E$ and $\Omega_C$. Panels (a) and (b) plot $\Delta\sigma_{h,E}^0$ and $\Delta\sigma_{h,C}^0$ (in Pa), respectively. The charge transfer equation is ECM coupled. Also, $y_{Li}^0 = 0.5$ and the discharge current density is 0.24 mA/cm².